\documentclass[aps, prb, twocolumn, titlepage, longbibliography,showpacs,superscriptaddress]{revtex4-2}
\usepackage{amsmath,amssymb,amsfonts,bm}
\usepackage{graphicx}
\usepackage{epstopdf}
\usepackage{dcolumn}
\usepackage{mathrsfs}
\usepackage{bbold}
\usepackage{dsfont}
\usepackage{float}
\usepackage[colorlinks=true,linkcolor=blue,citecolor=blue, urlcolor=blue,bookmarks=false]{hyperref}
\usepackage{url}
\usepackage{makecell}
\usepackage{tgtermes}
\usepackage{multirow}
\usepackage{natbib}
\usepackage{accents}
\usepackage{caption}
\usepackage{ulem}
\usepackage{siunitx}
\usepackage{overpic}    
\usepackage{subcaption}
\usepackage{textcase}

\makeatletter
\def\@bib@pub@addr#1#2{%
  #1% 出版社
  \ifx#2\@empty\else
    , #2% 将下划线改为逗号
  \fi}
\makeatother

\begin{document}
\title{Bilateral hydrogenation induced high-Chern-number quantum anomalous Hall state in monolayer Cr$_{\text{2}}$Ge$_{\text{2}}$Te$_{\text{6}}$}
\date{\today }
\author{Xiang Li}
\affiliation{School of Quantum, University of Chinese Academy of Sciences, Beijing 100190, China}
\author{Xin-Wei Yi}
\affiliation{Institute of Theoretical Physics, Chinese Academy of Sciences, Beijing, China}
\author{Jing-Yang You}
\affiliation{Peng Huanwu Collaborative Center for Research and Education, Beihang University, Beijing 100191, China}
\author{Jia-Wen Li}
\affiliation{Laboratory of Theoretical and Computational Nanoscience, National Center for Nanoscience and Technology, Chinese Academy of Sciences, Beijing 100190, China}
\author{Qing-Han Yang}
\affiliation{School of Quantum, University of Chinese Academy of Sciences, Beijing 100190, China}
\author{Gang Su}\email{gsu@ucas.ac.cn}
\affiliation{School of Quantum, University of Chinese Academy of Sciences, Beijing 100190, China}
\affiliation{Institute of Theoretical Physics, Chinese Academy of Sciences, Beijing, China}
\affiliation{School of Physical Sciences, University of Chinese Academy of Sciences, Beijing 100049, China}
\affiliation{Physical Science Laboratory, Huairou National Comprehensive Science Center, Beijing 101400, China}
\author{Bo Gu}\email{gubo@ucas.ac.cn}
\affiliation{School of Quantum, University of Chinese Academy of Sciences, Beijing 100190, China}
\affiliation{Physical Science Laboratory, Huairou National Comprehensive Science Center, Beijing 101400, China}
\begin{abstract}
The pursuit of high-temperature quantum anomalous Hall (QAH) insulators faces fundamental challenges, 
including narrow topological gaps and low Curie temperatures ($T_{\text{C}}$) in existing materials. 
Here, we propose a transformative strategy using bilateral hydrogenation to engineer a robust QAH state 
in the topologically trivial ferromagnetic 
semiconductor Cr$_{\text{2}}$Ge$_{\text{2}}$Te$_{\text{6}}$ via covalent orbital reconstruction. 
First-principles calculations reveal that 
by fundamentally rewiring the underlying orbital hybridization network, 
hydrogenation alters orbital occupations to shift preexisting
Dirac points—originally embedded in the conduction 
bands—to the vicinity of the Fermi level in Cr$_{\text{2}}$Ge$_{\text{2}}$Te$_{\text{6}}$H$_{\text{6}}$.
This electronic restructuring, coupled with spin-orbit coupling, 
opens a global topological gap of 118.1 meV, establishing a robust QAH state with Chern number $C=$ 3.
Concurrently, this same orbital reconstruction effectively tunes the energy difference 
between the ligand $p$ and transition metal $d$ orbitals. This specific energy shift enhances
ferromagnetic superexchange via the $d_{z^{2}}$-$p_{z}$-$d_{xz}$ channel, 
significantly strengthening the nearest-neighbor coupling $J_{\text{1}}$ by 3.06 times 
and switching $J_{\text{2}}$ from antiferromagnetic to ferromagnetic. 
Monte Carlo simulations based on the extracted exchange parameters indicate a 
pronounced enhancement of ferromagnetic stability compared with pristine Cr$_{\text{2}}$Ge$_{\text{2}}$Te$_{\text{6}}$,
we stress that the absolute Curie temperature depends on the mapping to an effective spin 
model and thus should be interpreted primarily in terms of relative trends.
The comparative enhancement of ferromagnetic stability 
after hydrogenation is a salient effect of this orbital tuning.
This work establishes targeted orbital reconstruction driven by surface hydrogenation as a powerful route to 
simultaneously control topology and magnetism in 2D materials, 
providing a general route to engineer QAH phases with large gaps and high Chern numbers in van der Waals ferromagnetic semiconductors.
\end{abstract}

\maketitle
\section{\NoCaseChange{INTRODUCTION}}
The quantum anomalous Hall (QAH) effect is a profound topological phenomenon offering dissipationless chiral edge states without 
an external magnetic field, driven by broken time-reversal symmetry 
and spin-orbit coupling (SOC) \cite{RMP-QAHE, HaldaneModel, Review-QAHE-QikunXue-2018, Review-QAHE-XiaoliangQi-2016}. 
This holds immense promise for low-power spintronic applications.
Initially observed in Cr-doped (Bi, Sb)$_{\text{2}}$Te$_{\text{3}}$ thin films at 30 mK \cite{EXP-CrBiSb2Te3-QAHE},
these systems suffer from magnetic disorder, limiting observation to below 2 K \cite{EXP-VBiSb2Te3-QAHE, EXP-CrBiSb2Te3-QAHE-2, 
EXP-CrBiSb2Te3-QAHE-3, EXP-CrBiSb2Te3-QAHE-5, EXP-CrBiSb2Te3-QAHE-6, EXP-CrBiSb2Te3-QAHE-4-up2K, 
EXP-CrVBiSb2Te3-QAHE-Codoping, Model-DFT-CrBiSb2Te3-QAHE}. 
Moiré materials like twisted graphene \cite{EXP-QAHE-moire-twistedGr, EXP-QAHE-TwistedGr-2, EXP-QAHE-TwistedGr-3}, 
MoTe$_{\text{2}}$/WSe$_{\text{2}}$ \cite{EXP-QAHE-moire-MoTe2WSe2}, 
and BN/graphene heterostructures \cite{EXP-QAHE-Gr+BN} also exhibit QAH effect but 
face challenges such as small topological gaps or low Curie temperatures $T_{\text{C}}$.
A significant advance is the discovery of QAH in the intrinsic magnetic van der Waals (vdW)
material MnBi$_{\text{2}}$Te$_{\text{4}}$ \cite{EXP-QAHE-MnBi2Te4, EXP-QAHE-MnBi2Te4-2, EXP-QAHE-MnBi2Te4-3}. 
Its ground state is intralayer ferromagnetism (FM) and interlayer antiferromagnetism (AFM), requiring an 
external magnetic field to align the layers ferromagnetically
to achieve a QAH state, yet its temperature remains below 10 K 
due to weak intralayer FM exchange \cite{EXP-QAHE-MnBi2Te4}. 
While theoretical calculations predict the potential of QAH heterostructures combining 
topological insulators with 2D ferromagnetic semiconductors \cite{QAHE-Model-QuantumWell, DFT-QAHE-Ge+CGT, 
DFT-QAHE-Ge+CGT-1}, 
even experimental examples like the MnBi$_{\text{2}}$Te$_{\text{4}}$/Bi$_{\text{2}}$Te$_{\text{3}}$ heterostructure \cite{EXP-MBT-Bi2Te3-Heterostructure-QAHE} 
are still limited by the low $T_{\text{C}}$ of the ferromagnetic component. 
This issue is pervasive, as currently available intrinsic ferromagnetic semiconductors also exhibit low $T_{\text{C}}$ in experiments 
\cite{Cr2Ge2Te6-FMsemiconductor-Tc30K, CrI3-FMsemiconductor-Tc45K, CrSBr-FMsemiconductor-Tc146K, CrBr3-FMsemiconductor-Tc34K, 
CrSiTe3-FMsemiconductor-Tc80K, CrCl3-FMsemiconductor-Tc17K, Cr2S3-FMsemiconductor-Tc75K-1, Cr2S3-FMsemiconductor-Tc75K-2}. 
Therefore, a central goal in the theoretical search for new QAH systems is to identify 
high-$T_{\text{C}}$ QAH insulators \cite{PtBr3-FM-QAHE, DFT-QAHE-PtCl3, DFT-LiFeSe-QAHE, 
DFT+Model-P-orbit-MagneticTI-QAHE, DFT-QAHE-Kagome, DFT-QAHE-2Dorganic}, 
or high-$T_{\text{C}}$ ferromagnetic semiconductors \cite{Cr2Ge2Te6-FMsemiconductor-EnhancementTc-Strain, 
Cr2Ge2Te2+PtSe2-FM-TcEnhancement, MnBi2Te4-FM-ElectricField-TcEnhancement, DFT-CGT-Strain-EnhancedTc, DFT-TST-family-Ferromagnetic-Semiconductors,
MagneticSemiconductor-Review-GroupPaper, Cr3O6-FM-Semiconductor-DFT, Review-EXP-StrainedDSM-JOS,EXP-StrainedCGT-Tc-Enhancd} for use in heterostructures.

Surface functionalization has emerged as a powerful tool for engineering the structural, electronic and magnetic
properties of 2D materials \cite{Review-Functionalization-Graphene, 2013-Review-Graphene-Hydrogenation, 
ReviewBook-Silicene-Functionalization, Review-Functionalization-Graphene, 
Review-Ge-Si-Sn-hydrogenation, Review-Silicene-Functionalization,
EXP-BN-Fluorinated-magnetism, EXP-Graphene-Fluorinated, EXP-TMDs-2Hto1T-metaltosemicon-highcoverage,
DFT-2HMoS2-Hydrogenation-2, DFT-2HMoS2-Hydrogenation-3,
DFT-Silicene-monolayer-single-surface-adsorp-F, DFT-Silicene-monolayer-two-side-Functionalization, DFT-Hydrogenation-Si-Ge}. 
By decorating their surfaces with adatoms, it is predicted to induce plenty phenomena, 
including enhanced FM \cite{DFT-CrI3-Tc-enhanced-Functionalization-O, DFT-cgt-MolecularAdsorption-Tc-enhanced, DFT-cgt-singlesite-adsorption-Tc-enhanced}, 
the emergence of magnetism in non-magnetic hosts \cite{DFT-Graphene-Semihydrogenation-Ferromagnetism, DFT-BN-Hydrogenation-monolayer}, 
the tuning of band gaps \cite{DFT-IIIBi-monolayer-Hydrogenated-fluorinated-TI, 
DFT-monolayer-Hydrogenated-TI, DFT-monolayer-GaBiCl2-TI-5strain,
DFT-Ti3N2F2-MXene-TI-monolayer, DFT-M2CO2-MXene-TI-monolayer, DFT-TI-monolayer-dumbbellStanene-Functionalization,
DFT-TI-monolayer-BiX/SbX-HFClBr, DFT-III-monochalcogenides-Functionalization-O-TI, DFT-BN-Hydrogenation-monolayer-2},
the creation of novel topological phases
\cite{DFT-NbSeH2-monolayer-magnetism,DFT-IIIBi-monolayer-QAHE-Functionalization-magnetism, DFT-IIIBi-monolayer-QSHE-Functionalization,
DFT-Bi-Sb-111-monolayer-QAHE-Functionalization-magnetism, DFT-Sn-monolayer-halfadsorp-HighTemp-QAHE-Functionalization-magnetism,
DFT-FunctionalizationBi-III-QAHE, DFT-C3N-monolayer-magnetism, DFT-MoSe2F2-LargeStrain, DFT-MoSe2F2-QAHE, DFT-Sn-monolayer-H-QAHE, 
DFT-MoTe2F2-QAHE-QuadraticBdCrossing} and strain \cite{DFT-monolayer-GaBiCl2-TI-5strain, DFT-MoSe2F2-LargeStrain}.
Hydrogenation is a particularly effective functionalization tool. Theoretical calculations predict 
hydrogenated graphene (Gr) monolayers exhibit FM \cite{DFT-Graphene-Semihydrogenation-Ferromagnetism}, 
structural distortion \cite{DFT-Graphene-Hydrogenation-magnetism-distortion} and a band gap \cite{DFT-Hydrogenation-graphene-gapopen},
which are strongly supported by experimental evidence, 
demonstrating that large-area and reversible surface hydrogenation of graphene is achievable \cite{2013-Review-Graphene-Hydrogenation, 
EXP-Hydrogenation-graphene, EXP-Hydrogenation-graphene-gapopen, EXP-Graphene-Adsorption-SingleSide, 
EXP-Graphene-Hydrogenation-EnhacnedSOC, EXP-Graphene-Hydrogenation-Passivation-RoomTemp, EXP-Graphene-Hydrogenation-magnetism}.
Similar effects, such as inducing ferromagnetism and triggering structural or electronic phase transitions, 
have been experimentally demonstrated in other 2D systems like boron nitride (BN) 
\cite{EXP-BN-Hydrogenation-Oneside, EXP-BN-Hydrogenation-OneSide-ClusterDiscussion, 
EXP-BN-Hydrogenation-OneSide-2020, EXP-BN-Hydrogenation-OneSide-OnNi111}, 
germanene (Ge) \cite{EXP-Germanene-Hydrogenation-Bulk-2, EXP-Germanene-Hydrogenation-Bulk-3, 
EXP-Germanene-Hydrogenation-FewLayers, EXP-Germanene-Hydrogenation-Bulk-4, 
EXP-Germanene-Hydrogenation-Bulk-5, EXP-Germanene-Hydrogenation-Bulk-6, 
EXP-Ge-Hydrogenation-lattice-change, EXP-Germanene-Hydrogenation-Bulk-6},
silicene (Si) \cite{EXP-Silicene-Hydrogenation-oneside-STM, 
EXP-Silicene-Hydrogenation-oneside-2, 
EXP-Silicene-Hydrogenation-Oneside-2016-APRES-HighCoverage50,
EXP-Silicene-Hydrogenation-Si6H6-twoside, EXP-Silicene-Hydrogenation-Twoside-1996, EXP-Silicene-Hydrogenation-Twoside-1997, 
EXP-Silicene-Hydrogenation-Bulk, EXP-Silicene-Hydrogenation-2022}, 
Bi$_{\text{2}}$Te$_{\text{3}}$ \cite{EXP-Hydrogenation-Bi2Te3-revisible-preventroxidation}, 
Sn$_{\text{2}}$Bi \cite{EXP-Sn2Bi-Hydrogenation-semiconductor}
and transition metal dichalcogenides (TMDs) monolayers 
\cite{EXP-2H-MoS2-Hydrogenation-Ferromagnetism, EXP-2H-MoS2-Hydrogenation-PhaseTransition-Twoside, 
EXP-1991-MoS2-Hydrogenation-uptake0.23, EXP-1980-MoS2-Hydrogenation-uptake0.37, 
EXP-MoS2-hydrogenation-np-type-Vacancy, EXP-MoSe2-Hydrogenation}.

Beyond chemical passivation, surface hydrogenation serves as a physical mechanism 
for covalent orbital reconstruction. By forming covalent bonds with surface atoms and displacing 
lone-pair electrons or unpassivated $p$ electrons, adatoms fundamentally rewire 
the underlying orbital hybridization network \cite{DFT-Graphene-Semihydrogenation-Ferromagnetism,
DFT-Graphene-Hydrogenation-magnetism-distortion,DFT-Hydrogenation-graphene-gapopen,
EXP-Hydrogenation-graphene, EXP-Hydrogenation-graphene-gapopen, 
EXP-Graphene-Hydrogenation-EnhacnedSOC, EXP-Graphene-Hydrogenation-Passivation-RoomTemp, EXP-Graphene-Hydrogenation-magnetism}. 
This alters orbital occupations and induces targeted charge transfer into the highly localized orbitals
(such as the $d$-orbitals in transition metals \cite{DFT-NbSeH2-monolayer-magnetism} or $p_z$-orbitals in main-group elements). 
While manipulating magnetic properties, such orbital reconstruction also holds the promise of inducing topological phase 
transitions \cite{DFT-NbSeH2-monolayer-magnetism, DFT-Bi-Sb-111-monolayer-QAHE-Functionalization-magnetism, 
DFT-Sn-monolayer-halfadsorp-HighTemp-QAHE-Functionalization-magnetism, DFT-Sn-monolayer-H-QAHE}. 
To demonstrate this principle, Cr$_{\text{2}}$Ge$_{\text{2}}$Te$_{\text{6}}$ 
is selected as a prototypical material platform. Structurally, its magnetic Cr atoms form a honeycomb 
lattice—an archetypal motif for hosting Dirac physics. 
Furthermore, its magnetic exchange depends on 
the energy difference between the ligand $p$ and transition metal $d$ orbitals \cite{Cr2Ge2Te2+PtSe2-FM-TcEnhancement}. 
Surface hydrogenation can deliberately tune these levels via covalent orbital 
reconstruction. Coupled with its clean semiconducting background, it provides an ideal 
platform to modify and map both magnetic and topological properties.

In this work, we employ first-principles calculations to demonstrate that bilateral hydrogenation of 
monolayer Cr$_{\text{2}}$Ge$_{\text{2}}$Te$_{\text{6}}$—initially a topologically trivial ferromagnetic semiconductor— 
drives a QAH insulating state with a high Chern number.
Rather than a conventional band-inversion scenario, this transition is directly governed by the aforementioned covalent orbital reconstruction. 
The resulting charge transfer lowers the chromium valence state (from Cr$^{3+}$ to Cr$^{2+}$), an electronic effect that shifts preexisting Dirac crossings, 
originally embedded in the spin-polarized conduction bands, toward the Fermi level. SOC then gaps these Dirac fermions and excavates the chiral edge states 
from the bulk band gap, yielding a global topological gap of 118.1 meV and a Chern number $C=$ 3.
Simultaneously, the deliberate tuning of the ligand $p$ and transition metal $d$ energy difference directly modifies the superexchange 
pathways (mainly the $d_{z^{2}}$-$p_{z}$-$d_{xz}$ channel). This manifests as drastic enhancements in the magnetic 
exchange interaction ($J_1$) and magnetic anisotropy, ultimately elevating the Curie temperature. Therefore, the magnetic enhancement 
and the topological phase transition are not separate phenomena, but concurrent consequences of the same fundamental covalent orbital reconstruction, 
highlighting surface functionalization as a powerful knob to 
simultaneously tailor magnetic interactions and topological electronic structures in two-dimensional van der Waals magnets.

\section{\NoCaseChange{COMPUTATIONAL METHODS}}
First-principles calculations in this work were performed with the projector augmented wave (PAW) method \cite{PAWmethod} based
on the DFT as implemented in the Vienna ab initio simulation package (VASP) \cite{VASP}. The choice of the electron exchange-correlation
functional was generalized gradient approximation (GGA) with the form of Perdew-Burke-Ernzerhof (PBE) realization \cite{PBE}. 
Lattice constants and atomic positions were fully relaxed until the maximum force acting on all atoms
was less than 1 $\times$ 10$^{-\text{3}}$ eV/$\mathring{\text{A}}$ and the 
total energy was converged to 1 $\times$ 10$^{-\text{7}}$ eV 
with the Gaussian smearing method. Calculations of exchange coupling $J_{i}$ and
magnetic single-ion anisotropy energy $E_{\text{SIA}}$
were performed by using the 2 $\times$ 2 $\times$ 1 supercell. 
The $\Gamma$-centered Monkhorst-Pack k-point mesh \cite{M-PKpoints} of 
size 15 $\times$ 15 $\times$ 1 (45 $\times$ 45 $\times$ 1 for density of states calculations) 
was used for the Brillouin zone (BZ) sampling 
in structure optimization and self-consistent processes of exchange coupling calculations.
The single-ion anisotropy energy $E_{\text{SIA}}$ was calculated with the inclusion of 
spin-orbit coupling and 5 $\times$ 5 $\times$ 1 $\Gamma$-centered Monkhorst-Pack k-point mesh.
The plane-wave cutoff energy was set to be 450 eV. The electron correlation of the 3$d$ transition atom Cr was considered by
using the DFT $+$ $U$ method introduced by Dudarev et al. \cite{LDAUTYPE=2}.
Results in the main text and Supplemental Material were obtained with $U=$ 4 eV otherwise mentioned particularly. 
The Monte Carlo simulations were performed on Heisenberg model with the single-ion anisotropy. The 20
$\times$ 20 $\times$ 1 supercells with periodic boundary conditions and magnetic sites 800 were adopted. 
Each temperature calculation used 10$^{\text{4}}$ Monte Carlo steps to achieve equilibrium \cite{MonteCarloMethod1, MonteCarloMethod2}. 
The Wannier90 package \cite{Wannier90, Review-Maximally-Localized-Wannier90} was utilized to construct maximally localized Wannier functions for 
superexchange analysis. Topological properties are calculated by using Wanniertools \cite{WannierTools}.

\section{\NoCaseChange{RESULTS}}

\subsection{\NoCaseChange{Adsorption energy of Cr$_{\text{2}}$Ge$_{\text{2}}$Te$_{\text{6}}$H$_{\text{6}}$}}
To assess the thermodynamic stability of hydrogenated materials, the adsorption energy $E_{\text{Ads}}$ 
is calculated by using the formula $E_{\text{Ads}}=[E(X+n\text{H})-nE(\text{H}_{2})/2-E(X)]/n$
\cite{DFT-III-monochalcogenides-Functionalization-O-TI, DFT-CrI3-Tc-enhanced-Functionalization-O,
DFT-cgt-singlesite-adsorption-Tc-enhanced, DFT-Ge-monolayer-adsorption-energy, DFT-MoS2-adsorption-energy},
where $E(X)$, $E(X+n\text{H})$, and $E(\text{H}_{2})$ \cite{DFT-H-H-Dissociation-energy-4.48eV}
represent the total energies of the pristine host compound $X$, 
the hydrogenated compound $X\text{H}_{n}$ with $n$ hydrogen atoms, and an isolated hydrogen molecule, respectively. 
A positive $E_{\text{Ads}}$ corresponds to an endothermic process, while a negative $E_{\text{Ads}}$ 
represents an exothermic process that is energetically more favorable for hydrogenation.

\begin{figure}[H]
	\captionsetup[figure]{name={FIG.},labelformat=simple,labelsep=period,singlelinecheck=off}
	\centering
	\includegraphics[width=1\columnwidth]{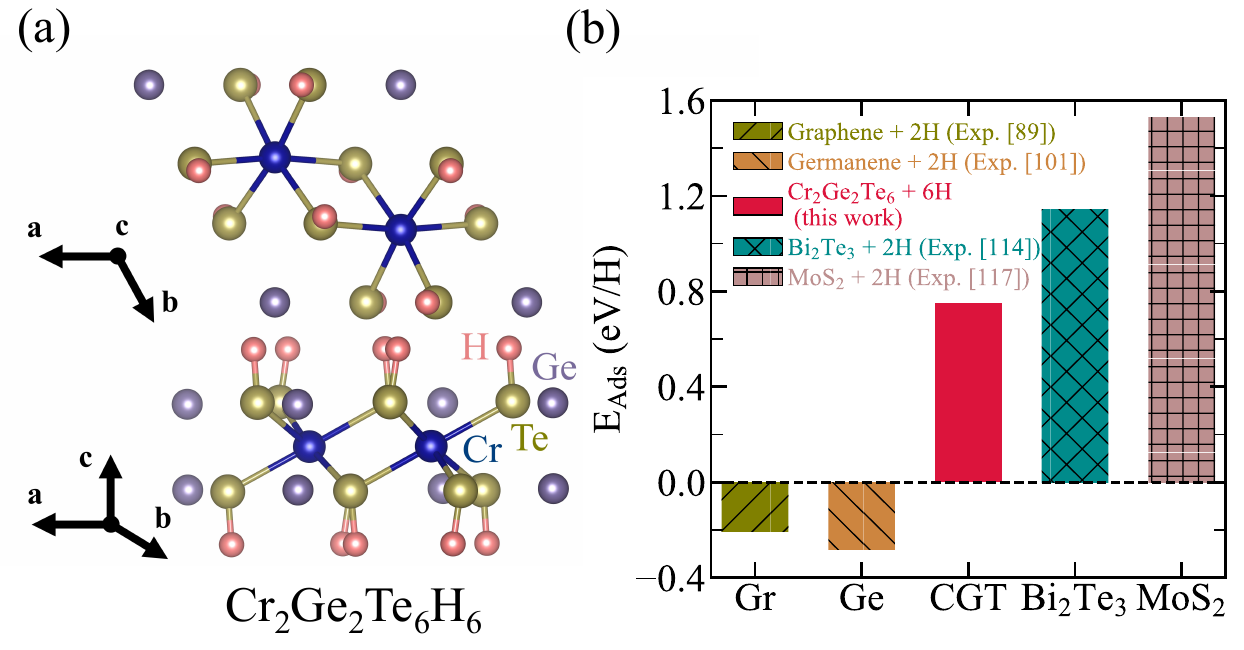}
	\caption{(a) Top view and side view of the lattice structure of the hydrogenated ferromagnetic semiconductor monolayer
	Cr$_{\text{2}}$Ge$_{\text{2}}$Te$_{\text{6}}$ + 6H (denoted as 
	Cr$_{\text{2}}$Ge$_{\text{2}}$Te$_{\text{6}}$H$_{\text{6}}$). 
	(b) Adsorption energies $E_{\text{Ads}}$ (eV/H) 
	for Cr$_{\text{2}}$Ge$_{\text{2}}$Te$_{\text{6}}$ (CGT) + 6H and 
	several other fully-hydrogenated monolayer materials, including graphene (Gr) + 2H, germanene (Ge) + 2H, 
	Bi$_{\text{2}}$Te$_{\text{3}}$ + 2H, and 2H-MoS$_{\text{2}}$ + 2H, 
	where more negative (lower) values indicate higher thermodynamic stability. 
	Experimental precedents demonstrating the structural feasibility of bilateral hydrogen 
	adsorption in these related materials (e.g., via the hydrogenation of 
	free-standing monolayers, bulk van der Waals crystals, or topmost structural layers) are 
	supported by Refs. \cite{EXP-Germanene-Hydrogenation-FewLayers, EXP-Hydrogenation-graphene,
	EXP-Hydrogenation-Bi2Te3-revisible-preventroxidation,
	EXP-2H-MoS2-Hydrogenation-PhaseTransition-Twoside}.}
	\phantomsubcaption{\label{fig: fig1-a}} 
    \phantomsubcaption{\label{fig: fig1-b}} 
\end{figure}

To identify the most favorable adsorption site on the Cr$_{\text{2}}$Ge$_{\text{2}}$Te$_{\text{6}}$ (CGT)
monolayer, the adsorption of a single hydrogen atom on three distinct sites including adsorption 
on the top of Cr, Ge, and Te atoms are investigated,
as depicted in the Supplemental Material \cite{Supplement}.
Our findings reveal that hydrogen preferentially forms a bond with the Te atom compared with other adsorption configurations. 
Given this preferential adsorption, for higher hydrogenation coverage, specifically the fully-hydrogenated 
Cr$_{\text{2}}$Ge$_{\text{2}}$Te$_{\text{6}}$ monolayer with six hydrogen atoms 
(denoted as Cr$_{\text{2}}$Ge$_{\text{2}}$Te$_{\text{6}}$H$_{\text{6}}$), 
all hydrogen atoms are considered to bond with the Te atoms,
as illustrated in Fig. \ref{fig: fig1-a}.
The H-Te bond lengths in Cr$_{\text{2}}$Ge$_{\text{2}}$Te$_{\text{6}}$H$_{\text{6}}$ are 
essentially equal at 1.693 $\mathring{\text{A}}$, demonstrating its ordered structure.
Hydrogenation significantly changes the lattice structure of Cr$_{\text{2}}$Ge$_{\text{2}}$Te$_{\text{6}}$. 
The optimized lattice constant of Cr$_{\text{2}}$Ge$_{\text{2}}$Te$_{\text{6}}$H$_{\text{6}}$ 
expands to $a = \text{8.072}$ $\mathring{\text{A}}$
compared to $a_{\text{0}} = \text{6.690}$ $\mathring{\text{A}}$ in Cr$_{\text{2}}$Ge$_{\text{2}}$Te$_{\text{6}}$.
This expansion is accompanied by a suppressed height of the Te atom plane.

For comparative analysis, the adsorption energies for other hydrogenated systems with experimental precedents, 
including graphene (Gr) + 2H \cite{EXP-Hydrogenation-graphene}, 
germanene (Ge) + 2H \cite{EXP-Germanene-Hydrogenation-FewLayers}, 
Bi$_{\text{2}}$Te$_{\text{3}}$ + 2H \cite{EXP-Hydrogenation-Bi2Te3-revisible-preventroxidation}, 
and 2H-MoS$_{\text{2}}$ + 2H \cite{EXP-2H-MoS2-Hydrogenation-PhaseTransition-Twoside}, 
are calculated, as depicted in Fig. \ref{fig: fig1-b}.
For these materials, hydrogen atoms are modeled as being adsorbed on both sides of the monolayers 
to provide a direct structural comparison.
The exothermic nature of fully-hydrogenated graphene and germanene 
is corroborated by their exceptional experimental thermal stability. For instance, breaking the C-H bonds 
in plasma-synthesized graphane requires extended annealing above 450$^{\circ}$C \cite{EXP-Hydrogenation-graphene}. 
Similarly, topochemically synthesized hydrogenated germanene only begins to decompose and release hydrogen 
above 200$^{\circ}$C to 250$^{\circ}$C \cite{EXP-Germanene-Hydrogenation-FewLayers}. 
These high desorption temperatures indicate thermodynamic energy minima, aligning 
with the negative adsorption energies ($-$0.21 eV/H for Gr + 2H and $-$0.28 eV/H for Ge + 2H) derived from our calculations.

Crucially, this thermodynamic picture also clarifies the synthesis 
feasibility of Cr$_{\text{2}}$Ge$_{\text{2}}$Te$_{\text{6}}$H$_{\text{6}}$. 
The endothermic nature of direct H$_{\text{2}}$ adsorption creates a dissociation barrier, 
precluding the use of ambient H$_{\text{2}}$ gas. While realizing free-standing monolayers 
with $100\%$ bilateral hydrogenation remains challenging, structural precedents exist: hydrogen 
can efficiently permeate the van der Waals gaps of bulk 2H-MoS$_{\text{2}}$ and Bi$_{\text{2}}$Te$_{\text{3}}$ to 
achieve bilateral adsorption \cite{EXP-2H-MoS2-Hydrogenation-PhaseTransition-Twoside, 
EXP-Hydrogenation-Bi2Te3-revisible-preventroxidation}. Notably, the thermodynamic barrier 
for Cr$_{\text{2}}$Ge$_{\text{2}}$Te$_{\text{6}}$ (0.75 eV/H) is significantly lower than 
those of Bi$_{\text{2}}$Te$_{\text{3}}$ (1.14 eV/H) and MoS$_{\text{2}}$ (1.53 eV/H). Since 
the latter materials have been successfully hydrogenated using high-energy techniques (e.g., 
plasma treatments or H$^{+}$ permeation in acids) to bypass the H$_{\text{2}}$ barrier, the 
experimental realization of bilaterally hydrogenated Cr$_{\text{2}}$Ge$_{\text{2}}$Te$_{\text{6}}$ is 
a realistic prospect.

\subsection{\NoCaseChange{Enhanced ferromagnetic superexchange in Cr$_{\text{2}}$Ge$_{\text{2}}$Te$_{\text{6}}$H$_{\text{6}}$}}
\begin{figure}[H]
	\captionsetup[figure]{name={FIG.},labelformat=simple,labelsep=period,singlelinecheck=off}
	\centering
	\includegraphics[width=1\columnwidth]{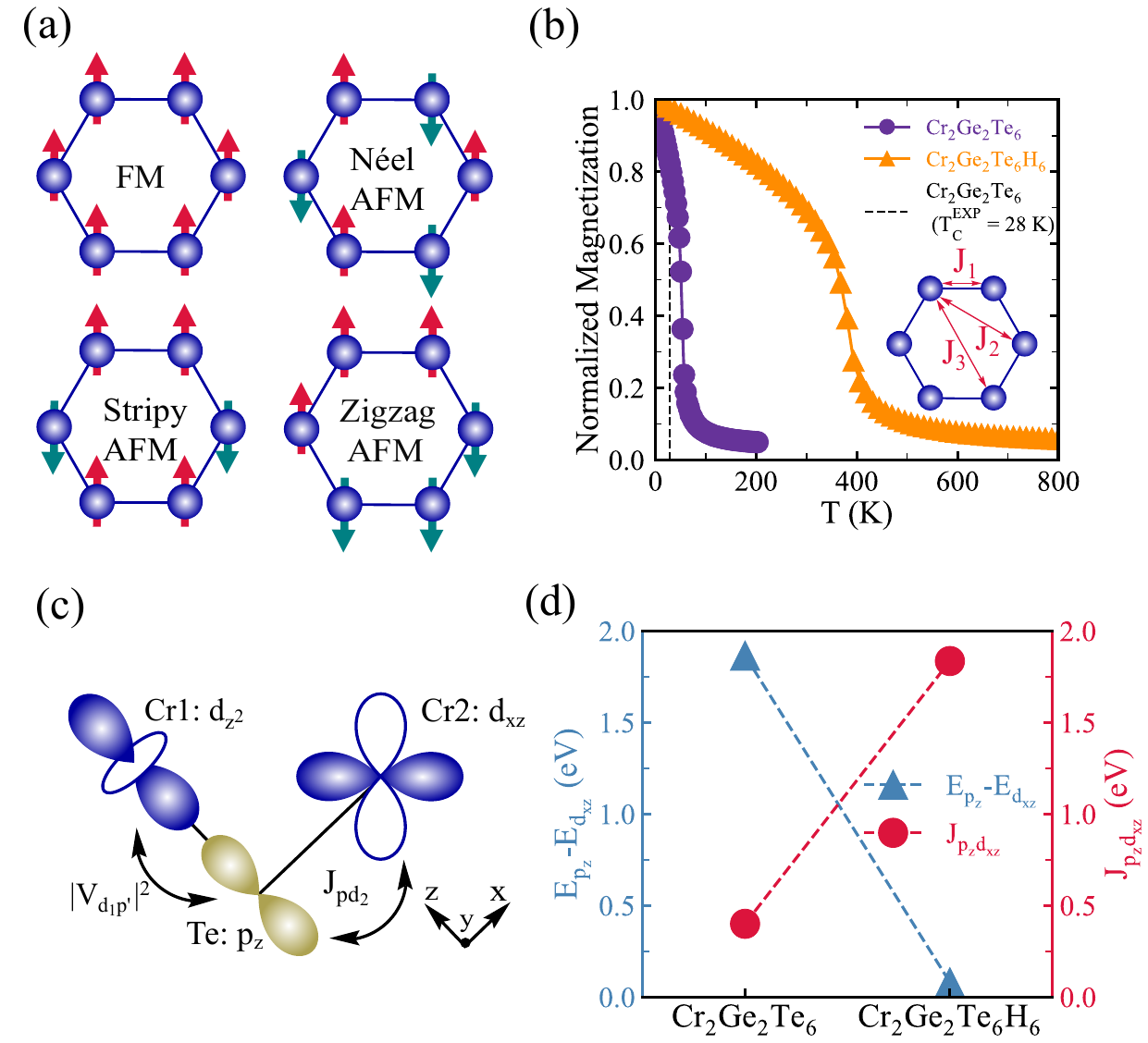}
	\caption{(a) Four magnetic configurations 
	including ferromagnetism (FM), Néel antiferromagnetism (NAFM), 
	stripy antiferromagnetism (SAFM) and zigzag antiferromagnetism (ZAFM)
	used for calculating the magnetic exchange couplings $J_{i}$ of Cr atoms 
	in Cr$_{\text{2}}$Ge$_{\text{2}}$Te$_{\text{6}}$ and 
	Cr$_{\text{2}}$Ge$_{\text{2}}$Te$_{\text{6}}$H$_{\text{6}}$.
	The indices $i=$ 1,2,3 correspond to the nearest, next-nearest neighbor, 
	and third-nearest neighbor, respectively, as illustrated in (b).
	The calculated $J_{i}$ results are listed in Table \ref{tab: J-cst-cgt}. 
	(b) The normalized magnetization as a function of temperature via Monte Carlo simulations, in which the experimental 
	Curie temperature value $T_{\text{C}}^{\text{EXP}}$ of Cr$_{\text{2}}$Ge$_{\text{2}}$Te$_{\text{6}}$ is labled \cite{Cr2Ge2Te6-FMsemiconductor-Tc30K}.
	(c) Illustration of the dominant superexchange interaction channel spin-up $d_{z^{\text{2}}}$-$p_{z}$-$d_{xz}$ 
	between Cr1 and Cr2 cations mediated by an intermediate ligand Te anion.
	$V_{d_{\text{1}}p^{\prime}}$ represents the hopping integral from Te-$p^{\prime}$ to Cr1-$d_{\text{1}}$ and 
	$J_{pd_{\text{2}}}$ is the direct exchange coupling of Te-$p$ and Cr2-$d_{\text{2}}$ according to Eq. \eqref{eq: J1super}.
	(d) Variations in the orbital energy difference $E_{p_{z}}-E_{d_{xz}}$ and direct exchange coupling $J_{p_{z}d_{xz}}$
	for Cr$_{\text{2}}$Ge$_{\text{2}}$Te$_{\text{6}}$ and
	Cr$_{\text{2}}$Ge$_{\text{2}}$Te$_{\text{6}}$H$_{\text{6}}$.}
	\phantomsubcaption{\label{fig: fig2-a}} 
	\phantomsubcaption{\label{fig: fig2-b}} 
	\phantomsubcaption{\label{fig: fig2-c}}
	\phantomsubcaption{\label{fig: fig2-d}}  
\end{figure}

The magnetism is investigated by the energy mapping 
method \cite{DFT-CGT-Strain-EnhancedTc, PtBr3-FM-QAHE, Cr2Ge2Te2+PtSe2-FM-TcEnhancement, 
Method-Energy-Mapping, Cr2Ge2Te6-FMsemiconductor-EnhancementTc-Strain},
magnetic Cr sublattice is mapped to a Heisenberg Hamiltonian with the single-ion anisotropy energy $E_{\text{SIA}}$
written as 
\begin{align}\label{eq: cgt-H-J1J2J3} 
	H=&-J_{1}\sum_{\langle i,j\rangle_{1}}\bm{S}_{i}\cdot\bm{S}_{j}
	-J_{2}\sum_{\langle i,j\rangle_{2}}\bm{S}_{i}\cdot\bm{S}_{j}\cr
	&-J_{3}\sum_{\langle i,j\rangle_{3}}\bm{S}_{i}\cdot\bm{S}_{j}
	-E_{\text{SIA}}\sum_{i}S_{i,z}^{2}+E_{0},
\end{align}
where $\bm{S}_{i}$ and $\bm{S}_{j}$ are unit vectors of
local magnetic moment at sites $i$ and $j$, 
$J_{\text{1}}$, $J_{\text{2}}$ and $J_{\text{3}}$ 
denote the exchange coupling for the nearest, next-nearest, and third-nearest neighbor pairs $\langle i,j\rangle_{p}$ ($p=\text{1, 2, 3}$), respectively, 
as indicated in Fig. \ref{fig: fig2-b},
$E_{0}$ is the non-magnetic energy component.
Four distinct magnetic configurations including ferromagnetism (FM), Néel antiferromagnetism (NAFM),
stripy antiferromagnetism (SAFM) and zigzag antiferromagnetism (ZAFM), as depicted in Fig. \ref{fig: fig2-a}, 
with corresponding energy formuae $E_{\text{FM}}=-12J_{1}-24J_{2}-12J_{3}+E_{0}$, 
$E_{\text{Neel}}=12J_{1}-24J_{2}+12J_{3}+E_{0}$,
$E_{\text{Stripy}}=4J_{1}+8J_{2}-12J_{3}+E_{0}$ and
$E_{\text{Zigzag}}=-4J_{1}+8J_{2}+12J_{3}+E_{0}$, incorporating total 
energies from first-principles calculations \cite{PAWmethod, VASP, PBE, M-PKpoints, LDAUTYPE=2}, 
are considered to derive exchange interaction couplings in Eq. \eqref{eq: cgt-H-J1J2J3}.
The single-ion anisotropy energy $E_{\text{SIA}}$, defined as $E_{\text{SIA}}\equiv (E_{\text{FM}\parallel}-E_{\text{FM}\perp})/8$,
is calculated by the total energy difference between in-plane and out-of-plane FM with the inclusion of SOC.

Both Cr$_{\text{2}}$Ge$_{\text{2}}$Te$_{\text{6}}$ and its hydrogenated derivative 
Cr$_{\text{2}}$Ge$_{\text{2}}$Te$_{\text{6}}$H$_{\text{6}}$ consistently exhibit a ferromagnetic ground state.
The exchange coupling parameters 
$J_{\text{1}}$, $J_{\text{2}}$ and $J_{\text{3}}$ are detailed in Table \ref{tab: J-cst-cgt},
where positive values denote ferromagnetic coupling 
and negative values indicate antiferromagnetic coupling.
The nearest coupling $J_{\text{1}}$ in Cr$_{\text{2}}$Ge$_{\text{2}}$Te$_{\text{6}}$H$_{\text{6}}$ 
dramatically increases by 3.06 times, soaring from 17.93 meV 
in Cr$_{\text{2}}$Ge$_{\text{2}}$Te$_{\text{6}}$ to 54.93 meV.
While long-range antiferromagnetic coupling $J_{\text{3}}$ is also strengthened, 
a crucial observation is the transition of the next-nearest coupling
$J_{\text{2}}$ from AFM to FM. Moreover, the $E_{\text{SIA}}$ values are 
0.70 and 5.30 meV/Cr for Cr$_{\text{2}}$Ge$_{\text{2}}$Te$_{\text{6}}$ and 
Cr$_{\text{2}}$Ge$_{\text{2}}$Te$_{\text{6}}$H$_{\text{6}}$, respectively,
while positive $E_{\text{SIA}}$ values indicate an out-of-plane easy magnetization axis for both materials.
This synergistic effect, driven by the dominant enhanced ferromagnetic 
$J_{\text{1}}$, transition from AFM to FM in $J_{\text{2}}$ and enhanced $E_{\text{SIA}}$, 
is directly responsible for the significantly enhanced FM in 
Cr$_{\text{2}}$Ge$_{\text{2}}$Te$_{\text{6}}$H$_{\text{6}}$ and 
indicates a higher Curie temperature $T_{\text{C}}$.

Fig. \ref{fig: fig2-b} illustrates the calculated normalized magnetization as a function of temperature, 
obtained via the Monte Carlo method \cite{MonteCarloMethod1, MonteCarloMethod2} based on model Eq. \eqref{eq: cgt-H-J1J2J3}.
A rescaling method is adopted to provide the improved estimation of Curie temperature.
The calculated Monte Carlo Curie temperature $T_{\text{C}}^{\text{MC}}$ for Cr$_{\text{2}}$Ge$_{\text{2}}$Te$_{\text{6}}$
is 54 K, which we correlate with the experimental value $T_{\text{C}}^{\text{EXP}}$ of 28 K \cite{Cr2Ge2Te6-FMsemiconductor-Tc30K}.
Based on the ratio of 1.93, all calculated $T_{\text{C}}^{\text{MC}}$ are divided by 
this factor. Both resacled and unscaled results are presented in Table \ref{tab: J-cst-cgt}.
$T_{\text{C}}$ of Cr$_{\text{2}}$Ge$_{\text{2}}$Te$_{\text{6}}$H$_{\text{6}}$
is dramatically enhanced to 198 K, which is approximately 7.06 times larger than
that of Cr$_{\text{2}}$Ge$_{\text{2}}$Te$_{\text{6}}$.
Notice the rescaled result is used only as a reference to compare relative trends and should
not be viewed as a quantitative prediction.

The interatomic distances between nearest-neighbor Cr atoms are relatively large
(4.02 $\mathring{\text{A}}$ in Cr$_{\text{2}}$Ge$_{\text{2}}$Te$_{\text{6}}$,
4.66 $\mathring{\text{A}}$ in Cr$_{\text{2}}$Ge$_{\text{2}}$Te$_{\text{6}}$H$_{\text{6}}$),
which strongly limits the direct spatial overlap of their $3d$ orbitals.
This is confirmed by our Wannier-based tight-binding Hamiltonian, 
showing nearly zero direct hopping parameters between adjacent Cr sites (see Supplemental Material \cite{Supplement}).
Consequently, direct exchange interactions between Cr atoms can be ruled 
out as the primary source of the enhanced ferromagnetism.
The great enhanced $T_{\text{C}}$ by hydrogenation in Cr$_{\text{2}}$Ge$_{\text{2}}$Te$_{\text{6}}$
could be understood by the superexchange interaction of Cr1-Te-Cr2 based on 
a four-electron picture ($d_{\text{1}}$, $p^{\prime}$, $p$, $d_{\text{2}}$) 
\cite{Model-Anderson-Superexchange-0, Model-Anderson-Superexchange, Model-Anderson-Superexchange-2, Model-Anderson-Superexchange-3}.
The superexchange interaction of two nearest
magnetic sites Cr1 and Cr2 is expressed as \cite{Model-Anderson-Superexchange-0, Qian2017Ferromagnetism, DFT-QAHE-Ge+CGT, Supplement, Wigner9j}
\begin{align}\label{eq: J1super}
	J_{\text{1}}^{\text{Super}}
	=\frac{1}{4A}\sum_{d_{\text{1}}d_{\text{2}}p^{\prime}p}
	|V_{d_{\text{1}}p^{\prime}}|^{2}J_{pd_{\text{2}}},
\end{align}
which includes two intermediate processes, one is the 
transfer from Te-$p^{\prime}$ to Cr1-$d_{\text{1}}$ orbitals
with hopping integral $V_{d_{\text{1}}p^{\prime}}$,
the other is the direct exchange coupling $J_{pd_{\text{2}}}$ between 
remaining Te-$p$ and Cr2-$d_{\text{2}}$ orbitals.
$A \equiv 1/(1/E_{\uparrow\uparrow}^{2}-1/E_{\uparrow\updownarrow}^{2})$ is taken as 
a pending constant \cite{Model-Anderson-SW}. 
$E_{\uparrow\downarrow}$ is the energy needed if 
$p^{\prime}$ and $d_{\text{1}}$ orbitals form a spin single state while 
$E_{\uparrow\uparrow}$ is the case for spin triplet state.
$J_{pd_{\text{2}}}$ can be derived from Schrieffer-Wolff transiformation 
and is expressed as \cite{Model-Anderson-SW}
\begin{align}
	J_{pd_{\text{2}}}
	=2|V_{pd_{\text{2}}}|^{2}\left(\frac{1}{E_{p}-E_{d_{\text{2}}}}+\frac{1}{E_{d_{\text{2}}}+U-E_{p}}\right),
\end{align}
where $V_{pd_{\text{2}}}$
is the hopping integral from Cr2-$d_{\text{2}}$ to Te-$p$
obtained by constructing tight-binding Hamiltonian with the maximally localized Wannier functions \cite{Wannier90, Review-Maximally-Localized-Wannier90},
$E_{p}$ and $E_{d_{2}}$ are orbital energies obtained by 
the integral of the first-principles electronic density of states, and
$U$ is the Hubbard correlation parameter \cite{LDAUTYPE=2}.
\begin{table}[H]
    \centering
	\caption{Lattice constants a ($\mathring{\text{A}}$), band gaps with the inclusion of 
	spin-orbit coupling $E_{\text{g}}$ (meV),
	the nearest, the next-nearest and the third-nearest 
	exchange couplings $J_{\text{1}}$, $J_{\text{2}}$ and $J_{\text{3}}$ (meV) of 
	two magnetic Cr atoms, single-ion anisotropy energy $E_{\text{SIA}}$ (meV/Cr),
	spin magnetic moments $M_{S}$ ($\mu_{\text{B}}$/Cr) 
	of $d$ orbitals of Cr atom, Monte Carlo Curie temperatures $T_{\text{C}}^{\text{MC}}$ and experimental 
	Curie temperature $T_{\text{C}}^{\text{EXP}}$ (K)
	of Cr$_{\text{2}}$Ge$_{\text{2}}$Te$_{\text{6}}$ and Cr$_{\text{2}}$Ge$_{\text{2}}$Te$_{\text{6}}$H$_{\text{6}}$.
	$E_{\text{SIA}}$ $>$ 0 ($<$ 0) indicates out-of-plane (in-plane) magnetic anisotropy.
	$J_{\text{1,2,3}}>$ 0 ($<$ 0) corresponds to ferromagnetic (antiferromagnetic) coupling.
	The scaling is used only as a reference to compare relative 
	trends and should not be viewed as a quantitative prediction. }\label{tab: J-cst-cgt}
    \renewcommand{\arraystretch}{1.2}
    \setlength{\tabcolsep}{5pt}  
    \small
    \begin{tabular}{lcc}
        \hline\hline
        & Cr$_{\text{2}}$Ge$_{\text{2}}$Te$_{\text{6}}$ 	
		& Cr$_{\text{2}}$Ge$_{\text{2}}$Te$_{\text{6}}$H$_{\text{6}}$\\
        \hline
        $a$ ($\mathring{\text{A}}$) & 6.690 & 8.072\\
        $E_{\text{g}}$ (meV) & 82.1 & 118.1\\
        $J_{\text{1}}$ (meV) & 17.93 & 54.93 \\
        $J_{\text{2}}$ (meV) & $-$1.08 & 5.57 \\
        $J_{\text{3}}$ (meV) & $-$1.40 & $-$3.87\\
        $E_{\text{SIA}}$ (meV/Cr) & 0.70 & 5.30 \\
        $M_{S}$ ($\mu_{\text{B}}$/Cr) & 3.5 & 4.0 \\
        $T_{\text{C}}^{\text{MC}}$ (K) & \makecell[c]{54 ($T_{\text{C}}^{\text{EXP}}$ = 28 K \\ \cite{Cr2Ge2Te6-FMsemiconductor-Tc30K})}& 381\\
		\makecell[l]{$T_{\text{C}}$ scaled to experiment\\
		 for pristine (K)} & 28 & 198\\
        \hline\hline
    \end{tabular}
\end{table}
The $d$ orbitals of Cr atom in Cr$_{\text{2}}$Ge$_{\text{2}}$Te$_{\text{6}}$ and Cr$_{\text{2}}$Ge$_{\text{2}}$Te$_{\text{6}}$H$_{\text{6}}$ 
are splitted into $e_{\text{g}}$ ($d_{z^{2}}$, $d_{x^{2}-y^{2}}$) 
and $t_{\text{2g}}$ ($d_{xz}$, $d_{yz}$, $d_{xy}$), 
owing to the crystal field of the distorted octahedron.
The Cr atom is in high-spin $e_{\text{g}}^{0}t_{\text{2g}}^{3}$ state (Cr$^{3+}$) in Cr$_{\text{2}}$Ge$_{\text{2}}$Te$_{\text{6}}$,
giving a spin magnetic moment $M_{S}$ of 3.5 $\mu_{\text{B}}$, while the hydrogenation
drives the Cr$^{3+}$ to Cr$^{2+}$ with high-spin $e_{\text{g}}^{1}t_{\text{2g}}^{3}$ state,
and enhances the $M_{S}$ to 4 $\mu_{\text{B}}$ in Cr$_{\text{2}}$Ge$_{\text{2}}$Te$_{\text{6}}$H$_{\text{6}}$. 
This change is explicitly reflected in the evolution of the $e_{\text{g}}$ orbital projections from 
the pristine [Fig. \ref{fig: fig3-b} and \ref{fig: fig3-d}] to the hydrogenated band structures 
[Fig. \ref{fig: fig3-f} and \ref{fig: fig3-h}].
The spin-up $e_{\text{g}}$-$p$-$t_{\text{2g}}$ channel, especially $d_{z^{\text{2}}}$-$p_{z}$-$d_{xz}$ channel,
as illustrated in Fig. \ref{fig: fig2-c},
dominants the superexchange interaction, according to fully-occupied spin-up $t_{\text{2g}}$ orbitals of Cr atoms 
for Cr$_{\text{2}}$Ge$_{\text{2}}$Te$_{\text{6}}$ and Cr$_{\text{2}}$Ge$_{\text{2}}$Te$_{\text{6}}$H$_{\text{6}}$ \cite{Supplement}.
The ratio of $J_{1}^{\text{Super}}$ from model Eq. \eqref{eq: J1super} 
is 3.57 for Cr$_{\text{2}}$Ge$_{\text{2}}$Te$_{\text{6}}$H$_{\text{6}}$
to Cr$_{\text{2}}$Ge$_{\text{2}}$Te$_{\text{6}}$, which is compared to 
the ratio of 3.06 from first-principles $J_{\text{1}}$ listed in Table \ref{tab: J-cst-cgt}.
$V_{d_{z^{\text{2}}}p_{z}}$ and $V_{p_{z}d_{xz}}$ for
Cr$_{\text{2}}$Ge$_{\text{2}}$Te$_{\text{6}}$ and Cr$_{\text{2}}$Ge$_{\text{2}}$Te$_{\text{6}}$H$_{\text{6}}$
(see in the Supplemental Material \cite{Supplement}) exhibit no significant difference. 
In contrast, 
the orbital energy difference of $E_{p_{z}}-E_{d_{xz}}$ 
decreases drastically, as depicted in Fig. \ref{fig: fig2-d}.
This dramatic reduction signifies that the $d_{xz}$ orbital becomes exceptionally close to the $p_{z}$ orbital, which, in turn,
enhances the direct exchange interaction $J_{p_{z}d_{xz}}$ and $J_{1}^{\text{Super}}$. Consequently, 
the decrease of the energy difference of $E_{p_{z}}-E_{d_{xz}}$ leads to an increase in the 
Curie temperature in Cr$_{\text{2}}$Ge$_{\text{2}}$Te$_{\text{6}}$H$_{\text{6}}$.

\subsection{\NoCaseChange{Electronic structure and covalent hybridization mechanism in Cr$_{2}$Ge$_{2}$Te$_{6}$H$_{6}$}}
\begin{figure}[H]
	\captionsetup[subfigure]{labelformat=simple}
	\centering
	\includegraphics[width=1\columnwidth]{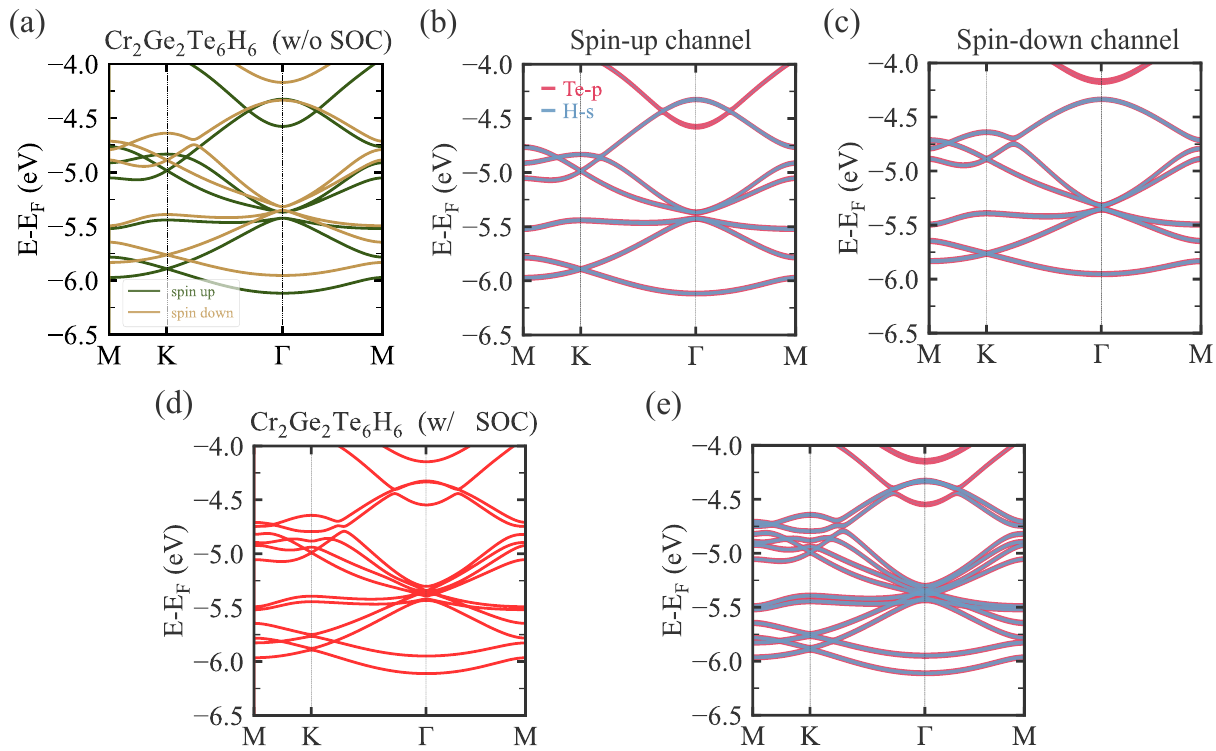}
	\caption{Deep-energy band structures and orbital-resolved fatbands of 
	Cr$_{\text{2}}$Ge$_{\text{2}}$Te$_{\text{6}}$H$_{\text{6}}$. (a) Spin-polarized band structure without SOC. 
	(b), (c) Fatband plots for the spin-up and spin-down channels without SOC (w/o SOC). The red and blue colors 
	highlight the contributions from Te-$p$ and H-$s$ orbitals, respectively. The significant overlap of 
	their weights confirms the strong hybridization and the formation of deep $\sigma_{\text{Te-H}}$ covalent bonds. 
	(d), (e) The band structure and corresponding Te-$p$/H-$s$ fatbands calculated with SOC (w/ SOC).}
	\label{fig: CGTTepHs}
\end{figure}
Hydrogenation in Cr$_{\text{2}}$Ge$_{\text{2}}$Te$_{\text{6}}$ induces a fundamental reconstruction of 
the electronic structure driven by strong covalent hybridization, instead of following a simple rigid-band ionic doping model.
The projected band structure (fatband) analysis depicted in Fig. \ref{fig: CGTTepHs} clearly shows that the occupied 
H-$s$ orbitals do not reside in the conduction band. Instead, they strongly overlap with the Te-$p$ orbitals, pushing the
bonding states $\sigma_{\text{Te-H}}$ deep within the valence band between -4.0 eV and -6.0 eV. As dictated by molecular orbital theory, 
the corresponding unoccupied anti-bonding states $\sigma_{\text{Te-H}}^{*}$ are pushed up into the conduction band. 
This deep energy splitting provides direct evidence that the H adatoms do not simply donate their electrons to the conduction band 
as free carriers; rather, their electrons are fully engaged in forming highly localized, strongly bound covalent bonds $\sigma_{\text{Te-H}}$. 
Because the H-$s$ electrons are locked in these deep bonding states, they are not the ones filling the lower conduction bands.
\begin{figure}[H]
	\captionsetup[subfigure]{labelformat=simple}
	\centering
	\includegraphics[width=1\columnwidth]{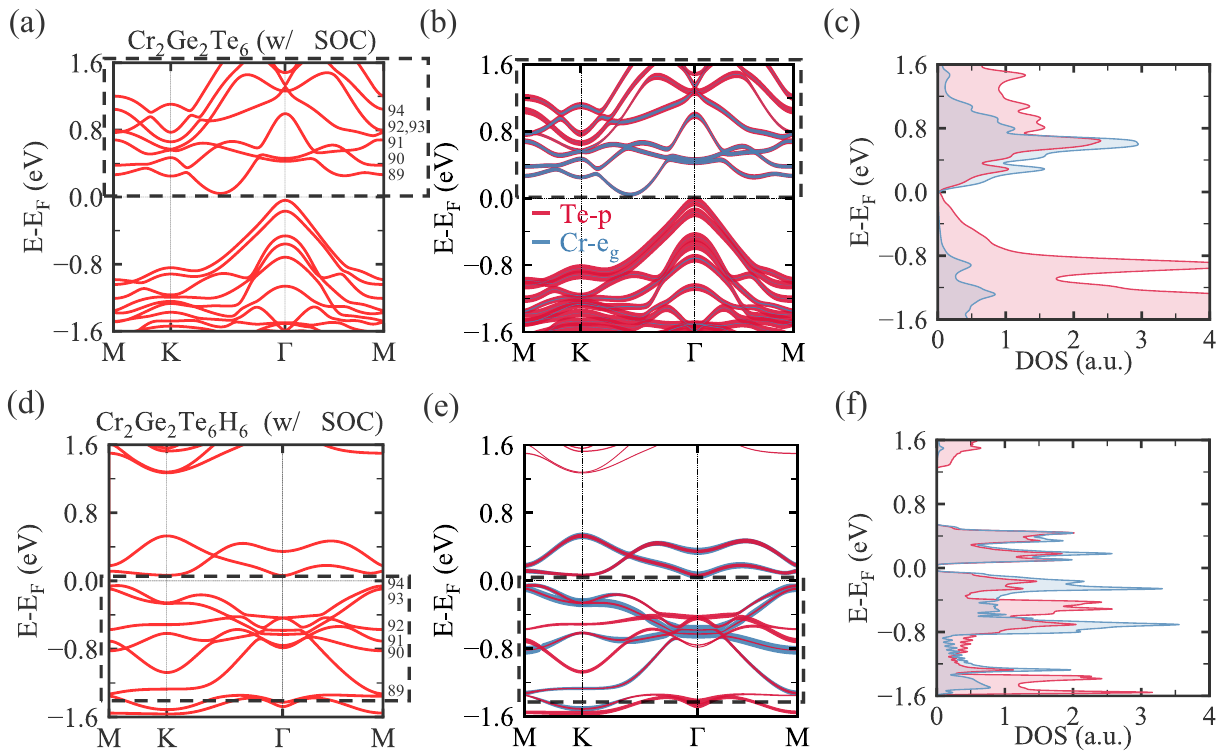}
	\caption{Evolution of the electronic structure and orbital hybridization driven 
	by hydrogenation with spin-orbit coupling (w/ SOC). (a) Band structure of pristine Cr$_{\text{2}}$Ge$_{\text{2}}$Te$_{\text{6}}$. 
	The dashed box highlights the pristine lowest six conduction bands (labeled 89 to 94), which are completely empty and located 
	above the Fermi level. (b) Orbital-resolved fatband structure and (c) partial density of states (PDOS) for 
	pristine Cr$_{\text{2}}$Ge$_{\text{2}}$Te$_{\text{6}}$. The red and blue colors denote the projected weights 
	of Te-$p$ and Cr-$e_{\text{g}}$ orbitals, respectively. (d)-(f) The corresponding band structure, fatband plot, 
	and PDOS for hydrogenated Cr$_{\text{2}}$Ge$_{\text{2}}$Te$_{\text{6}}$H$_{\text{6}}$. Notably, the original 
	conduction bands 89-94 are pulled below the Fermi level to accommodate the extra displaced electrons. 
	The highly overlapping resonant peaks of the Cr-$e_{\text{g}}$ 
	and Te-$p$ orbitals in (e) and (f) demonstrate that these occupied states form a strongly coupled 
	covalent anti-bonding network rather than simple isolated states.}
	\label{fig: CGT6HfbdosTepCreg}
	\refstepcounter{subfigure}\label{fig: CGT6HfbdosTepCreg-a} 
    \refstepcounter{subfigure}\label{fig: CGT6HfbdosTepCreg-b} 
    \refstepcounter{subfigure}\label{fig: CGT6HfbdosTepCreg-c} 
    \refstepcounter{subfigure}\label{fig: CGT6HfbdosTepCreg-d} 
    \refstepcounter{subfigure}\label{fig: CGT6HfbdosTepCreg-e} 
    \refstepcounter{subfigure}\label{fig: CGT6HfbdosTepCreg-f}
\end{figure}
Before hydrogenation, the 6 surface Te atoms possess fully occupied non-bonding lone pairs, contributing 12 electrons to the upper valence band. 
The formation of the 6 deep $\sigma_{\text{Te-H}}$ covalent bonds requires 12 electrons in total (6 from H, and 6 from Te). Consequently, 
6 of the original Te lone-pair electrons are displaced from their stable states. These 6 displaced electrons are forced to occupy the lowest available 
empty states. This is explicitly visualized in Fig. \ref{fig: CGT6HfbdosTepCreg-a} and \ref{fig: CGT6HfbdosTepCreg-d}, where the six lowest pristine conduction bands (labeled 89 to 94) are 
completely pulled below the Fermi level to accommodate these extra electrons.

As shown in fatband plots [Fig. \ref{fig: CGT6HfbdosTepCreg-b} and \ref{fig: CGT6HfbdosTepCreg-e}] and further 
corroborated by the partial density of states (PDOS) 
profiles [Fig. \ref{fig: CGT6HfbdosTepCreg-c} and \ref{fig: CGT6HfbdosTepCreg-f}], bands 89 to 94 are strongly hybridized states. 
In the PDOS plots, the highly overlapping 
resonant peaks of the Cr-$e_{\text{g}}$ (blue) and Te-$p$ (red) orbitals right below the Fermi level provide definitive proof that these 
are not isolated states, but rather a strongly coupled Cr-Te anti-bonding network. Therefore, the 6 displaced electrons are 
shared across this covalent network. The filling of the Cr-$e_{\text{g}}$ components of these hybridized bands accounts for the formal 
reduction of Cr (3$+$ to 2$+$), while the simultaneous filling of the Te-p components accommodates the remaining 4 electrons. 
All 6 extra electrons are localized together within these specific hybridized bands (89-94).

\begin{figure}[H]
	\captionsetup[subfigure]{labelformat=simple}
	\centering
	\includegraphics[width=1\columnwidth]{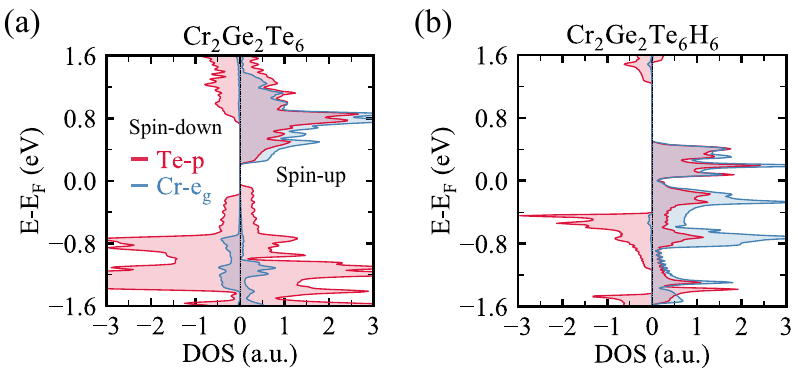}
	\caption{Spin-polarized partial density of states (PDOS) without 
	spin-orbit coupling for (a) pristine Cr$_{\text{2}}$Ge$_{\text{2}}$Te$_{\text{6}}$ and 
	(b) hydrogenated Cr$_{\text{2}}$Ge$_{\text{2}}$Te$_{\text{6}}$H$_{\text{6}}$. The positive and 
	negative values represent the spin-up and spin-down channels, respectively. The blue and red shaded 
	areas highlight the projected weights of the Cr-$e_{\text{g}}$ and Te-$p$ orbitals. The comparison clearly
	visualizes the strong exchange splitting and the specific occupation of the spin-up Cr-$e_{\text{g}}$ states right below the Fermi level after hydrogenation.}
	\label{fig: DOSwoSOC}
	\refstepcounter{subfigure}\label{fig: DOSwoSOC-a} 
    \refstepcounter{subfigure}\label{fig: DOSwoSOC-b} 
\end{figure}

\begin{figure*}[htbp]
	\captionsetup[figure]{name={FIG.},labelformat=simple,labelsep=period,singlelinecheck=off}
	\centering
	\includegraphics[width=2\columnwidth]{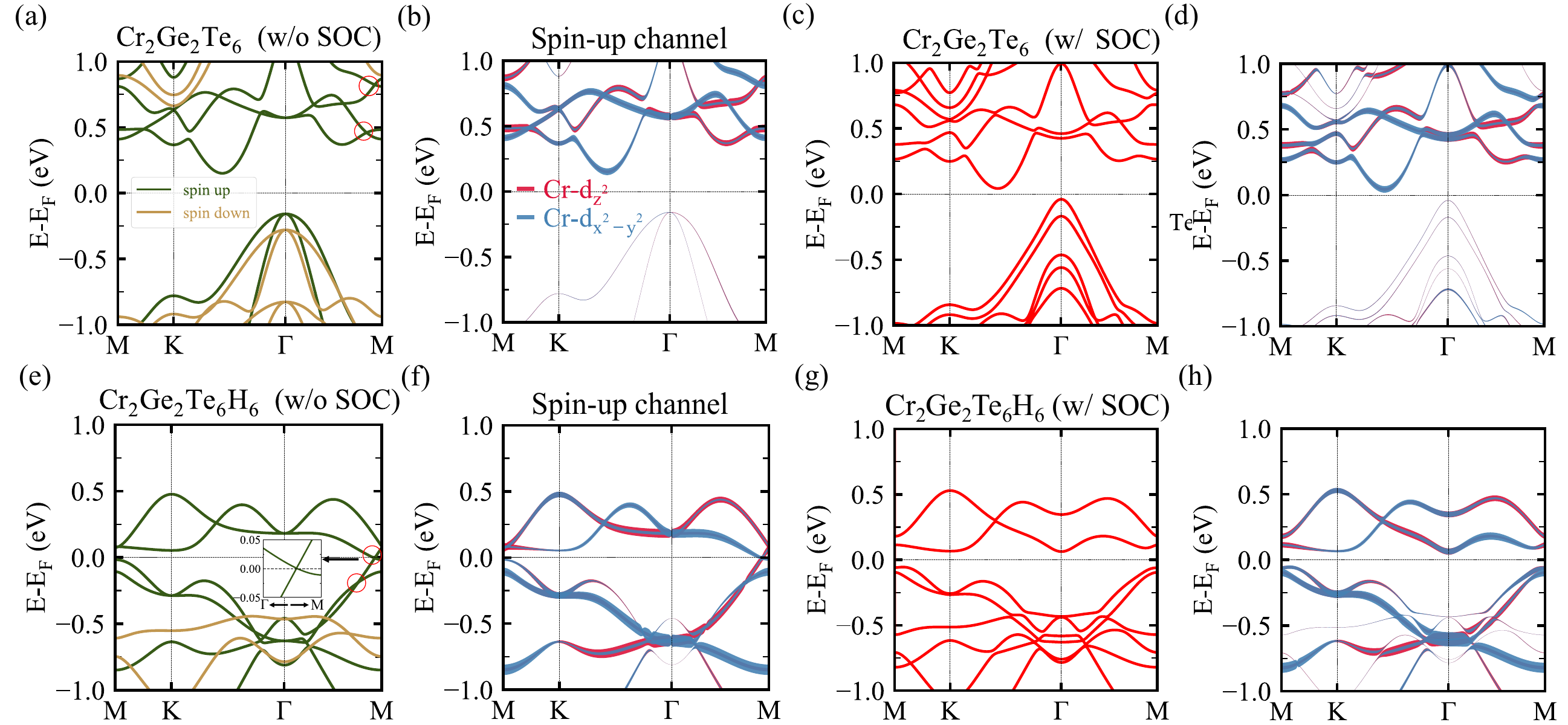}
	\caption{\raggedright(a, c) Band structures without 
	the inclusion of spin-orbit coupling (w/o SOC)
	and with the inclusion of spin-orbit coupl- 
	-ing (w/ SOC) for 
	Cr$_{\text{2}}$Ge$_{\text{2}}$Te$_{\text{6}}$,
	respectively, (b, d) show the corresponding projected band structures. 
	Orbital contributions from Cr-$d_{z^{2}}$ and Cr-$d_{x^{2}-y^{2}}$ 
	orbitals are explicitly highlighted.
	(e-h) show the electronic structures for 
	Cr$_{\text{2}}$Ge$_{\text{2}}$Te$_{\text{6}}$H$_{\text{6}}$.
	Hydrogenation drives the topological Dirac points, labeled by red circles, 
	in conduction bands on the $\Gamma$-M path in 
	Cr$_{\text{2}}$Ge$_{\text{2}}$Te$_{\text{6}}$
	evolves to Dirac points near the Fermi level in (e) for 
	Cr$_{\text{2}}$Ge$_{\text{2}}$Te$_{\text{6}}$H$_{\text{6}}$.
	For w/ SOC cases, the magnetic moments are considered to 
	align with the out-of-plane easy axis.
	}\label{fig: fig3}
	\phantomsubcaption{\label{fig: fig3-a}} 
	\phantomsubcaption{\label{fig: fig3-b}} 
	\phantomsubcaption{\label{fig: fig3-c}} 
	\phantomsubcaption{\label{fig: fig3-d}} 
	\phantomsubcaption{\label{fig: fig3-e}}
	\phantomsubcaption{\label{fig: fig3-f}}
	\phantomsubcaption{\label{fig: fig3-g}}
	\phantomsubcaption{\label{fig: fig3-h}}    
\end{figure*}

To provide a more direct and quantitative visualization of the exchange 
splitting and the occupation of these $e_{\text{g}}$ states, we present the 
spin-polarized PDOS without SOC in Fig. \ref{fig: DOSwoSOC}. In pristine 
Cr$_{\text{2}}$Ge$_{\text{2}}$Te$_{\text{6}}$ [Fig. \ref{fig: DOSwoSOC-a}], the Cr-$e_{\text{g}}$ states 
reside above the Fermi level, corresponding to the Cr$^{3+}$ ($t_{\text{2g}}^{\text{3}} e_{\text{g}}^{\text{0}}$) 
configuration. Upon hydrogenation [Fig. \ref{fig: DOSwoSOC-b}], the strong exchange splitting leaves the 
spin-down states empty, while a pronounced spin-up Cr-$e_{\text{g}}$ peak emerges 
below the Fermi level. This explicitly visualizes the selective occupation of the spin-majority $e_{\text{g}}$ state 
by the displaced electrons. Consequently, the formal valence state of Cr is lowered 
to 2$+$ ($t_{\text{2g}}^{\text{3}} e_{\text{g}}^{\text{1}}$), 
which accounts for the enhanced localized magnetic moment of 4 $\mu_{\text{B}}$ per Cr atom.
Beyond merely increasing the localized magnetic moment, this fundamental covalent orbital reconstruction 
is precisely the driving force behind the enhanced ferromagnetism. Because the displaced electrons 
are injected into a strongly coupled Cr-Te anti-bonding network rather than isolated atomic orbitals, the energy difference 
between the ligand Te-$p$ and transition metal Cr-$d$ orbitals is effectively minimized [as evidenced in Fig. \ref{fig: fig2-d}]. 
As established in our earlier superexchange analysis, this specific energy tuning induced by covalent 
hybridization optimizes the $d_{z^{2}}$-$p_{z}$-$d_{xz}$ channel, resulting in the 
dramatic enhancement of the nearest-neighbor coupling $J_{\text{1}}$ and solidifying the robust ferromagnetic ground state.

\subsection{\NoCaseChange{Quantum anomalous Hall effect in Cr$_{\text{2}}$Ge$_{\text{2}}$Te$_{\text{6}}$H$_{\text{6}}$}}
Fig. \ref{fig: fig3} displays the overall electronic structures alongside their $e_{\text{g}}$ orbital-projected 
components for monolayers Cr$_{\text{2}}$Ge$_{\text{2}}$Te$_{\text{6}}$ [Fig. \ref{fig: fig3-a}-\ref{fig: fig3-d}] 
and Cr$_{\text{2}}$Ge$_{\text{2}}$Te$_{\text{6}}$H$_{\text{6}}$ [Fig. \ref{fig: fig3-e}-\ref{fig: fig3-h}], 
both with and without SOC. As shown in Fig. \ref{fig: fig3-c} and \ref{fig: fig3-g}, 
both systems are ferromagnetic semiconductors. 
Dirac points above the Fermi level along the $\Gamma$-M path are identified 
by red circles in Fig. \ref{fig: fig3-a}, 
suggesting the presence of hidden topological properties in Cr$_{\text{2}}$Ge$_{\text{2}}$Te$_{\text{6}}$.
The targeted filling of the Cr-Te anti-bonding network after hydrogenation pushes the Fermi level 
upward suggested by Fig. \ref{fig: CGT6HfbdosTepCreg}.
Concurrently, the electronic structure of Cr$_{\text{2}}$Ge$_{\text{2}}$Te$_{\text{6}}$ is modified, 
the topological Dirac points originally 
located in the conduction bands move to the vicinity of the Fermi level 
in Cr$_{\text{2}}$Ge$_{\text{2}}$Te$_{\text{6}}$H$_{\text{6}}$, as depicted in Fig. \ref{fig: fig3-e}.
\begin{figure}[H]
	\captionsetup[figure]{name={FIG.},labelformat=simple,labelsep=period,singlelinecheck=off}
	\centering
	\includegraphics[width=1\columnwidth]{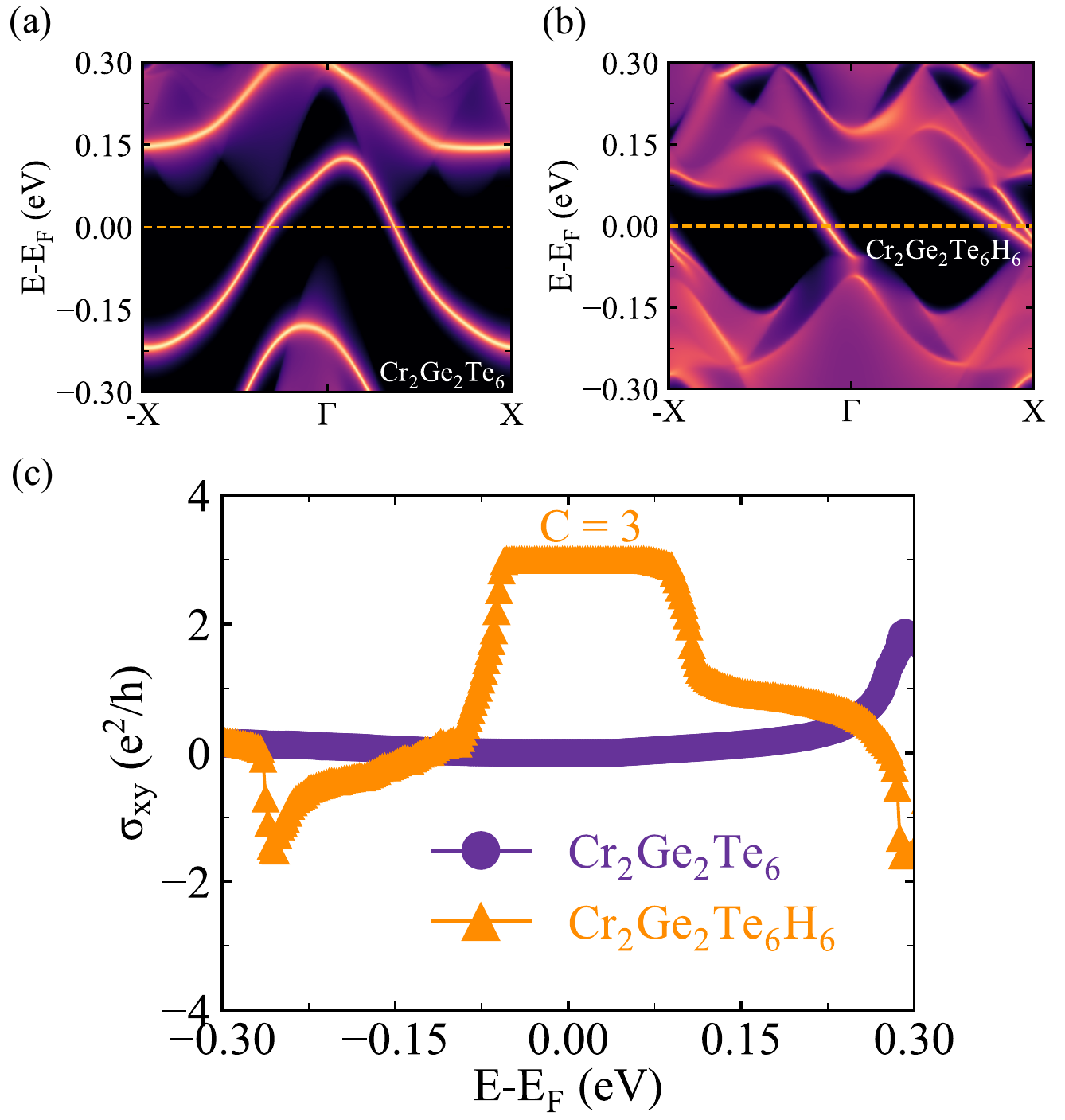}
	\caption{
	(a-b) Surface band structures along the (100) direction for (a)  	
	Cr$_{\text{2}}$Ge$_{\text{2}}$Te$_{\text{6}}$ and (b) Cr$_{\text{2}}$Ge$_{\text{2}}$Te$_{\text{6}}$H$_{\text{6}}$, respectively.
	Cr$_{\text{2}}$Ge$_{\text{2}}$Te$_{\text{6}}$H$_{\text{6}}$ exhibits three topological chiral edge states connecting valance bands and conduction bands,
	while Cr$_{\text{2}}$Ge$_{\text{2}}$Te$_{\text{6}}$ hosts trivial edge states.
	(c) Anomalous Hall conductance $\sigma_{xy}$ ($e^{2}/h$) for Cr$_{\text{2}}$Ge$_{\text{2}}$Te$_{\text{6}}$
	and Cr$_{\text{2}}$Ge$_{\text{2}}$Te$_{\text{6}}$H$_{\text{6}}$.
	Cr$_{\text{2}}$Ge$_{\text{2}}$Te$_{\text{6}}$H$_{\text{6}}$ features a
	QAH plateau with a Chern number $C$ $=$ 3 in the bulk band gap.}
	\phantomsubcaption{\label{fig: fig4-a}} 
	\phantomsubcaption{\label{fig: fig4-b}} 
	\phantomsubcaption{\label{fig: fig4-c}}
	\phantomsubcaption{\label{fig: fig4-d}}  
\end{figure}
When SOC is included, the Dirac point opens a global band 
gap of 118.1 meV in Fig. \ref{fig: fig3-g}, which exceeds the thermal 
energy at room temperature. Hydrogenation
triggers the topological electronic phase transition 
in Cr$_{\text{2}}$Ge$_{\text{2}}$Te$_{\text{6}}$.
The emergence of topological properties is furthered evidenced by the calculated surface states
by using Wanniertools \cite{WannierTools}, as
depicted in Fig. \ref{fig: fig4-a} for 
Cr$_{\text{2}}$Ge$_{\text{2}}$Te$_{\text{6}}$ and Fig. \ref{fig: fig4-b} for 
Cr$_{\text{2}}$Ge$_{\text{2}}$Te$_{\text{6}}$H$_{\text{6}}$. 
The Cr$_{\text{2}}$Ge$_{\text{2}}$Te$_{\text{6}}$ monolayer
exhibits trivial edge states within its bulk band gap, while
Cr$_{\text{2}}$Ge$_{\text{2}}$Te$_{\text{6}}$H$_{\text{6}}$ 
hosts three pairs of chiral topological edge states that robustly connect the bulk valence and conduction bands, 
which corresponds to a QAH
plateau with a high Chern number $C=3$, as demonstrated in Fig. \ref{fig: fig4-c} by the 
quantized Hall conductance $\sigma_{xy}=Ce^{2}/h$. 

The structural symmetry remains unchanged upon hydrogenation, belonging to 
the space group $P\bar{3}1m$ (point group $D_{3d}$). In the absence of SOC, 
the purely spin-polarized channel respects a spinless time-reversal symmetry $T^* = \mathcal{K}$. 
The composite $PT^*$ symmetry enforces the Berry phase along any closed loop to be strictly quantized to 0 or $\pi$, 
topologically protecting the Dirac crossings. Furthermore, these crossings are pinned on the $\Gamma$-M paths 
because the two intersecting bands possess opposite mirror eigenvalues 
(e.g., $+1$ and $-1$ under the $\sigma_v$ mirror plane of the $D_{3d}$ point group), which forbids their hybridization 
along this mirror-symmetric line. When SOC is included, a band gap opens, transforming these crossings into massive
Dirac cones. The six massive Dirac cones in the Brillouin zone are related by spatial symmetries ($C_{3z}$ and $P$). 
Since the Berry curvature $\Omega(\boldsymbol{k})$ transforms as $\Omega(C_{3z}\boldsymbol{k}) = \Omega(\boldsymbol{k})$ 
and $\Omega(P\boldsymbol{k}) = \Omega(\boldsymbol{k})$, these spatial symmetries 
guarantee that all six cones possess the same sign of Berry curvature. Here, time-reversal breaking by 
ferromagnetism permits the net Berry curvature to be non-zero. Because each massive Dirac cone contributes 
exactly $1/2$ to the topological invariant, the Brillouin-zone integral yields a total Chern number $C = 6 \times (1/2) = 3$, 
consistent with the three chiral edge modes in the ribbon spectrum and a quantized anomalous Hall plateau $\sigma_{xy}=3e^2/h$.

\section{\NoCaseChange{DISCUSSION AND CONCLUSION}}
Surface functionalization provides a symmetry-preserving way 
to drive covalent orbital reconstruction, tuning both orbital 
occupation and hybridization in van der Waals magnets. In the present system, 
hydrogenation plays a dual role driven by this singular physical mechanism: 
it alters orbital occupations to shifts preexisting Dirac crossings toward 
the Fermi level, excavating the hidden topological phase in Cr$_{\text{2}}$Ge$_{\text{2}}$Te$_{\text{6}}$,
and simultaneously tunes the energy difference between the ligand $p$ and 
transition metal $d$ orbitals, thereby strengthening the superexchange network that 
stabilizes ferromagnetism. While the experimental realization of functionalization-induced QAH phases 
remains challenging in general, the large SOC-induced gap of 118.1 meV and the clear 
edge-state signatures identified here make Cr$_{\text{2}}$Ge$_{\text{2}}$Te$_{\text{6}}$H$_{\text{6}}$
a useful platform for exploring high-Chern-number QAH 
physics driven by covalent orbital electronic reconstruction.

By establishing this connection between covalent orbital reconstruction and the effective spin system,
we emphasize that Curie temperatures extracted from Monte Carlo simulations depend on 
the underlying effective spin model and its parameterization from first-principles calculations.
In this work, we therefore use Monte Carlo results primarily 
to assess the comparative ferromagnetic stability before and after hydrogenation,
which serves as a salient effect of this orbital tuning.
The substantial enhancement of $J_{\text{1}}$
can be understood as a direct consequence of the modified 
orbital and electronic structure, consistently indicating strengthened ferromagnetism in
Cr$_{\text{2}}$Ge$_{\text{2}}$Te$_{\text{6}}$H$_{\text{6}}$ across a broad range of Hubbard 
U values, and the robustness of the high-Chern-number phase is validated for these Hubbard 
U values (see Supplemental Material \cite{Supplement}).
In summary, our results bridge microscopic surface chemistry 
and macroscopic quantum phases, suggesting a general route to engineer 
high-Chern-number QAH insulators in magnetic semiconductors 
by activating symmetry-related Dirac fermions via targeted orbital reconstruction.

\section{\NoCaseChange{ACKNOWLEDGMENTS}}
This work is supported by National Key R\&D Program
of China (Grant No. 2022YFA1405100), Chinese
Academy of Sciences Project for Young Scientists in Basic
Research (Grant No. YSBR-030), and Basic Research
Program of the Chinese Academy of Sciences Based on
Major Scientific Infrastructures (Grant No. JZHKYPT-
2021-08). GS was supported in part by the Innovation
Program for Quantum Science and Technology under
Grant No. 2024ZD0300500, NSFC Nos. 12534009, 12447101 and the
Strategic Priority Research Program of Chinese Academy
of Sciences (Grant No. XDB1270000) and Computing Power Support from 
the Supercomputing Platform of the University of Chinese Academy of Sciences.

\bibliographystyle{apsrev4-2}

\begin{thebibliography}{143}%
\makeatletter
\providecommand \@ifxundefined [1]{%
 \@ifx{#1\undefined}
}%
\providecommand \@ifnum [1]{%
 \ifnum #1\expandafter \@firstoftwo
 \else \expandafter \@secondoftwo
 \fi
}%
\providecommand \@ifx [1]{%
 \ifx #1\expandafter \@firstoftwo
 \else \expandafter \@secondoftwo
 \fi
}%
\providecommand \natexlab [1]{#1}%
\providecommand \enquote  [1]{``#1''}%
\providecommand \bibnamefont  [1]{#1}%
\providecommand \bibfnamefont [1]{#1}%
\providecommand \citenamefont [1]{#1}%
\providecommand \href@noop [0]{\@secondoftwo}%
\providecommand \href [0]{\begingroup \@sanitize@url \@href}%
\providecommand \@href[1]{\@@startlink{#1}\@@href}%
\providecommand \@@href[1]{\endgroup#1\@@endlink}%
\providecommand \@sanitize@url [0]{\catcode `\\12\catcode `\$12\catcode
  `\&12\catcode `\#12\catcode `\^12\catcode `\_12\catcode `\%12\relax}%
\providecommand \@@startlink[1]{}%
\providecommand \@@endlink[0]{}%
\providecommand \url  [0]{\begingroup\@sanitize@url \@url }%
\providecommand \@url [1]{\endgroup\@href {#1}{\urlprefix }}%
\providecommand \urlprefix  [0]{URL }%
\providecommand \Eprint [0]{\href }%
\providecommand \doibase [0]{https://doi.org/}%
\providecommand \selectlanguage [0]{\@gobble}%
\providecommand \bibinfo  [0]{\@secondoftwo}%
\providecommand \bibfield  [0]{\@secondoftwo}%
\providecommand \translation [1]{[#1]}%
\providecommand \BibitemOpen [0]{}%
\providecommand \bibitemStop [0]{}%
\providecommand \bibitemNoStop [0]{.\EOS\space}%
\providecommand \EOS [0]{\spacefactor3000\relax}%
\providecommand \BibitemShut  [1]{\csname bibitem#1\endcsname}%
\let\auto@bib@innerbib\@empty
%</preamble>
\bibitem [{\citenamefont {Chang}\ \emph {et~al.}(2023)\citenamefont {Chang},
  \citenamefont {Liu},\ and\ \citenamefont {MacDonald}}]{RMP-QAHE}%
  \BibitemOpen
  \bibfield  {author} {\bibinfo {author} {\bibfnamefont {C.-Z.}\ \bibnamefont
  {Chang}}, \bibinfo {author} {\bibfnamefont {C.-X.}\ \bibnamefont {Liu}},\
  and\ \bibinfo {author} {\bibfnamefont {A.~H.}\ \bibnamefont {MacDonald}},\
  }\bibinfo {title} {Colloquium: Quantum anomalous Hall effect},\ \href
  {https://doi.org/10.1103/RevModPhys.95.011002} {\bibfield  {journal}
  {\bibinfo  {journal} {Rev. Mod. Phys.}\ }\textbf {\bibinfo {volume} {95}},\
  \bibinfo {pages} {011002} (\bibinfo {year} {2023})}\BibitemShut {NoStop}%
\bibitem [{\citenamefont {Haldane}(1988)}]{HaldaneModel}%
  \BibitemOpen
  \bibfield  {author} {\bibinfo {author} {\bibfnamefont {F.~D.~M.}\
  \bibnamefont {Haldane}},\ }\bibinfo {title} {Model for a Quantum Hall effect
  without Landau levels: Condensed-matter realization of the "Parity
  anomaly"},\ \href {https://doi.org/10.1103/PhysRevLett.61.2015} {\bibfield
  {journal} {\bibinfo  {journal} {Phys. Rev. Lett.}\ }\textbf {\bibinfo
  {volume} {61}},\ \bibinfo {pages} {2015} (\bibinfo {year}
  {1988})}\BibitemShut {NoStop}%
\bibitem [{\citenamefont {He}\ \emph {et~al.}(2018)\citenamefont {He},
  \citenamefont {Wang},\ and\ \citenamefont {Xue}}]{Review-QAHE-QikunXue-2018}%
  \BibitemOpen
  \bibfield  {author} {\bibinfo {author} {\bibfnamefont {K.}~\bibnamefont
  {He}}, \bibinfo {author} {\bibfnamefont {Y.}~\bibnamefont {Wang}},\ and\
  \bibinfo {author} {\bibfnamefont {Q.-K.}\ \bibnamefont {Xue}},\ }\bibinfo
  {title} {Topological materials: Quantum anomalous Hall system},\ \href
  {https://doi.org/https://doi.org/10.1146/annurev-conmatphys-033117-054144}
  {\bibfield  {journal} {\bibinfo  {journal} {Annu. Rev. Condens. Matter
  Phys.}\ }\textbf {\bibinfo {volume} {9}},\ \bibinfo {pages} {329} (\bibinfo
  {year} {2018})}\BibitemShut {NoStop}%
\bibitem [{\citenamefont {Liu}\ \emph {et~al.}(2016)\citenamefont {Liu},
  \citenamefont {Zhang},\ and\ \citenamefont
  {Qi}}]{Review-QAHE-XiaoliangQi-2016}%
  \BibitemOpen
  \bibfield  {author} {\bibinfo {author} {\bibfnamefont {C.-X.}\ \bibnamefont
  {Liu}}, \bibinfo {author} {\bibfnamefont {S.-C.}\ \bibnamefont {Zhang}},\
  and\ \bibinfo {author} {\bibfnamefont {X.-L.}\ \bibnamefont {Qi}},\ }\bibinfo
  {title} {The quantum anomalous Hall effect: Theory and experiment},\ \href
  {https://doi.org/https://doi.org/10.1146/annurev-conmatphys-031115-011417}
  {\bibfield  {journal} {\bibinfo  {journal} {Annu. Rev. Condens. Matter
  Phys.}\ }\textbf {\bibinfo {volume} {7}},\ \bibinfo {pages} {301} (\bibinfo
  {year} {2016})}\BibitemShut {NoStop}%
\bibitem [{\citenamefont {Chang}\ \emph
  {et~al.}(2013{\natexlab{a}})\citenamefont {Chang}, \citenamefont {Zhang},
  \citenamefont {Feng}, \citenamefont {Shen}, \citenamefont {Zhang},
  \citenamefont {Guo}, \citenamefont {Li}, \citenamefont {Ou}, \citenamefont
  {Wei}, \citenamefont {Wang}, \citenamefont {Ji}, \citenamefont {Feng},
  \citenamefont {Ji}, \citenamefont {Chen}, \citenamefont {Jia}, \citenamefont
  {Dai}, \citenamefont {Fang}, \citenamefont {Zhang}, \citenamefont {He},
  \citenamefont {Wang}, \citenamefont {Lu}, \citenamefont {Ma},\ and\
  \citenamefont {Xue}}]{EXP-CrBiSb2Te3-QAHE}%
  \BibitemOpen
  \bibfield  {author} {\bibinfo {author} {\bibfnamefont {C.-Z.}\ \bibnamefont
  {Chang}}, \bibinfo {author} {\bibfnamefont {J.}~\bibnamefont {Zhang}},
  \bibinfo {author} {\bibfnamefont {X.}~\bibnamefont {Feng}}, \bibinfo {author}
  {\bibfnamefont {J.}~\bibnamefont {Shen}}, \bibinfo {author} {\bibfnamefont
  {Z.}~\bibnamefont {Zhang}}, \bibinfo {author} {\bibfnamefont
  {M.}~\bibnamefont {Guo}}, \bibinfo {author} {\bibfnamefont {K.}~\bibnamefont
  {Li}}, \bibinfo {author} {\bibfnamefont {Y.}~\bibnamefont {Ou}}, \bibinfo
  {author} {\bibfnamefont {P.}~\bibnamefont {Wei}}, \bibinfo {author}
  {\bibfnamefont {L.-L.}\ \bibnamefont {Wang}}, \bibinfo {author}
  {\bibfnamefont {Z.-Q.}\ \bibnamefont {Ji}}, \bibinfo {author} {\bibfnamefont
  {Y.}~\bibnamefont {Feng}}, \bibinfo {author} {\bibfnamefont {S.}~\bibnamefont
  {Ji}}, \bibinfo {author} {\bibfnamefont {X.}~\bibnamefont {Chen}}, \bibinfo
  {author} {\bibfnamefont {J.}~\bibnamefont {Jia}}, \bibinfo {author}
  {\bibfnamefont {X.}~\bibnamefont {Dai}}, \bibinfo {author} {\bibfnamefont
  {Z.}~\bibnamefont {Fang}}, \bibinfo {author} {\bibfnamefont {S.-C.}\
  \bibnamefont {Zhang}}, \bibinfo {author} {\bibfnamefont {K.}~\bibnamefont
  {He}}, \bibinfo {author} {\bibfnamefont {Y.}~\bibnamefont {Wang}}, \bibinfo
  {author} {\bibfnamefont {L.}~\bibnamefont {Lu}}, \bibinfo {author}
  {\bibfnamefont {X.-C.}\ \bibnamefont {Ma}},\ and\ \bibinfo {author}
  {\bibfnamefont {Q.-K.}\ \bibnamefont {Xue}},\ }\bibinfo {title} {Experimental
  observation of the quantum anomalous Hall effect in a magnetic topological
  insulator},\ \href {https://doi.org/10.1126/science.1234414} {\bibfield
  {journal} {\bibinfo  {journal} {Science}\ }\textbf {\bibinfo {volume}
  {340}},\ \bibinfo {pages} {167} (\bibinfo {year}
  {2013}{\natexlab{a}})}\BibitemShut {NoStop}%
\bibitem [{\citenamefont {Chang}\ \emph {et~al.}(2015)\citenamefont {Chang},
  \citenamefont {Zhao}, \citenamefont {Kim}, \citenamefont {Zhang},
  \citenamefont {Assaf}, \citenamefont {Heiman}, \citenamefont {Zhang},
  \citenamefont {Liu}, \citenamefont {Chan},\ and\ \citenamefont
  {Moodera}}]{EXP-VBiSb2Te3-QAHE}%
  \BibitemOpen
  \bibfield  {author} {\bibinfo {author} {\bibfnamefont {C.-Z.}\ \bibnamefont
  {Chang}}, \bibinfo {author} {\bibfnamefont {W.}~\bibnamefont {Zhao}},
  \bibinfo {author} {\bibfnamefont {D.~Y.}\ \bibnamefont {Kim}}, \bibinfo
  {author} {\bibfnamefont {H.}~\bibnamefont {Zhang}}, \bibinfo {author}
  {\bibfnamefont {B.~A.}\ \bibnamefont {Assaf}}, \bibinfo {author}
  {\bibfnamefont {D.}~\bibnamefont {Heiman}}, \bibinfo {author} {\bibfnamefont
  {S.-C.}\ \bibnamefont {Zhang}}, \bibinfo {author} {\bibfnamefont
  {C.}~\bibnamefont {Liu}}, \bibinfo {author} {\bibfnamefont {M.~H.~W.}\
  \bibnamefont {Chan}},\ and\ \bibinfo {author} {\bibfnamefont {J.~S.}\
  \bibnamefont {Moodera}},\ }\bibinfo {title} {High-precision realization of
  robust quantum anomalous Hall state in a hard ferromagnetic topological
  insulator},\ \href {https://doi.org/10.1038/nmat4204} {\bibfield  {journal}
  {\bibinfo  {journal} {Nat. Mater.}\ }\textbf {\bibinfo {volume} {14}},\
  \bibinfo {pages} {473} (\bibinfo {year} {2015})}\BibitemShut {NoStop}%
\bibitem [{\citenamefont {Checkelsky}\ \emph {et~al.}(2014)\citenamefont
  {Checkelsky}, \citenamefont {Yoshimi}, \citenamefont {Tsukazaki},
  \citenamefont {Takahashi}, \citenamefont {Kozuka}, \citenamefont {Falson},
  \citenamefont {Kawasaki},\ and\ \citenamefont
  {Tokura}}]{EXP-CrBiSb2Te3-QAHE-2}%
  \BibitemOpen
  \bibfield  {author} {\bibinfo {author} {\bibfnamefont {J.~G.}\ \bibnamefont
  {Checkelsky}}, \bibinfo {author} {\bibfnamefont {R.}~\bibnamefont {Yoshimi}},
  \bibinfo {author} {\bibfnamefont {A.}~\bibnamefont {Tsukazaki}}, \bibinfo
  {author} {\bibfnamefont {K.~S.}\ \bibnamefont {Takahashi}}, \bibinfo {author}
  {\bibfnamefont {Y.}~\bibnamefont {Kozuka}}, \bibinfo {author} {\bibfnamefont
  {J.}~\bibnamefont {Falson}}, \bibinfo {author} {\bibfnamefont
  {M.}~\bibnamefont {Kawasaki}},\ and\ \bibinfo {author} {\bibfnamefont
  {Y.}~\bibnamefont {Tokura}},\ }\bibinfo {title} {Trajectory of the anomalous
  Hall effect towards the quantized state in a ferromagnetic topological
  insulator},\ \href {https://doi.org/10.1038/nphys3053} {\bibfield  {journal}
  {\bibinfo  {journal} {Nat. Phys.}\ }\textbf {\bibinfo {volume} {10}},\
  \bibinfo {pages} {731} (\bibinfo {year} {2014})}\BibitemShut {NoStop}%
\bibitem [{\citenamefont {Kou}\ \emph {et~al.}(2014)\citenamefont {Kou},
  \citenamefont {Guo}, \citenamefont {Fan}, \citenamefont {Pan}, \citenamefont
  {Lang}, \citenamefont {Jiang}, \citenamefont {Shao}, \citenamefont {Nie},
  \citenamefont {Murata}, \citenamefont {Tang}, \citenamefont {Wang},
  \citenamefont {He}, \citenamefont {Lee}, \citenamefont {Lee},\ and\
  \citenamefont {Wang}}]{EXP-CrBiSb2Te3-QAHE-3}%
  \BibitemOpen
  \bibfield  {author} {\bibinfo {author} {\bibfnamefont {X.}~\bibnamefont
  {Kou}}, \bibinfo {author} {\bibfnamefont {S.-T.}\ \bibnamefont {Guo}},
  \bibinfo {author} {\bibfnamefont {Y.}~\bibnamefont {Fan}}, \bibinfo {author}
  {\bibfnamefont {L.}~\bibnamefont {Pan}}, \bibinfo {author} {\bibfnamefont
  {M.}~\bibnamefont {Lang}}, \bibinfo {author} {\bibfnamefont {Y.}~\bibnamefont
  {Jiang}}, \bibinfo {author} {\bibfnamefont {Q.}~\bibnamefont {Shao}},
  \bibinfo {author} {\bibfnamefont {T.}~\bibnamefont {Nie}}, \bibinfo {author}
  {\bibfnamefont {K.}~\bibnamefont {Murata}}, \bibinfo {author} {\bibfnamefont
  {J.}~\bibnamefont {Tang}}, \bibinfo {author} {\bibfnamefont {Y.}~\bibnamefont
  {Wang}}, \bibinfo {author} {\bibfnamefont {L.}~\bibnamefont {He}}, \bibinfo
  {author} {\bibfnamefont {T.-K.}\ \bibnamefont {Lee}}, \bibinfo {author}
  {\bibfnamefont {W.-L.}\ \bibnamefont {Lee}},\ and\ \bibinfo {author}
  {\bibfnamefont {K.~L.}\ \bibnamefont {Wang}},\ }\bibinfo {title}
  {Scale-invariant quantum anomalous Hall effect in magnetic topological
  insulators beyond the two-dimensional limit},\ \href
  {https://doi.org/10.1103/PhysRevLett.113.137201} {\bibfield  {journal}
  {\bibinfo  {journal} {Phys. Rev. Lett.}\ }\textbf {\bibinfo {volume} {113}},\
  \bibinfo {pages} {137201} (\bibinfo {year} {2014})}\BibitemShut {NoStop}%
\bibitem [{\citenamefont {Kou}\ \emph {et~al.}(2015)\citenamefont {Kou},
  \citenamefont {Pan}, \citenamefont {Wang}, \citenamefont {Fan}, \citenamefont
  {Choi}, \citenamefont {Lee}, \citenamefont {Nie}, \citenamefont {Murata},
  \citenamefont {Shao}, \citenamefont {Zhang},\ and\ \citenamefont
  {Wang}}]{EXP-CrBiSb2Te3-QAHE-5}%
  \BibitemOpen
  \bibfield  {author} {\bibinfo {author} {\bibfnamefont {X.}~\bibnamefont
  {Kou}}, \bibinfo {author} {\bibfnamefont {L.}~\bibnamefont {Pan}}, \bibinfo
  {author} {\bibfnamefont {J.}~\bibnamefont {Wang}}, \bibinfo {author}
  {\bibfnamefont {Y.}~\bibnamefont {Fan}}, \bibinfo {author} {\bibfnamefont
  {E.~S.}\ \bibnamefont {Choi}}, \bibinfo {author} {\bibfnamefont {W.-L.}\
  \bibnamefont {Lee}}, \bibinfo {author} {\bibfnamefont {T.}~\bibnamefont
  {Nie}}, \bibinfo {author} {\bibfnamefont {K.}~\bibnamefont {Murata}},
  \bibinfo {author} {\bibfnamefont {Q.}~\bibnamefont {Shao}}, \bibinfo {author}
  {\bibfnamefont {S.-C.}\ \bibnamefont {Zhang}},\ and\ \bibinfo {author}
  {\bibfnamefont {K.~L.}\ \bibnamefont {Wang}},\ }\bibinfo {title}
  {Metal-to-insulator switching in quantum anomalous Hall states},\ \href
  {https://doi.org/10.1038/ncomms9474} {\bibfield  {journal} {\bibinfo
  {journal} {Nat. Commun.}\ }\textbf {\bibinfo {volume} {6}},\ \bibinfo {pages}
  {8474} (\bibinfo {year} {2015})}\BibitemShut {NoStop}%
\bibitem [{\citenamefont {Chang}\ \emph
  {et~al.}(2013{\natexlab{b}})\citenamefont {Chang}, \citenamefont {Zhang},
  \citenamefont {Liu}, \citenamefont {Zhang}, \citenamefont {Feng},
  \citenamefont {Li}, \citenamefont {Wang}, \citenamefont {Chen}, \citenamefont
  {Dai}, \citenamefont {Fang}, \citenamefont {Qi}, \citenamefont {Zhang},
  \citenamefont {Wang}, \citenamefont {He}, \citenamefont {Ma},\ and\
  \citenamefont {Xue}}]{EXP-CrBiSb2Te3-QAHE-6}%
  \BibitemOpen
  \bibfield  {author} {\bibinfo {author} {\bibfnamefont {C.-Z.}\ \bibnamefont
  {Chang}}, \bibinfo {author} {\bibfnamefont {J.}~\bibnamefont {Zhang}},
  \bibinfo {author} {\bibfnamefont {M.}~\bibnamefont {Liu}}, \bibinfo {author}
  {\bibfnamefont {Z.}~\bibnamefont {Zhang}}, \bibinfo {author} {\bibfnamefont
  {X.}~\bibnamefont {Feng}}, \bibinfo {author} {\bibfnamefont {K.}~\bibnamefont
  {Li}}, \bibinfo {author} {\bibfnamefont {L.-L.}\ \bibnamefont {Wang}},
  \bibinfo {author} {\bibfnamefont {X.}~\bibnamefont {Chen}}, \bibinfo {author}
  {\bibfnamefont {X.}~\bibnamefont {Dai}}, \bibinfo {author} {\bibfnamefont
  {Z.}~\bibnamefont {Fang}}, \bibinfo {author} {\bibfnamefont {X.-L.}\
  \bibnamefont {Qi}}, \bibinfo {author} {\bibfnamefont {S.-C.}\ \bibnamefont
  {Zhang}}, \bibinfo {author} {\bibfnamefont {Y.}~\bibnamefont {Wang}},
  \bibinfo {author} {\bibfnamefont {K.}~\bibnamefont {He}}, \bibinfo {author}
  {\bibfnamefont {X.-C.}\ \bibnamefont {Ma}},\ and\ \bibinfo {author}
  {\bibfnamefont {Q.-K.}\ \bibnamefont {Xue}},\ }\bibinfo {title} {Thin films
  of magnetically doped topological insulator with carrier-independent
  long-Range ferromagnetic order},\ \href
  {https://doi.org/https://doi.org/10.1002/adma.201203493} {\bibfield
  {journal} {\bibinfo  {journal} {Adv. Mater.}\ }\textbf {\bibinfo {volume}
  {25}},\ \bibinfo {pages} {1065} (\bibinfo {year}
  {2013}{\natexlab{b}})}\BibitemShut {NoStop}%
\bibitem [{\citenamefont {Mogi}\ \emph {et~al.}(2015)\citenamefont {Mogi},
  \citenamefont {Yoshimi}, \citenamefont {Tsukazaki}, \citenamefont {Yasuda},
  \citenamefont {Kozuka}, \citenamefont {Takahashi}, \citenamefont {Kawasaki},\
  and\ \citenamefont {Tokura}}]{EXP-CrBiSb2Te3-QAHE-4-up2K}%
  \BibitemOpen
  \bibfield  {author} {\bibinfo {author} {\bibfnamefont {M.}~\bibnamefont
  {Mogi}}, \bibinfo {author} {\bibfnamefont {R.}~\bibnamefont {Yoshimi}},
  \bibinfo {author} {\bibfnamefont {A.}~\bibnamefont {Tsukazaki}}, \bibinfo
  {author} {\bibfnamefont {K.}~\bibnamefont {Yasuda}}, \bibinfo {author}
  {\bibfnamefont {Y.}~\bibnamefont {Kozuka}}, \bibinfo {author} {\bibfnamefont
  {K.~S.}\ \bibnamefont {Takahashi}}, \bibinfo {author} {\bibfnamefont
  {M.}~\bibnamefont {Kawasaki}},\ and\ \bibinfo {author} {\bibfnamefont
  {Y.}~\bibnamefont {Tokura}},\ }\bibinfo {title} {Magnetic modulation doping
  in topological insulators toward higher-temperature quantum anomalous Hall
  effect},\ \href {https://doi.org/10.1063/1.4935075} {\bibfield  {journal}
  {\bibinfo  {journal} {Appl. Phys. Lett.}\ }\textbf {\bibinfo {volume}
  {107}},\ \bibinfo {pages} {182401} (\bibinfo {year} {2015})}\BibitemShut
  {NoStop}%
\bibitem [{\citenamefont {Ou}\ \emph {et~al.}(2018)\citenamefont {Ou},
  \citenamefont {Liu}, \citenamefont {Jiang}, \citenamefont {Feng},
  \citenamefont {Zhao}, \citenamefont {Wu}, \citenamefont {Wang}, \citenamefont
  {Li}, \citenamefont {Song}, \citenamefont {Wang}, \citenamefont {Wang},
  \citenamefont {Wu}, \citenamefont {Wang}, \citenamefont {He}, \citenamefont
  {Ma},\ and\ \citenamefont {Xue}}]{EXP-CrVBiSb2Te3-QAHE-Codoping}%
  \BibitemOpen
  \bibfield  {author} {\bibinfo {author} {\bibfnamefont {Y.}~\bibnamefont
  {Ou}}, \bibinfo {author} {\bibfnamefont {C.}~\bibnamefont {Liu}}, \bibinfo
  {author} {\bibfnamefont {G.}~\bibnamefont {Jiang}}, \bibinfo {author}
  {\bibfnamefont {Y.}~\bibnamefont {Feng}}, \bibinfo {author} {\bibfnamefont
  {D.}~\bibnamefont {Zhao}}, \bibinfo {author} {\bibfnamefont {W.}~\bibnamefont
  {Wu}}, \bibinfo {author} {\bibfnamefont {X.-X.}\ \bibnamefont {Wang}},
  \bibinfo {author} {\bibfnamefont {W.}~\bibnamefont {Li}}, \bibinfo {author}
  {\bibfnamefont {C.}~\bibnamefont {Song}}, \bibinfo {author} {\bibfnamefont
  {L.-L.}\ \bibnamefont {Wang}}, \bibinfo {author} {\bibfnamefont
  {W.}~\bibnamefont {Wang}}, \bibinfo {author} {\bibfnamefont {W.}~\bibnamefont
  {Wu}}, \bibinfo {author} {\bibfnamefont {Y.}~\bibnamefont {Wang}}, \bibinfo
  {author} {\bibfnamefont {K.}~\bibnamefont {He}}, \bibinfo {author}
  {\bibfnamefont {X.-C.}\ \bibnamefont {Ma}},\ and\ \bibinfo {author}
  {\bibfnamefont {Q.-K.}\ \bibnamefont {Xue}},\ }\bibinfo {title} {Enhancing
  the quantum anomalous Hall effect by magnetic codoping in a topological
  insulator},\ \href {https://doi.org/https://doi.org/10.1002/adma.201703062}
  {\bibfield  {journal} {\bibinfo  {journal} {Adv. Mater.}\ }\textbf {\bibinfo
  {volume} {30}},\ \bibinfo {pages} {1703062} (\bibinfo {year}
  {2018})}\BibitemShut {NoStop}%
\bibitem [{\citenamefont {Yu}\ \emph {et~al.}(2010)\citenamefont {Yu},
  \citenamefont {Zhang}, \citenamefont {Zhang}, \citenamefont {Zhang},
  \citenamefont {Dai},\ and\ \citenamefont {Fang}}]{Model-DFT-CrBiSb2Te3-QAHE}%
  \BibitemOpen
  \bibfield  {author} {\bibinfo {author} {\bibfnamefont {R.}~\bibnamefont
  {Yu}}, \bibinfo {author} {\bibfnamefont {W.}~\bibnamefont {Zhang}}, \bibinfo
  {author} {\bibfnamefont {H.-J.}\ \bibnamefont {Zhang}}, \bibinfo {author}
  {\bibfnamefont {S.-C.}\ \bibnamefont {Zhang}}, \bibinfo {author}
  {\bibfnamefont {X.}~\bibnamefont {Dai}},\ and\ \bibinfo {author}
  {\bibfnamefont {Z.}~\bibnamefont {Fang}},\ }\bibinfo {title} {Quantized
  anomalous Hall effect in magnetic topological insulators},\ \href
  {https://doi.org/10.1126/science.1187485} {\bibfield  {journal} {\bibinfo
  {journal} {Science}\ }\textbf {\bibinfo {volume} {329}},\ \bibinfo {pages}
  {61} (\bibinfo {year} {2010})}\BibitemShut {NoStop}%
\bibitem [{\citenamefont {Serlin}\ \emph {et~al.}(2020)\citenamefont {Serlin},
  \citenamefont {Tschirhart}, \citenamefont {Polshyn}, \citenamefont {Zhang},
  \citenamefont {Zhu}, \citenamefont {Watanabe}, \citenamefont {Taniguchi},
  \citenamefont {Balents},\ and\ \citenamefont
  {Young}}]{EXP-QAHE-moire-twistedGr}%
  \BibitemOpen
  \bibfield  {author} {\bibinfo {author} {\bibfnamefont {M.}~\bibnamefont
  {Serlin}}, \bibinfo {author} {\bibfnamefont {C.~L.}\ \bibnamefont
  {Tschirhart}}, \bibinfo {author} {\bibfnamefont {H.}~\bibnamefont {Polshyn}},
  \bibinfo {author} {\bibfnamefont {Y.}~\bibnamefont {Zhang}}, \bibinfo
  {author} {\bibfnamefont {J.}~\bibnamefont {Zhu}}, \bibinfo {author}
  {\bibfnamefont {K.}~\bibnamefont {Watanabe}}, \bibinfo {author}
  {\bibfnamefont {T.}~\bibnamefont {Taniguchi}}, \bibinfo {author}
  {\bibfnamefont {L.}~\bibnamefont {Balents}},\ and\ \bibinfo {author}
  {\bibfnamefont {A.~F.}\ \bibnamefont {Young}},\ }\bibinfo {title} {Intrinsic
  quantized anomalous Hall effect in a moiré heterostructure},\ \href
  {https://doi.org/10.1126/science.aay5533} {\bibfield  {journal} {\bibinfo
  {journal} {Science}\ }\textbf {\bibinfo {volume} {367}},\ \bibinfo {pages}
  {900} (\bibinfo {year} {2020})}\BibitemShut {NoStop}%
\bibitem [{\citenamefont {Sharpe}\ \emph {et~al.}(2019)\citenamefont {Sharpe},
  \citenamefont {Fox}, \citenamefont {Barnard}, \citenamefont {Finney},
  \citenamefont {Watanabe}, \citenamefont {Taniguchi}, \citenamefont
  {Kastner},\ and\ \citenamefont {Goldhaber-Gordon}}]{EXP-QAHE-TwistedGr-2}%
  \BibitemOpen
  \bibfield  {author} {\bibinfo {author} {\bibfnamefont {A.~L.}\ \bibnamefont
  {Sharpe}}, \bibinfo {author} {\bibfnamefont {E.~J.}\ \bibnamefont {Fox}},
  \bibinfo {author} {\bibfnamefont {A.~W.}\ \bibnamefont {Barnard}}, \bibinfo
  {author} {\bibfnamefont {J.}~\bibnamefont {Finney}}, \bibinfo {author}
  {\bibfnamefont {K.}~\bibnamefont {Watanabe}}, \bibinfo {author}
  {\bibfnamefont {T.}~\bibnamefont {Taniguchi}}, \bibinfo {author}
  {\bibfnamefont {M.~A.}\ \bibnamefont {Kastner}},\ and\ \bibinfo {author}
  {\bibfnamefont {D.}~\bibnamefont {Goldhaber-Gordon}},\ }\bibinfo {title}
  {Emergent ferromagnetism near three-quarters filling in twisted bilayer
  graphene},\ \href {https://doi.org/10.1126/science.aaw3780} {\bibfield
  {journal} {\bibinfo  {journal} {Science}\ }\textbf {\bibinfo {volume}
  {365}},\ \bibinfo {pages} {605} (\bibinfo {year} {2019})}\BibitemShut
  {NoStop}%
\bibitem [{\citenamefont {Tschirhart}\ \emph {et~al.}(2021)\citenamefont
  {Tschirhart}, \citenamefont {Serlin}, \citenamefont {Polshyn}, \citenamefont
  {Shragai}, \citenamefont {Xia}, \citenamefont {Zhu}, \citenamefont {Zhang},
  \citenamefont {Watanabe}, \citenamefont {Taniguchi}, \citenamefont {Huber},\
  and\ \citenamefont {Young}}]{EXP-QAHE-TwistedGr-3}%
  \BibitemOpen
  \bibfield  {author} {\bibinfo {author} {\bibfnamefont {C.~L.}\ \bibnamefont
  {Tschirhart}}, \bibinfo {author} {\bibfnamefont {M.}~\bibnamefont {Serlin}},
  \bibinfo {author} {\bibfnamefont {H.}~\bibnamefont {Polshyn}}, \bibinfo
  {author} {\bibfnamefont {A.}~\bibnamefont {Shragai}}, \bibinfo {author}
  {\bibfnamefont {Z.}~\bibnamefont {Xia}}, \bibinfo {author} {\bibfnamefont
  {J.}~\bibnamefont {Zhu}}, \bibinfo {author} {\bibfnamefont {Y.}~\bibnamefont
  {Zhang}}, \bibinfo {author} {\bibfnamefont {K.}~\bibnamefont {Watanabe}},
  \bibinfo {author} {\bibfnamefont {T.}~\bibnamefont {Taniguchi}}, \bibinfo
  {author} {\bibfnamefont {M.~E.}\ \bibnamefont {Huber}},\ and\ \bibinfo
  {author} {\bibfnamefont {A.~F.}\ \bibnamefont {Young}},\ }\bibinfo {title}
  {Imaging orbital ferromagnetism in a moiré Chern insulator},\ \href
  {https://doi.org/10.1126/science.abd3190} {\bibfield  {journal} {\bibinfo
  {journal} {Science}\ }\textbf {\bibinfo {volume} {372}},\ \bibinfo {pages}
  {1323} (\bibinfo {year} {2021})}\BibitemShut {NoStop}%
\bibitem [{\citenamefont {Li}\ \emph {et~al.}(2021)\citenamefont {Li},
  \citenamefont {Jiang}, \citenamefont {Shen}, \citenamefont {Zhang},
  \citenamefont {Li}, \citenamefont {Tao}, \citenamefont {Devakul},
  \citenamefont {Watanabe}, \citenamefont {Taniguchi}, \citenamefont {Fu},
  \citenamefont {Shan},\ and\ \citenamefont {Mak}}]{EXP-QAHE-moire-MoTe2WSe2}%
  \BibitemOpen
  \bibfield  {author} {\bibinfo {author} {\bibfnamefont {T.}~\bibnamefont
  {Li}}, \bibinfo {author} {\bibfnamefont {S.}~\bibnamefont {Jiang}}, \bibinfo
  {author} {\bibfnamefont {B.}~\bibnamefont {Shen}}, \bibinfo {author}
  {\bibfnamefont {Y.}~\bibnamefont {Zhang}}, \bibinfo {author} {\bibfnamefont
  {L.}~\bibnamefont {Li}}, \bibinfo {author} {\bibfnamefont {Z.}~\bibnamefont
  {Tao}}, \bibinfo {author} {\bibfnamefont {T.}~\bibnamefont {Devakul}},
  \bibinfo {author} {\bibfnamefont {K.}~\bibnamefont {Watanabe}}, \bibinfo
  {author} {\bibfnamefont {T.}~\bibnamefont {Taniguchi}}, \bibinfo {author}
  {\bibfnamefont {L.}~\bibnamefont {Fu}}, \bibinfo {author} {\bibfnamefont
  {J.}~\bibnamefont {Shan}},\ and\ \bibinfo {author} {\bibfnamefont {K.~F.}\
  \bibnamefont {Mak}},\ }\bibinfo {title} {Quantum anomalous Hall effect from
  intertwined moiré bands},\ \href
  {https://doi.org/10.1038/s41586-021-04171-1} {\bibfield  {journal} {\bibinfo
  {journal} {Nature}\ }\textbf {\bibinfo {volume} {600}},\ \bibinfo {pages}
  {641} (\bibinfo {year} {2021})}\BibitemShut {NoStop}%
\bibitem [{\citenamefont {Lu}\ \emph {et~al.}(2025)\citenamefont {Lu},
  \citenamefont {Han}, \citenamefont {Yao}, \citenamefont {Hadjri},
  \citenamefont {Yang}, \citenamefont {Seo}, \citenamefont {Shi}, \citenamefont
  {Ye}, \citenamefont {Watanabe}, \citenamefont {Taniguchi},\ and\
  \citenamefont {Ju}}]{EXP-QAHE-Gr+BN}%
  \BibitemOpen
  \bibfield  {author} {\bibinfo {author} {\bibfnamefont {Z.}~\bibnamefont
  {Lu}}, \bibinfo {author} {\bibfnamefont {T.}~\bibnamefont {Han}}, \bibinfo
  {author} {\bibfnamefont {Y.}~\bibnamefont {Yao}}, \bibinfo {author}
  {\bibfnamefont {Z.}~\bibnamefont {Hadjri}}, \bibinfo {author} {\bibfnamefont
  {J.}~\bibnamefont {Yang}}, \bibinfo {author} {\bibfnamefont {J.}~\bibnamefont
  {Seo}}, \bibinfo {author} {\bibfnamefont {L.}~\bibnamefont {Shi}}, \bibinfo
  {author} {\bibfnamefont {S.}~\bibnamefont {Ye}}, \bibinfo {author}
  {\bibfnamefont {K.}~\bibnamefont {Watanabe}}, \bibinfo {author}
  {\bibfnamefont {T.}~\bibnamefont {Taniguchi}},\ and\ \bibinfo {author}
  {\bibfnamefont {L.}~\bibnamefont {Ju}},\ }\bibinfo {title} {Extended quantum
  anomalous Hall states in graphene/hBN moiré superlattices},\ \href
  {https://doi.org/10.1038/s41586-024-08470-1} {\bibfield  {journal} {\bibinfo
  {journal} {Nature}\ }\textbf {\bibinfo {volume} {637}},\ \bibinfo {pages}
  {1090} (\bibinfo {year} {2025})}\BibitemShut {NoStop}%
\bibitem [{\citenamefont {Deng}\ \emph {et~al.}(2020)\citenamefont {Deng},
  \citenamefont {Yu}, \citenamefont {Shi}, \citenamefont {Guo}, \citenamefont
  {Xu}, \citenamefont {Wang}, \citenamefont {Chen},\ and\ \citenamefont
  {Zhang}}]{EXP-QAHE-MnBi2Te4}%
  \BibitemOpen
  \bibfield  {author} {\bibinfo {author} {\bibfnamefont {Y.}~\bibnamefont
  {Deng}}, \bibinfo {author} {\bibfnamefont {Y.}~\bibnamefont {Yu}}, \bibinfo
  {author} {\bibfnamefont {M.~Z.}\ \bibnamefont {Shi}}, \bibinfo {author}
  {\bibfnamefont {Z.}~\bibnamefont {Guo}}, \bibinfo {author} {\bibfnamefont
  {Z.}~\bibnamefont {Xu}}, \bibinfo {author} {\bibfnamefont {J.}~\bibnamefont
  {Wang}}, \bibinfo {author} {\bibfnamefont {X.~H.}\ \bibnamefont {Chen}},\
  and\ \bibinfo {author} {\bibfnamefont {Y.}~\bibnamefont {Zhang}},\ }\bibinfo
  {title} {Quantum anomalous Hall effect in intrinsic magnetic topological
  insulator MnBi$_{\text{2}}$Te$_{\text{4}}$},\ \href
  {https://doi.org/10.1126/science.aax8156} {\bibfield  {journal} {\bibinfo
  {journal} {Science}\ }\textbf {\bibinfo {volume} {367}},\ \bibinfo {pages}
  {895} (\bibinfo {year} {2020})}\BibitemShut {NoStop}%
\bibitem [{\citenamefont {Liu}\ \emph {et~al.}(2020)\citenamefont {Liu},
  \citenamefont {Wang}, \citenamefont {Li}, \citenamefont {Wu}, \citenamefont
  {Li}, \citenamefont {Li}, \citenamefont {He}, \citenamefont {Xu},
  \citenamefont {Zhang},\ and\ \citenamefont {Wang}}]{EXP-QAHE-MnBi2Te4-2}%
  \BibitemOpen
  \bibfield  {author} {\bibinfo {author} {\bibfnamefont {C.}~\bibnamefont
  {Liu}}, \bibinfo {author} {\bibfnamefont {Y.}~\bibnamefont {Wang}}, \bibinfo
  {author} {\bibfnamefont {H.}~\bibnamefont {Li}}, \bibinfo {author}
  {\bibfnamefont {Y.}~\bibnamefont {Wu}}, \bibinfo {author} {\bibfnamefont
  {Y.}~\bibnamefont {Li}}, \bibinfo {author} {\bibfnamefont {J.}~\bibnamefont
  {Li}}, \bibinfo {author} {\bibfnamefont {K.}~\bibnamefont {He}}, \bibinfo
  {author} {\bibfnamefont {Y.}~\bibnamefont {Xu}}, \bibinfo {author}
  {\bibfnamefont {J.}~\bibnamefont {Zhang}},\ and\ \bibinfo {author}
  {\bibfnamefont {Y.}~\bibnamefont {Wang}},\ }\bibinfo {title} {Robust axion
  insulator and Chern insulator phases in a two-dimensional antiferromagnetic
  topological insulator},\ \href {https://doi.org/10.1038/s41563-019-0573-3}
  {\bibfield  {journal} {\bibinfo  {journal} {Nat. Mater.}\ }\textbf {\bibinfo
  {volume} {19}},\ \bibinfo {pages} {522} (\bibinfo {year} {2020})}\BibitemShut
  {NoStop}%
\bibitem [{\citenamefont {Ge}\ \emph {et~al.}(2020)\citenamefont {Ge},
  \citenamefont {Liu}, \citenamefont {Li}, \citenamefont {Li}, \citenamefont
  {Luo}, \citenamefont {Wu}, \citenamefont {Xu},\ and\ \citenamefont
  {Wang}}]{EXP-QAHE-MnBi2Te4-3}%
  \BibitemOpen
  \bibfield  {author} {\bibinfo {author} {\bibfnamefont {J.}~\bibnamefont
  {Ge}}, \bibinfo {author} {\bibfnamefont {Y.}~\bibnamefont {Liu}}, \bibinfo
  {author} {\bibfnamefont {J.}~\bibnamefont {Li}}, \bibinfo {author}
  {\bibfnamefont {H.}~\bibnamefont {Li}}, \bibinfo {author} {\bibfnamefont
  {T.}~\bibnamefont {Luo}}, \bibinfo {author} {\bibfnamefont {Y.}~\bibnamefont
  {Wu}}, \bibinfo {author} {\bibfnamefont {Y.}~\bibnamefont {Xu}},\ and\
  \bibinfo {author} {\bibfnamefont {J.}~\bibnamefont {Wang}},\ }\bibinfo
  {title} {High-Chern-number and high-temperature quantum Hall effect without
  Landau levels},\ \href {https://doi.org/10.1093/nsr/nwaa089} {\bibfield
  {journal} {\bibinfo  {journal} {Natl. Sci. Rev.}\ }\textbf {\bibinfo {volume}
  {7}},\ \bibinfo {pages} {1280} (\bibinfo {year} {2020})}\BibitemShut
  {NoStop}%
\bibitem [{\citenamefont {Liu}\ \emph {et~al.}(2008)\citenamefont {Liu},
  \citenamefont {Qi}, \citenamefont {Dai}, \citenamefont {Fang},\ and\
  \citenamefont {Zhang}}]{QAHE-Model-QuantumWell}%
  \BibitemOpen
  \bibfield  {author} {\bibinfo {author} {\bibfnamefont {C.-X.}\ \bibnamefont
  {Liu}}, \bibinfo {author} {\bibfnamefont {X.-L.}\ \bibnamefont {Qi}},
  \bibinfo {author} {\bibfnamefont {X.}~\bibnamefont {Dai}}, \bibinfo {author}
  {\bibfnamefont {Z.}~\bibnamefont {Fang}},\ and\ \bibinfo {author}
  {\bibfnamefont {S.-C.}\ \bibnamefont {Zhang}},\ }\bibinfo {title} {Quantum
  anomalous Hall effect in Hg$_{\text{1-y}}$Mn$_{\text{y}}$Te quantum wells},\
  \href {https://doi.org/10.1103/PhysRevLett.101.146802} {\bibfield  {journal}
  {\bibinfo  {journal} {Phys. Rev. Lett.}\ }\textbf {\bibinfo {volume} {101}},\
  \bibinfo {pages} {146802} (\bibinfo {year} {2008})}\BibitemShut {NoStop}%
\bibitem [{\citenamefont {Yang}\ \emph {et~al.}(2025)\citenamefont {Yang},
  \citenamefont {Li}, \citenamefont {Yi}, \citenamefont {Li}, \citenamefont
  {You}, \citenamefont {Su},\ and\ \citenamefont {Gu}}]{DFT-QAHE-Ge+CGT}%
  \BibitemOpen
  \bibfield  {author} {\bibinfo {author} {\bibfnamefont {Q.-H.}\ \bibnamefont
  {Yang}}, \bibinfo {author} {\bibfnamefont {J.-W.}\ \bibnamefont {Li}},
  \bibinfo {author} {\bibfnamefont {X.-W.}\ \bibnamefont {Yi}}, \bibinfo
  {author} {\bibfnamefont {X.}~\bibnamefont {Li}}, \bibinfo {author}
  {\bibfnamefont {J.-Y.}\ \bibnamefont {You}}, \bibinfo {author} {\bibfnamefont
  {G.}~\bibnamefont {Su}},\ and\ \bibinfo {author} {\bibfnamefont
  {B.}~\bibnamefont {Gu}},\ }\bibinfo {title} {Enhancement of temperature of
  the quantum anomalous Hall effect in two-dimensional
  germanene/magnetic-semiconductor heterostructures},\ \href
  {https://doi.org/10.1103/PhysRevB.111.184422} {\bibfield  {journal} {\bibinfo
   {journal} {Phys. Rev. B}\ }\textbf {\bibinfo {volume} {111}},\ \bibinfo
  {pages} {184422} (\bibinfo {year} {2025})}\BibitemShut {NoStop}%
\bibitem [{\citenamefont {Zou}\ \emph {et~al.}(2020)\citenamefont {Zou},
  \citenamefont {Zhan}, \citenamefont {Zheng}, \citenamefont {Wu},
  \citenamefont {Fan},\ and\ \citenamefont {Wang}}]{DFT-QAHE-Ge+CGT-1}%
  \BibitemOpen
  \bibfield  {author} {\bibinfo {author} {\bibfnamefont {R.}~\bibnamefont
  {Zou}}, \bibinfo {author} {\bibfnamefont {F.}~\bibnamefont {Zhan}}, \bibinfo
  {author} {\bibfnamefont {B.}~\bibnamefont {Zheng}}, \bibinfo {author}
  {\bibfnamefont {X.}~\bibnamefont {Wu}}, \bibinfo {author} {\bibfnamefont
  {J.}~\bibnamefont {Fan}},\ and\ \bibinfo {author} {\bibfnamefont
  {R.}~\bibnamefont {Wang}},\ }\bibinfo {title} {Intrinsic quantum anomalous
  Hall phase induced by proximity in the van der Waals heterostructure
  germanene/Cr$_{\text{2}}$Ge$_{\text{2}}$Te$_{\text{6}}$},\ \href
  {https://doi.org/10.1103/PhysRevB.101.161108} {\bibfield  {journal} {\bibinfo
   {journal} {Phys. Rev. B}\ }\textbf {\bibinfo {volume} {101}},\ \bibinfo
  {pages} {161108} (\bibinfo {year} {2020})}\BibitemShut {NoStop}%
\bibitem [{\citenamefont {Deng}\ \emph {et~al.}(2021)\citenamefont {Deng},
  \citenamefont {Chen}, \citenamefont {Wołoś}, \citenamefont {Konczykowski},
  \citenamefont {Sobczak}, \citenamefont {Sitnicka}, \citenamefont
  {Fedorchenko}, \citenamefont {Borysiuk}, \citenamefont {Heider},
  \citenamefont {Pluciński}, \citenamefont {Park}, \citenamefont {Georgescu},
  \citenamefont {Cano},\ and\ \citenamefont
  {Krusin-Elbaum}}]{EXP-MBT-Bi2Te3-Heterostructure-QAHE}%
  \BibitemOpen
  \bibfield  {author} {\bibinfo {author} {\bibfnamefont {H.}~\bibnamefont
  {Deng}}, \bibinfo {author} {\bibfnamefont {Z.}~\bibnamefont {Chen}}, \bibinfo
  {author} {\bibfnamefont {A.}~\bibnamefont {Wołoś}}, \bibinfo {author}
  {\bibfnamefont {M.}~\bibnamefont {Konczykowski}}, \bibinfo {author}
  {\bibfnamefont {K.}~\bibnamefont {Sobczak}}, \bibinfo {author} {\bibfnamefont
  {J.}~\bibnamefont {Sitnicka}}, \bibinfo {author} {\bibfnamefont {I.~V.}\
  \bibnamefont {Fedorchenko}}, \bibinfo {author} {\bibfnamefont
  {J.}~\bibnamefont {Borysiuk}}, \bibinfo {author} {\bibfnamefont
  {T.}~\bibnamefont {Heider}}, \bibinfo {author} {\bibfnamefont
  {L.}~\bibnamefont {Pluciński}}, \bibinfo {author} {\bibfnamefont
  {K.}~\bibnamefont {Park}}, \bibinfo {author} {\bibfnamefont {A.~B.}\
  \bibnamefont {Georgescu}}, \bibinfo {author} {\bibfnamefont {J.}~\bibnamefont
  {Cano}},\ and\ \bibinfo {author} {\bibfnamefont {L.}~\bibnamefont
  {Krusin-Elbaum}},\ }\bibinfo {title} {High-temperature quantum anomalous Hall
  regime in a MnBi$_{\text{2}}$Te$_{\text{4}}$/Bi$_{\text{2}}$Te$_{\text{3}}$
  superlattice},\ \href {https://doi.org/10.1038/s41567-020-0998-2} {\bibfield
  {journal} {\bibinfo  {journal} {Nat. Phys.}\ }\textbf {\bibinfo {volume}
  {17}},\ \bibinfo {pages} {36} (\bibinfo {year} {2021})}\BibitemShut {NoStop}%
\bibitem [{\citenamefont {Gong}\ \emph {et~al.}(2017)\citenamefont {Gong},
  \citenamefont {Li}, \citenamefont {Li}, \citenamefont {Ji}, \citenamefont
  {Stern}, \citenamefont {Xia}, \citenamefont {Cao}, \citenamefont {Bao},
  \citenamefont {Wang}, \citenamefont {Wang}, \citenamefont {Qiu},
  \citenamefont {Cava}, \citenamefont {Louie}, \citenamefont {Xia},\ and\
  \citenamefont {Zhang}}]{Cr2Ge2Te6-FMsemiconductor-Tc30K}%
  \BibitemOpen
  \bibfield  {author} {\bibinfo {author} {\bibfnamefont {C.}~\bibnamefont
  {Gong}}, \bibinfo {author} {\bibfnamefont {L.}~\bibnamefont {Li}}, \bibinfo
  {author} {\bibfnamefont {Z.}~\bibnamefont {Li}}, \bibinfo {author}
  {\bibfnamefont {H.}~\bibnamefont {Ji}}, \bibinfo {author} {\bibfnamefont
  {A.}~\bibnamefont {Stern}}, \bibinfo {author} {\bibfnamefont
  {Y.}~\bibnamefont {Xia}}, \bibinfo {author} {\bibfnamefont {T.}~\bibnamefont
  {Cao}}, \bibinfo {author} {\bibfnamefont {W.}~\bibnamefont {Bao}}, \bibinfo
  {author} {\bibfnamefont {C.}~\bibnamefont {Wang}}, \bibinfo {author}
  {\bibfnamefont {Y.}~\bibnamefont {Wang}}, \bibinfo {author} {\bibfnamefont
  {Z.~Q.}\ \bibnamefont {Qiu}}, \bibinfo {author} {\bibfnamefont {R.~J.}\
  \bibnamefont {Cava}}, \bibinfo {author} {\bibfnamefont {S.~G.}\ \bibnamefont
  {Louie}}, \bibinfo {author} {\bibfnamefont {J.}~\bibnamefont {Xia}},\ and\
  \bibinfo {author} {\bibfnamefont {X.}~\bibnamefont {Zhang}},\ }\bibinfo
  {title} {Discovery of intrinsic ferromagnetism in two-dimensional {van der
  Waals} crystals},\ \href {https://doi.org/10.1038/nature22060} {\bibfield
  {journal} {\bibinfo  {journal} {Nature}\ }\textbf {\bibinfo {volume} {546}},\
  \bibinfo {pages} {265} (\bibinfo {year} {2017})}\BibitemShut {NoStop}%
\bibitem [{\citenamefont {Huang}\ \emph {et~al.}(2017)\citenamefont {Huang},
  \citenamefont {Clark}, \citenamefont {Navarro-Moratalla}, \citenamefont
  {Klein}, \citenamefont {Cheng}, \citenamefont {Seyler}, \citenamefont
  {Zhong}, \citenamefont {Schmidgall}, \citenamefont {McGuire}, \citenamefont
  {Cobden}, \citenamefont {Yao}, \citenamefont {Xiao}, \citenamefont
  {Jarillo-Herrero},\ and\ \citenamefont {Xu}}]{CrI3-FMsemiconductor-Tc45K}%
  \BibitemOpen
  \bibfield  {author} {\bibinfo {author} {\bibfnamefont {B.}~\bibnamefont
  {Huang}}, \bibinfo {author} {\bibfnamefont {G.}~\bibnamefont {Clark}},
  \bibinfo {author} {\bibfnamefont {E.}~\bibnamefont {Navarro-Moratalla}},
  \bibinfo {author} {\bibfnamefont {D.~R.}\ \bibnamefont {Klein}}, \bibinfo
  {author} {\bibfnamefont {R.}~\bibnamefont {Cheng}}, \bibinfo {author}
  {\bibfnamefont {K.~L.}\ \bibnamefont {Seyler}}, \bibinfo {author}
  {\bibfnamefont {D.}~\bibnamefont {Zhong}}, \bibinfo {author} {\bibfnamefont
  {E.}~\bibnamefont {Schmidgall}}, \bibinfo {author} {\bibfnamefont {M.~A.}\
  \bibnamefont {McGuire}}, \bibinfo {author} {\bibfnamefont {D.~H.}\
  \bibnamefont {Cobden}}, \bibinfo {author} {\bibfnamefont {W.}~\bibnamefont
  {Yao}}, \bibinfo {author} {\bibfnamefont {D.}~\bibnamefont {Xiao}}, \bibinfo
  {author} {\bibfnamefont {P.}~\bibnamefont {Jarillo-Herrero}},\ and\ \bibinfo
  {author} {\bibfnamefont {X.}~\bibnamefont {Xu}},\ }\bibinfo {title}
  {Layer-dependent ferromagnetism in a {van der Waals} crystal down to the
  monolayer limit},\ \href {https://doi.org/10.1038/nature22391} {\bibfield
  {journal} {\bibinfo  {journal} {Nature}\ }\textbf {\bibinfo {volume} {546}},\
  \bibinfo {pages} {270} (\bibinfo {year} {2017})}\BibitemShut {NoStop}%
\bibitem [{\citenamefont {Lee}\ \emph {et~al.}(2021)\citenamefont {Lee},
  \citenamefont {Dismukes}, \citenamefont {Telford}, \citenamefont {Wiscons},
  \citenamefont {Wang}, \citenamefont {Xu}, \citenamefont {Nuckolls},
  \citenamefont {Dean}, \citenamefont {Roy},\ and\ \citenamefont
  {Zhu}}]{CrSBr-FMsemiconductor-Tc146K}%
  \BibitemOpen
  \bibfield  {author} {\bibinfo {author} {\bibfnamefont {K.}~\bibnamefont
  {Lee}}, \bibinfo {author} {\bibfnamefont {A.~H.}\ \bibnamefont {Dismukes}},
  \bibinfo {author} {\bibfnamefont {E.~J.}\ \bibnamefont {Telford}}, \bibinfo
  {author} {\bibfnamefont {R.~A.}\ \bibnamefont {Wiscons}}, \bibinfo {author}
  {\bibfnamefont {J.}~\bibnamefont {Wang}}, \bibinfo {author} {\bibfnamefont
  {X.}~\bibnamefont {Xu}}, \bibinfo {author} {\bibfnamefont {C.}~\bibnamefont
  {Nuckolls}}, \bibinfo {author} {\bibfnamefont {C.~R.}\ \bibnamefont {Dean}},
  \bibinfo {author} {\bibfnamefont {X.}~\bibnamefont {Roy}},\ and\ \bibinfo
  {author} {\bibfnamefont {X.}~\bibnamefont {Zhu}},\ }\bibinfo {title}
  {Magnetic order and symmetry in the {2D} semiconductor {CrSBr}},\ \href
  {https://doi.org/10.1021/acs.nanolett.1c00219} {\bibfield  {journal}
  {\bibinfo  {journal} {Nano Lett.}\ }\textbf {\bibinfo {volume} {21}},\
  \bibinfo {pages} {3511} (\bibinfo {year} {2021})}\BibitemShut {NoStop}%
\bibitem [{\citenamefont {Zhang}\ \emph {et~al.}(2019)\citenamefont {Zhang},
  \citenamefont {Shang}, \citenamefont {Jiang}, \citenamefont {Rasmita},
  \citenamefont {Gao},\ and\ \citenamefont {Yu}}]{CrBr3-FMsemiconductor-Tc34K}%
  \BibitemOpen
  \bibfield  {author} {\bibinfo {author} {\bibfnamefont {Z.}~\bibnamefont
  {Zhang}}, \bibinfo {author} {\bibfnamefont {J.}~\bibnamefont {Shang}},
  \bibinfo {author} {\bibfnamefont {C.}~\bibnamefont {Jiang}}, \bibinfo
  {author} {\bibfnamefont {A.}~\bibnamefont {Rasmita}}, \bibinfo {author}
  {\bibfnamefont {W.}~\bibnamefont {Gao}},\ and\ \bibinfo {author}
  {\bibfnamefont {T.}~\bibnamefont {Yu}},\ }\bibinfo {title} {Direct
  photoluminescence probing of ferromagnetism in monolayer two-Dimensional
  {CrBr$_{\text{3}}$}},\ \href {https://doi.org/10.1021/acs.nanolett.9b00553}
  {\bibfield  {journal} {\bibinfo  {journal} {Nano Lett.}\ }\textbf {\bibinfo
  {volume} {19}},\ \bibinfo {pages} {3138} (\bibinfo {year}
  {2019})}\BibitemShut {NoStop}%
\bibitem [{\citenamefont {Achinuq}\ \emph {et~al.}(2022)\citenamefont
  {Achinuq}, \citenamefont {Fujita}, \citenamefont {Xia}, \citenamefont {Guo},
  \citenamefont {Bencok}, \citenamefont {van~der Laan},\ and\ \citenamefont
  {Hesjedal}}]{CrSiTe3-FMsemiconductor-Tc80K}%
  \BibitemOpen
  \bibfield  {author} {\bibinfo {author} {\bibfnamefont {B.}~\bibnamefont
  {Achinuq}}, \bibinfo {author} {\bibfnamefont {R.}~\bibnamefont {Fujita}},
  \bibinfo {author} {\bibfnamefont {W.}~\bibnamefont {Xia}}, \bibinfo {author}
  {\bibfnamefont {Y.}~\bibnamefont {Guo}}, \bibinfo {author} {\bibfnamefont
  {P.}~\bibnamefont {Bencok}}, \bibinfo {author} {\bibfnamefont
  {G.}~\bibnamefont {van~der Laan}},\ and\ \bibinfo {author} {\bibfnamefont
  {T.}~\bibnamefont {Hesjedal}},\ }\bibinfo {title} {Covalent mixing in the
  {2D} ferromagnet {CrSiTe$_{\text{3}}$} evidenced by magnetic {X}-ray circular
  dichroism},\ \href {https://doi.org/https://doi.org/10.1002/pssr.202100566}
  {\bibfield  {journal} {\bibinfo  {journal} {Phys. Status Solidi RRL}\
  }\textbf {\bibinfo {volume} {16}},\ \bibinfo {pages} {2100566} (\bibinfo
  {year} {2022})}\BibitemShut {NoStop}%
\bibitem [{\citenamefont {Cai}\ \emph {et~al.}(2019)\citenamefont {Cai},
  \citenamefont {Song}, \citenamefont {Wilson}, \citenamefont {Clark},
  \citenamefont {He}, \citenamefont {Zhang}, \citenamefont {Taniguchi},
  \citenamefont {Watanabe}, \citenamefont {Yao}, \citenamefont {Xiao},
  \citenamefont {McGuire}, \citenamefont {Cobden},\ and\ \citenamefont
  {Xu}}]{CrCl3-FMsemiconductor-Tc17K}%
  \BibitemOpen
  \bibfield  {author} {\bibinfo {author} {\bibfnamefont {X.}~\bibnamefont
  {Cai}}, \bibinfo {author} {\bibfnamefont {T.}~\bibnamefont {Song}}, \bibinfo
  {author} {\bibfnamefont {N.~P.}\ \bibnamefont {Wilson}}, \bibinfo {author}
  {\bibfnamefont {G.}~\bibnamefont {Clark}}, \bibinfo {author} {\bibfnamefont
  {M.}~\bibnamefont {He}}, \bibinfo {author} {\bibfnamefont {X.}~\bibnamefont
  {Zhang}}, \bibinfo {author} {\bibfnamefont {T.}~\bibnamefont {Taniguchi}},
  \bibinfo {author} {\bibfnamefont {K.}~\bibnamefont {Watanabe}}, \bibinfo
  {author} {\bibfnamefont {W.}~\bibnamefont {Yao}}, \bibinfo {author}
  {\bibfnamefont {D.}~\bibnamefont {Xiao}}, \bibinfo {author} {\bibfnamefont
  {M.~A.}\ \bibnamefont {McGuire}}, \bibinfo {author} {\bibfnamefont {D.~H.}\
  \bibnamefont {Cobden}},\ and\ \bibinfo {author} {\bibfnamefont
  {X.}~\bibnamefont {Xu}},\ }\bibinfo {title} {Atomically thin
  {CrCl$_{\text{3}}$}: An in-plane layered antiferromagnetic insulator},\ \href
  {https://doi.org/10.1021/acs.nanolett.9b01317} {\bibfield  {journal}
  {\bibinfo  {journal} {Nano Lett.}\ }\textbf {\bibinfo {volume} {19}},\
  \bibinfo {pages} {3993} (\bibinfo {year} {2019})}\BibitemShut {NoStop}%
\bibitem [{\citenamefont {Cui}\ \emph {et~al.}(2020)\citenamefont {Cui},
  \citenamefont {Zhao}, \citenamefont {Xu}, \citenamefont {Tang}, \citenamefont
  {Shang}, \citenamefont {Shi}, \citenamefont {Huan}, \citenamefont {Liao},
  \citenamefont {Chen}, \citenamefont {Hou}, \citenamefont {Zhang},
  \citenamefont {Pennycook},\ and\ \citenamefont
  {Zhang}}]{Cr2S3-FMsemiconductor-Tc75K-1}%
  \BibitemOpen
  \bibfield  {author} {\bibinfo {author} {\bibfnamefont {F.}~\bibnamefont
  {Cui}}, \bibinfo {author} {\bibfnamefont {X.}~\bibnamefont {Zhao}}, \bibinfo
  {author} {\bibfnamefont {J.}~\bibnamefont {Xu}}, \bibinfo {author}
  {\bibfnamefont {B.}~\bibnamefont {Tang}}, \bibinfo {author} {\bibfnamefont
  {Q.}~\bibnamefont {Shang}}, \bibinfo {author} {\bibfnamefont
  {J.}~\bibnamefont {Shi}}, \bibinfo {author} {\bibfnamefont {Y.}~\bibnamefont
  {Huan}}, \bibinfo {author} {\bibfnamefont {J.}~\bibnamefont {Liao}}, \bibinfo
  {author} {\bibfnamefont {Q.}~\bibnamefont {Chen}}, \bibinfo {author}
  {\bibfnamefont {Y.}~\bibnamefont {Hou}}, \bibinfo {author} {\bibfnamefont
  {Q.}~\bibnamefont {Zhang}}, \bibinfo {author} {\bibfnamefont {S.~J.}\
  \bibnamefont {Pennycook}},\ and\ \bibinfo {author} {\bibfnamefont
  {Y.}~\bibnamefont {Zhang}},\ }\bibinfo {title} {Controlled growth and
  thickness-dependent conduction-type transition of {2D} ferrimagnetic
  {Cr$_{\text{2}}$S$_{\text{3}}$} semiconductors},\ \href
  {https://doi.org/https://doi.org/10.1002/adma.201905896} {\bibfield
  {journal} {\bibinfo  {journal} {Adv. Mater.}\ }\textbf {\bibinfo {volume}
  {32}},\ \bibinfo {pages} {1905896} (\bibinfo {year} {2020})}\BibitemShut
  {NoStop}%
\bibitem [{\citenamefont {Chu}\ \emph {et~al.}(2019)\citenamefont {Chu},
  \citenamefont {Zhang}, \citenamefont {Wen}, \citenamefont {Qiao},
  \citenamefont {Wu}, \citenamefont {He}, \citenamefont {Yin}, \citenamefont
  {Cheng}, \citenamefont {Wang}, \citenamefont {Wang}, \citenamefont {Xiong},
  \citenamefont {Li},\ and\ \citenamefont
  {He}}]{Cr2S3-FMsemiconductor-Tc75K-2}%
  \BibitemOpen
  \bibfield  {author} {\bibinfo {author} {\bibfnamefont {J.}~\bibnamefont
  {Chu}}, \bibinfo {author} {\bibfnamefont {Y.}~\bibnamefont {Zhang}}, \bibinfo
  {author} {\bibfnamefont {Y.}~\bibnamefont {Wen}}, \bibinfo {author}
  {\bibfnamefont {R.}~\bibnamefont {Qiao}}, \bibinfo {author} {\bibfnamefont
  {C.}~\bibnamefont {Wu}}, \bibinfo {author} {\bibfnamefont {P.}~\bibnamefont
  {He}}, \bibinfo {author} {\bibfnamefont {L.}~\bibnamefont {Yin}}, \bibinfo
  {author} {\bibfnamefont {R.}~\bibnamefont {Cheng}}, \bibinfo {author}
  {\bibfnamefont {F.}~\bibnamefont {Wang}}, \bibinfo {author} {\bibfnamefont
  {Z.}~\bibnamefont {Wang}}, \bibinfo {author} {\bibfnamefont {J.}~\bibnamefont
  {Xiong}}, \bibinfo {author} {\bibfnamefont {Y.}~\bibnamefont {Li}},\ and\
  \bibinfo {author} {\bibfnamefont {J.}~\bibnamefont {He}},\ }\bibinfo {title}
  {Sub-millimeter-scale growth of one-unit-cell-thick ferrimagnetic
  {Cr$_{\text{2}}$S$_{\text{3}}$} nanosheets},\ \href
  {https://doi.org/10.1021/acs.nanolett.9b00386} {\bibfield  {journal}
  {\bibinfo  {journal} {Nano Lett.}\ }\textbf {\bibinfo {volume} {19}},\
  \bibinfo {pages} {2154} (\bibinfo {year} {2019})}\BibitemShut {NoStop}%
\bibitem [{\citenamefont {You}\ \emph {et~al.}(2019{\natexlab{a}})\citenamefont
  {You}, \citenamefont {Zhang}, \citenamefont {Gu},\ and\ \citenamefont
  {Su}}]{PtBr3-FM-QAHE}%
  \BibitemOpen
  \bibfield  {author} {\bibinfo {author} {\bibfnamefont {J.-Y.}\ \bibnamefont
  {You}}, \bibinfo {author} {\bibfnamefont {Z.}~\bibnamefont {Zhang}}, \bibinfo
  {author} {\bibfnamefont {B.}~\bibnamefont {Gu}},\ and\ \bibinfo {author}
  {\bibfnamefont {G.}~\bibnamefont {Su}},\ }\bibinfo {title} {Two-dimensional
  room-temperature ferromagnetic semiconductors with quantum anomalous Hall
  effect},\ \href {https://doi.org/10.1103/PhysRevApplied.12.024063} {\bibfield
   {journal} {\bibinfo  {journal} {Phys. Rev. Appl.}\ }\textbf {\bibinfo
  {volume} {12}},\ \bibinfo {pages} {024063} (\bibinfo {year}
  {2019}{\natexlab{a}})}\BibitemShut {NoStop}%
\bibitem [{\citenamefont {You}\ \emph {et~al.}(2019{\natexlab{b}})\citenamefont
  {You}, \citenamefont {Chen}, \citenamefont {Zhang}, \citenamefont {Sheng},
  \citenamefont {Yang},\ and\ \citenamefont {Su}}]{DFT-QAHE-PtCl3}%
  \BibitemOpen
  \bibfield  {author} {\bibinfo {author} {\bibfnamefont {J.-Y.}\ \bibnamefont
  {You}}, \bibinfo {author} {\bibfnamefont {C.}~\bibnamefont {Chen}}, \bibinfo
  {author} {\bibfnamefont {Z.}~\bibnamefont {Zhang}}, \bibinfo {author}
  {\bibfnamefont {X.-L.}\ \bibnamefont {Sheng}}, \bibinfo {author}
  {\bibfnamefont {S.~A.}\ \bibnamefont {Yang}},\ and\ \bibinfo {author}
  {\bibfnamefont {G.}~\bibnamefont {Su}},\ }\bibinfo {title} {Two-dimensional
  Weyl half-semimetal and tunable quantum anomalous Hall effect},\ \href
  {https://doi.org/10.1103/PhysRevB.100.064408} {\bibfield  {journal} {\bibinfo
   {journal} {Phys. Rev. B}\ }\textbf {\bibinfo {volume} {100}},\ \bibinfo
  {pages} {064408} (\bibinfo {year} {2019}{\natexlab{b}})}\BibitemShut
  {NoStop}%
\bibitem [{\citenamefont {Li}\ \emph {et~al.}(2020)\citenamefont {Li},
  \citenamefont {Li}, \citenamefont {Li}, \citenamefont {Ye}, \citenamefont
  {Zheng}, \citenamefont {Zhang}, \citenamefont {Fu}, \citenamefont {Duan},\
  and\ \citenamefont {Xu}}]{DFT-LiFeSe-QAHE}%
  \BibitemOpen
  \bibfield  {author} {\bibinfo {author} {\bibfnamefont {Y.}~\bibnamefont
  {Li}}, \bibinfo {author} {\bibfnamefont {J.}~\bibnamefont {Li}}, \bibinfo
  {author} {\bibfnamefont {Y.}~\bibnamefont {Li}}, \bibinfo {author}
  {\bibfnamefont {M.}~\bibnamefont {Ye}}, \bibinfo {author} {\bibfnamefont
  {F.}~\bibnamefont {Zheng}}, \bibinfo {author} {\bibfnamefont
  {Z.}~\bibnamefont {Zhang}}, \bibinfo {author} {\bibfnamefont
  {J.}~\bibnamefont {Fu}}, \bibinfo {author} {\bibfnamefont {W.}~\bibnamefont
  {Duan}},\ and\ \bibinfo {author} {\bibfnamefont {Y.}~\bibnamefont {Xu}},\
  }\bibinfo {title} {High-temperature quantum anomalous Hall insulators in
  lithium-decorated iron-based superconductor materials},\ \href
  {https://doi.org/10.1103/PhysRevLett.125.086401} {\bibfield  {journal}
  {\bibinfo  {journal} {Phys. Rev. Lett.}\ }\textbf {\bibinfo {volume} {125}},\
  \bibinfo {pages} {086401} (\bibinfo {year} {2020})}\BibitemShut {NoStop}%
\bibitem [{\citenamefont {You}\ \emph {et~al.}(2022)\citenamefont {You},
  \citenamefont {Gu},\ and\ \citenamefont
  {Su}}]{DFT+Model-P-orbit-MagneticTI-QAHE}%
  \BibitemOpen
  \bibfield  {author} {\bibinfo {author} {\bibfnamefont {J.-Y.}\ \bibnamefont
  {You}}, \bibinfo {author} {\bibfnamefont {B.}~\bibnamefont {Gu}},\ and\
  \bibinfo {author} {\bibfnamefont {G.}~\bibnamefont {Su}},\ }\bibinfo {title}
  {The p-orbital magnetic topological states on a square lattice},\ \href
  {https://doi.org/10.1093/nsr/nwab114} {\bibfield  {journal} {\bibinfo
  {journal} {Natl. Sci. Rev.}\ }\textbf {\bibinfo {volume} {9}},\ \bibinfo
  {pages} {nwab114} (\bibinfo {year} {2022})}\BibitemShut {NoStop}%
\bibitem [{\citenamefont {Zhang}\ \emph {et~al.}(2021)\citenamefont {Zhang},
  \citenamefont {You}, \citenamefont {Ma}, \citenamefont {Gu},\ and\
  \citenamefont {Su}}]{DFT-QAHE-Kagome}%
  \BibitemOpen
  \bibfield  {author} {\bibinfo {author} {\bibfnamefont {Z.}~\bibnamefont
  {Zhang}}, \bibinfo {author} {\bibfnamefont {J.-Y.}\ \bibnamefont {You}},
  \bibinfo {author} {\bibfnamefont {X.-Y.}\ \bibnamefont {Ma}}, \bibinfo
  {author} {\bibfnamefont {B.}~\bibnamefont {Gu}},\ and\ \bibinfo {author}
  {\bibfnamefont {G.}~\bibnamefont {Su}},\ }\bibinfo {title} {Kagome quantum
  anomalous Hall effect with high Chern number and large band gap},\ \href
  {https://doi.org/10.1103/PhysRevB.103.014410} {\bibfield  {journal} {\bibinfo
   {journal} {Phys. Rev. B}\ }\textbf {\bibinfo {volume} {103}},\ \bibinfo
  {pages} {014410} (\bibinfo {year} {2021})}\BibitemShut {NoStop}%
\bibitem [{\citenamefont {Wang}\ \emph {et~al.}(2013)\citenamefont {Wang},
  \citenamefont {Liu},\ and\ \citenamefont {Liu}}]{DFT-QAHE-2Dorganic}%
  \BibitemOpen
  \bibfield  {author} {\bibinfo {author} {\bibfnamefont {Z.~F.}\ \bibnamefont
  {Wang}}, \bibinfo {author} {\bibfnamefont {Z.}~\bibnamefont {Liu}},\ and\
  \bibinfo {author} {\bibfnamefont {F.}~\bibnamefont {Liu}},\ }\bibinfo {title}
  {Quantum anomalous Hall effect in 2D organic topological insulators},\ \href
  {https://doi.org/10.1103/PhysRevLett.110.196801} {\bibfield  {journal}
  {\bibinfo  {journal} {Phys. Rev. Lett.}\ }\textbf {\bibinfo {volume} {110}},\
  \bibinfo {pages} {196801} (\bibinfo {year} {2013})}\BibitemShut {NoStop}%
\bibitem [{\citenamefont {Dong}\ \emph {et~al.}(2019)\citenamefont {Dong},
  \citenamefont {You}, \citenamefont {Gu},\ and\ \citenamefont
  {Su}}]{Cr2Ge2Te6-FMsemiconductor-EnhancementTc-Strain}%
  \BibitemOpen
  \bibfield  {author} {\bibinfo {author} {\bibfnamefont {X.-J.}\ \bibnamefont
  {Dong}}, \bibinfo {author} {\bibfnamefont {J.-Y.}\ \bibnamefont {You}},
  \bibinfo {author} {\bibfnamefont {B.}~\bibnamefont {Gu}},\ and\ \bibinfo
  {author} {\bibfnamefont {G.}~\bibnamefont {Su}},\ }\bibinfo {title}
  {Strain-induced room-temperature ferromagnetic semiconductors with large
  anomalous {Hall} conductivity in two-dimensional
  {Cr$_{\text{2}}$Ge$_{\text{2}}$Se$_{\text{6}}$}},\ \href
  {https://doi.org/10.1103/PhysRevApplied.12.014020} {\bibfield  {journal}
  {\bibinfo  {journal} {Phys. Rev. Appl.}\ }\textbf {\bibinfo {volume} {12}},\
  \bibinfo {pages} {014020} (\bibinfo {year} {2019})}\BibitemShut {NoStop}%
\bibitem [{\citenamefont {Dong}\ \emph {et~al.}(2020)\citenamefont {Dong},
  \citenamefont {You}, \citenamefont {Zhang}, \citenamefont {Gu},\ and\
  \citenamefont {Su}}]{Cr2Ge2Te2+PtSe2-FM-TcEnhancement}%
  \BibitemOpen
  \bibfield  {author} {\bibinfo {author} {\bibfnamefont {X.-J.}\ \bibnamefont
  {Dong}}, \bibinfo {author} {\bibfnamefont {J.-Y.}\ \bibnamefont {You}},
  \bibinfo {author} {\bibfnamefont {Z.}~\bibnamefont {Zhang}}, \bibinfo
  {author} {\bibfnamefont {B.}~\bibnamefont {Gu}},\ and\ \bibinfo {author}
  {\bibfnamefont {G.}~\bibnamefont {Su}},\ }\bibinfo {title} {Great enhancement
  of {Curie} temperature and magnetic anisotropy in two-dimensional {van der
  Waals} magnetic semiconductor heterostructures},\ \href
  {https://doi.org/10.1103/PhysRevB.102.144443} {\bibfield  {journal} {\bibinfo
   {journal} {Phys. Rev. B}\ }\textbf {\bibinfo {volume} {102}},\ \bibinfo
  {pages} {144443} (\bibinfo {year} {2020})}\BibitemShut {NoStop}%
\bibitem [{\citenamefont {You}\ \emph {et~al.}(2021)\citenamefont {You},
  \citenamefont {Dong}, \citenamefont {Gu},\ and\ \citenamefont
  {Su}}]{MnBi2Te4-FM-ElectricField-TcEnhancement}%
  \BibitemOpen
  \bibfield  {author} {\bibinfo {author} {\bibfnamefont {J.-Y.}\ \bibnamefont
  {You}}, \bibinfo {author} {\bibfnamefont {X.-J.}\ \bibnamefont {Dong}},
  \bibinfo {author} {\bibfnamefont {B.}~\bibnamefont {Gu}},\ and\ \bibinfo
  {author} {\bibfnamefont {G.}~\bibnamefont {Su}},\ }\bibinfo {title} {Electric
  field induced topological phase transition and large enhancements of
  spin-orbit coupling and {Curie} temperature in two-dimensional ferromagnetic
  semiconductors},\ \href {https://doi.org/10.1103/PhysRevB.103.104403}
  {\bibfield  {journal} {\bibinfo  {journal} {Phys. Rev. B}\ }\textbf {\bibinfo
  {volume} {103}},\ \bibinfo {pages} {104403} (\bibinfo {year}
  {2021})}\BibitemShut {NoStop}%
\bibitem [{\citenamefont {Huang}\ \emph {et~al.}(2018)\citenamefont {Huang},
  \citenamefont {Feng}, \citenamefont {Wu}, \citenamefont {Ahmed},
  \citenamefont {Huang}, \citenamefont {Xiang}, \citenamefont {Deng},\ and\
  \citenamefont {Kan}}]{DFT-CGT-Strain-EnhancedTc}%
  \BibitemOpen
  \bibfield  {author} {\bibinfo {author} {\bibfnamefont {C.}~\bibnamefont
  {Huang}}, \bibinfo {author} {\bibfnamefont {J.}~\bibnamefont {Feng}},
  \bibinfo {author} {\bibfnamefont {F.}~\bibnamefont {Wu}}, \bibinfo {author}
  {\bibfnamefont {D.}~\bibnamefont {Ahmed}}, \bibinfo {author} {\bibfnamefont
  {B.}~\bibnamefont {Huang}}, \bibinfo {author} {\bibfnamefont
  {H.}~\bibnamefont {Xiang}}, \bibinfo {author} {\bibfnamefont
  {K.}~\bibnamefont {Deng}},\ and\ \bibinfo {author} {\bibfnamefont
  {E.}~\bibnamefont {Kan}},\ }\bibinfo {title} {Toward intrinsic
  room-temperature ferromagnetism in two-dimensional semiconductors},\ \href
  {https://doi.org/10.1021/jacs.8b07879} {\bibfield  {journal} {\bibinfo
  {journal} {J. Am. Chem. Soc.}\ }\textbf {\bibinfo {volume} {140}},\ \bibinfo
  {pages} {11519} (\bibinfo {year} {2018})}\BibitemShut {NoStop}%
\bibitem [{\citenamefont {You}\ \emph {et~al.}(2020)\citenamefont {You},
  \citenamefont {Zhang}, \citenamefont {Dong}, \citenamefont {Gu},\ and\
  \citenamefont {Su}}]{DFT-TST-family-Ferromagnetic-Semiconductors}%
  \BibitemOpen
  \bibfield  {author} {\bibinfo {author} {\bibfnamefont {J.-Y.}\ \bibnamefont
  {You}}, \bibinfo {author} {\bibfnamefont {Z.}~\bibnamefont {Zhang}}, \bibinfo
  {author} {\bibfnamefont {X.-J.}\ \bibnamefont {Dong}}, \bibinfo {author}
  {\bibfnamefont {B.}~\bibnamefont {Gu}},\ and\ \bibinfo {author}
  {\bibfnamefont {G.}~\bibnamefont {Su}},\ }\bibinfo {title} {Two-dimensional
  magnetic semiconductors with room Curie temperatures},\ \href
  {https://doi.org/10.1103/PhysRevResearch.2.013002} {\bibfield  {journal}
  {\bibinfo  {journal} {Phys. Rev. Res.}\ }\textbf {\bibinfo {volume} {2}},\
  \bibinfo {pages} {013002} (\bibinfo {year} {2020})}\BibitemShut {NoStop}%
\bibitem [{\citenamefont {You}\ \emph {et~al.}(2023)\citenamefont {You},
  \citenamefont {Dong}, \citenamefont {Gu},\ and\ \citenamefont
  {Su}}]{MagneticSemiconductor-Review-GroupPaper}%
  \BibitemOpen
  \bibfield  {author} {\bibinfo {author} {\bibfnamefont {J.-Y.}\ \bibnamefont
  {You}}, \bibinfo {author} {\bibfnamefont {X.-J.}\ \bibnamefont {Dong}},
  \bibinfo {author} {\bibfnamefont {B.}~\bibnamefont {Gu}},\ and\ \bibinfo
  {author} {\bibfnamefont {G.}~\bibnamefont {Su}},\ }\bibinfo {title} {Possible
  room-temperature ferromagnetic semiconductors},\ \href
  {https://doi.org/10.1088/0256-307X/40/6/067502} {\bibfield  {journal}
  {\bibinfo  {journal} {Chin. Phys. Lett.}\ }\textbf {\bibinfo {volume} {40}},\
  \bibinfo {pages} {067502} (\bibinfo {year} {2023})}\BibitemShut {NoStop}%
\bibitem [{\citenamefont {Li}\ \emph {et~al.}(2023)\citenamefont {Li},
  \citenamefont {Zhang}, \citenamefont {You}, \citenamefont {Gu},\ and\
  \citenamefont {Su}}]{Cr3O6-FM-Semiconductor-DFT}%
  \BibitemOpen
  \bibfield  {author} {\bibinfo {author} {\bibfnamefont {J.-W.}\ \bibnamefont
  {Li}}, \bibinfo {author} {\bibfnamefont {Z.}~\bibnamefont {Zhang}}, \bibinfo
  {author} {\bibfnamefont {J.-Y.}\ \bibnamefont {You}}, \bibinfo {author}
  {\bibfnamefont {B.}~\bibnamefont {Gu}},\ and\ \bibinfo {author}
  {\bibfnamefont {G.}~\bibnamefont {Su}},\ }\bibinfo {title} {Two-dimensional
  Heisenberg model with material-dependent superexchange interactions},\ \href
  {https://doi.org/10.1103/PhysRevB.107.224411} {\bibfield  {journal} {\bibinfo
   {journal} {Phys. Rev. B}\ }\textbf {\bibinfo {volume} {107}},\ \bibinfo
  {pages} {224411} (\bibinfo {year} {2023})}\BibitemShut {NoStop}%
\bibitem [{\citenamefont {Wei}(2023)}]{Review-EXP-StrainedDSM-JOS}%
  \BibitemOpen
  \bibfield  {author} {\bibinfo {author} {\bibfnamefont {D.}~\bibnamefont
  {Wei}},\ }\bibinfo {title} {The room temperature ferromagnetism in highly
  strained two-dimensional magnetic semiconductors},\ \href
  {https://doi.org/10.1088/1674-4926/44/4/040401} {\bibfield  {journal}
  {\bibinfo  {journal} {J. Semicond.}\ }\textbf {\bibinfo {volume} {44}},\
  \bibinfo {pages} {040401} (\bibinfo {year} {2023})}\BibitemShut {NoStop}%
\bibitem [{\citenamefont {O'Neill}\ \emph {et~al.}(2023)\citenamefont
  {O'Neill}, \citenamefont {Rahman}, \citenamefont {Zhang}, \citenamefont
  {Schoenherr}, \citenamefont {Yildirim}, \citenamefont {Gu}, \citenamefont
  {Su}, \citenamefont {Lu},\ and\ \citenamefont
  {Seidel}}]{EXP-StrainedCGT-Tc-Enhancd}%
  \BibitemOpen
  \bibfield  {author} {\bibinfo {author} {\bibfnamefont {A.}~\bibnamefont
  {O'Neill}}, \bibinfo {author} {\bibfnamefont {S.}~\bibnamefont {Rahman}},
  \bibinfo {author} {\bibfnamefont {Z.}~\bibnamefont {Zhang}}, \bibinfo
  {author} {\bibfnamefont {P.}~\bibnamefont {Schoenherr}}, \bibinfo {author}
  {\bibfnamefont {T.}~\bibnamefont {Yildirim}}, \bibinfo {author}
  {\bibfnamefont {B.}~\bibnamefont {Gu}}, \bibinfo {author} {\bibfnamefont
  {G.}~\bibnamefont {Su}}, \bibinfo {author} {\bibfnamefont {Y.}~\bibnamefont
  {Lu}},\ and\ \bibinfo {author} {\bibfnamefont {J.}~\bibnamefont {Seidel}},\
  }\bibinfo {title} {Enhanced room temperature ferromagnetism in highly
  strained 2D semiconductor Cr$_{\text{2}}$Ge$_{\text{2}}$Te$_{\text{6}}$},\
  \href {https://doi.org/10.1021/acsnano.2c10209} {\bibfield  {journal}
  {\bibinfo  {journal} {ACS Nano}\ }\textbf {\bibinfo {volume} {17}},\ \bibinfo
  {pages} {735} (\bibinfo {year} {2023})}\BibitemShut {NoStop}%
\bibitem [{\citenamefont {Kuila}\ \emph {et~al.}(2012)\citenamefont {Kuila},
  \citenamefont {Bose}, \citenamefont {Mishra}, \citenamefont {Khanra},
  \citenamefont {Kim},\ and\ \citenamefont
  {Lee}}]{Review-Functionalization-Graphene}%
  \BibitemOpen
  \bibfield  {author} {\bibinfo {author} {\bibfnamefont {T.}~\bibnamefont
  {Kuila}}, \bibinfo {author} {\bibfnamefont {S.}~\bibnamefont {Bose}},
  \bibinfo {author} {\bibfnamefont {A.~K.}\ \bibnamefont {Mishra}}, \bibinfo
  {author} {\bibfnamefont {P.}~\bibnamefont {Khanra}}, \bibinfo {author}
  {\bibfnamefont {N.~H.}\ \bibnamefont {Kim}},\ and\ \bibinfo {author}
  {\bibfnamefont {J.~H.}\ \bibnamefont {Lee}},\ }\bibinfo {title} {Chemical
  functionalization of graphene and its applications},\ \href
  {https://doi.org/https://doi.org/10.1016/j.pmatsci.2012.03.002} {\bibfield
  {journal} {\bibinfo  {journal} {Prog. Mater. Sci.}\ }\textbf {\bibinfo
  {volume} {57}},\ \bibinfo {pages} {1061} (\bibinfo {year}
  {2012})}\BibitemShut {NoStop}%
\bibitem [{\citenamefont {Tozzini}\ and\ \citenamefont
  {Pellegrini}(2013)}]{2013-Review-Graphene-Hydrogenation}%
  \BibitemOpen
  \bibfield  {author} {\bibinfo {author} {\bibfnamefont {V.}~\bibnamefont
  {Tozzini}}\ and\ \bibinfo {author} {\bibfnamefont {V.}~\bibnamefont
  {Pellegrini}},\ }\bibinfo {title} {Prospects for hydrogen storage in
  graphene},\ \href {https://doi.org/10.1039/C2CP42538F} {\bibfield  {journal}
  {\bibinfo  {journal} {Phys. Chem. Chem. Phys.}\ }\textbf {\bibinfo {volume}
  {15}},\ \bibinfo {pages} {80} (\bibinfo {year} {2013})}\BibitemShut {NoStop}%
\bibitem [{\citenamefont {Zhao}\ and\ \citenamefont
  {Wu}(2018)}]{ReviewBook-Silicene-Functionalization}%
  \BibitemOpen
  \bibfield  {author} {\bibinfo {author} {\bibfnamefont {J.}~\bibnamefont
  {Zhao}}\ and\ \bibinfo {author} {\bibfnamefont {K.}~\bibnamefont {Wu}},\
  }\href {https://doi.org/10.1007/978-3-319-99964-7_11} {\bibinfo {title}
  {Silicene: Prediction, Synthesis, Application}},\ edited by\ \bibinfo
  {editor} {\bibfnamefont {P.}~\bibnamefont {Vogt}}\ and\ \bibinfo {editor}
  {\bibfnamefont {G.}~\bibnamefont {Le~Lay}}\ (\bibinfo  {publisher} {Springer
  International Publishing},\ \bibinfo {address} {Cham},\ \bibinfo {year}
  {2018})\ pp.\ \bibinfo {pages} {211--233}\BibitemShut {NoStop}%
\bibitem [{\citenamefont {Hartman}\ and\ \citenamefont
  {Sofer}(2019)}]{Review-Ge-Si-Sn-hydrogenation}%
  \BibitemOpen
  \bibfield  {author} {\bibinfo {author} {\bibfnamefont {T.}~\bibnamefont
  {Hartman}}\ and\ \bibinfo {author} {\bibfnamefont {Z.}~\bibnamefont
  {Sofer}},\ }\bibinfo {title} {Beyond graphene: Chemistry of group 14 graphene
  analogues: Silicene, germanene, and stanene},\ \href
  {https://doi.org/10.1021/acsnano.9b04466} {\bibfield  {journal} {\bibinfo
  {journal} {ACS Nano}\ }\textbf {\bibinfo {volume} {13}},\ \bibinfo {pages}
  {8566} (\bibinfo {year} {2019})}\BibitemShut {NoStop}%
\bibitem [{\citenamefont {Wang}\ \emph {et~al.}(2015)\citenamefont {Wang},
  \citenamefont {Xu},\ and\ \citenamefont
  {Pi}}]{Review-Silicene-Functionalization}%
  \BibitemOpen
  \bibfield  {author} {\bibinfo {author} {\bibfnamefont {R.}~\bibnamefont
  {Wang}}, \bibinfo {author} {\bibfnamefont {M.-S.}\ \bibnamefont {Xu}},\ and\
  \bibinfo {author} {\bibfnamefont {X.-D.}\ \bibnamefont {Pi}},\ }\bibinfo
  {title} {Chemical modification of silicene},\ \href
  {https://doi.org/10.1088/1674-1056/24/8/086807} {\bibfield  {journal}
  {\bibinfo  {journal} {Chin. Phys. B.}\ }\textbf {\bibinfo {volume} {24}},\
  \bibinfo {pages} {086807} (\bibinfo {year} {2015})}\BibitemShut {NoStop}%
\bibitem [{\citenamefont {Radhakrishnan}\ \emph {et~al.}(2017)\citenamefont
  {Radhakrishnan}, \citenamefont {Das}, \citenamefont {Samanta}, \citenamefont
  {de~los Reyes}, \citenamefont {Deng}, \citenamefont {Alemany}, \citenamefont
  {Weldeghiorghis}, \citenamefont {Khabashesku}, \citenamefont {Kochat},
  \citenamefont {Jin}, \citenamefont {Sudeep}, \citenamefont {Martí},
  \citenamefont {Chu}, \citenamefont {Roy}, \citenamefont {Tiwary},
  \citenamefont {Singh},\ and\ \citenamefont
  {Ajayan}}]{EXP-BN-Fluorinated-magnetism}%
  \BibitemOpen
  \bibfield  {author} {\bibinfo {author} {\bibfnamefont {S.}~\bibnamefont
  {Radhakrishnan}}, \bibinfo {author} {\bibfnamefont {D.}~\bibnamefont {Das}},
  \bibinfo {author} {\bibfnamefont {A.}~\bibnamefont {Samanta}}, \bibinfo
  {author} {\bibfnamefont {C.~A.}\ \bibnamefont {de~los Reyes}}, \bibinfo
  {author} {\bibfnamefont {L.}~\bibnamefont {Deng}}, \bibinfo {author}
  {\bibfnamefont {L.~B.}\ \bibnamefont {Alemany}}, \bibinfo {author}
  {\bibfnamefont {T.~K.}\ \bibnamefont {Weldeghiorghis}}, \bibinfo {author}
  {\bibfnamefont {V.~N.}\ \bibnamefont {Khabashesku}}, \bibinfo {author}
  {\bibfnamefont {V.}~\bibnamefont {Kochat}}, \bibinfo {author} {\bibfnamefont
  {Z.}~\bibnamefont {Jin}}, \bibinfo {author} {\bibfnamefont {P.~M.}\
  \bibnamefont {Sudeep}}, \bibinfo {author} {\bibfnamefont {A.~A.}\
  \bibnamefont {Martí}}, \bibinfo {author} {\bibfnamefont {C.-W.}\
  \bibnamefont {Chu}}, \bibinfo {author} {\bibfnamefont {A.}~\bibnamefont
  {Roy}}, \bibinfo {author} {\bibfnamefont {C.~S.}\ \bibnamefont {Tiwary}},
  \bibinfo {author} {\bibfnamefont {A.~K.}\ \bibnamefont {Singh}},\ and\
  \bibinfo {author} {\bibfnamefont {P.~M.}\ \bibnamefont {Ajayan}},\ }\bibinfo
  {title} {Fluorinated h-BN as a magnetic semiconductor},\ \href
  {https://doi.org/10.1126/sciadv.1700842} {\bibfield  {journal} {\bibinfo
  {journal} {Sci. Adv.}\ }\textbf {\bibinfo {volume} {3}},\ \bibinfo {pages}
  {e1700842} (\bibinfo {year} {2017})}\BibitemShut {NoStop}%
\bibitem [{\citenamefont {Robinson}\ \emph {et~al.}(2010)\citenamefont
  {Robinson}, \citenamefont {Burgess}, \citenamefont {Junkermeier},
  \citenamefont {Badescu}, \citenamefont {Reinecke}, \citenamefont {Perkins},
  \citenamefont {Zalalutdniov}, \citenamefont {Baldwin}, \citenamefont
  {Culbertson}, \citenamefont {Sheehan},\ and\ \citenamefont
  {Snow}}]{EXP-Graphene-Fluorinated}%
  \BibitemOpen
  \bibfield  {author} {\bibinfo {author} {\bibfnamefont {J.~T.}\ \bibnamefont
  {Robinson}}, \bibinfo {author} {\bibfnamefont {J.~S.}\ \bibnamefont
  {Burgess}}, \bibinfo {author} {\bibfnamefont {C.~E.}\ \bibnamefont
  {Junkermeier}}, \bibinfo {author} {\bibfnamefont {S.~C.}\ \bibnamefont
  {Badescu}}, \bibinfo {author} {\bibfnamefont {T.~L.}\ \bibnamefont
  {Reinecke}}, \bibinfo {author} {\bibfnamefont {F.~K.}\ \bibnamefont
  {Perkins}}, \bibinfo {author} {\bibfnamefont {M.~K.}\ \bibnamefont
  {Zalalutdniov}}, \bibinfo {author} {\bibfnamefont {J.~W.}\ \bibnamefont
  {Baldwin}}, \bibinfo {author} {\bibfnamefont {J.~C.}\ \bibnamefont
  {Culbertson}}, \bibinfo {author} {\bibfnamefont {P.~E.}\ \bibnamefont
  {Sheehan}},\ and\ \bibinfo {author} {\bibfnamefont {E.~S.}\ \bibnamefont
  {Snow}},\ }\bibinfo {title} {Properties of fluorinated graphene films},\
  \href {https://doi.org/10.1021/nl101437p} {\bibfield  {journal} {\bibinfo
  {journal} {Nano Lett.}\ }\textbf {\bibinfo {volume} {10}},\ \bibinfo {pages}
  {3001} (\bibinfo {year} {2010})}\BibitemShut {NoStop}%
\bibitem [{\citenamefont {Voiry}\ \emph {et~al.}(2015)\citenamefont {Voiry},
  \citenamefont {Goswami}, \citenamefont {Kappera}, \citenamefont {Silva},
  \citenamefont {Kaplan}, \citenamefont {Fujita}, \citenamefont {Chen},
  \citenamefont {Asefa},\ and\ \citenamefont
  {Chhowalla}}]{EXP-TMDs-2Hto1T-metaltosemicon-highcoverage}%
  \BibitemOpen
  \bibfield  {author} {\bibinfo {author} {\bibfnamefont {D.}~\bibnamefont
  {Voiry}}, \bibinfo {author} {\bibfnamefont {A.}~\bibnamefont {Goswami}},
  \bibinfo {author} {\bibfnamefont {R.}~\bibnamefont {Kappera}}, \bibinfo
  {author} {\bibfnamefont {C.~d. C. C.~e.}\ \bibnamefont {Silva}}, \bibinfo
  {author} {\bibfnamefont {D.}~\bibnamefont {Kaplan}}, \bibinfo {author}
  {\bibfnamefont {T.}~\bibnamefont {Fujita}}, \bibinfo {author} {\bibfnamefont
  {M.}~\bibnamefont {Chen}}, \bibinfo {author} {\bibfnamefont {T.}~\bibnamefont
  {Asefa}},\ and\ \bibinfo {author} {\bibfnamefont {M.}~\bibnamefont
  {Chhowalla}},\ }\bibinfo {title} {Covalent functionalization of monolayered
  transition metal dichalcogenides by phase engineering},\ \href
  {https://doi.org/10.1038/nchem.2108} {\bibfield  {journal} {\bibinfo
  {journal} {Nat. Chem.}\ }\textbf {\bibinfo {volume} {7}},\ \bibinfo {pages}
  {45} (\bibinfo {year} {2015})}\BibitemShut {NoStop}%
\bibitem [{\citenamefont {Yue}\ \emph {et~al.}(2013)\citenamefont {Yue},
  \citenamefont {Chang}, \citenamefont {Qin},\ and\ \citenamefont
  {Li}}]{DFT-2HMoS2-Hydrogenation-2}%
  \BibitemOpen
  \bibfield  {author} {\bibinfo {author} {\bibfnamefont {Q.}~\bibnamefont
  {Yue}}, \bibinfo {author} {\bibfnamefont {S.}~\bibnamefont {Chang}}, \bibinfo
  {author} {\bibfnamefont {S.}~\bibnamefont {Qin}},\ and\ \bibinfo {author}
  {\bibfnamefont {J.}~\bibnamefont {Li}},\ }\bibinfo {title} {Functionalization
  of monolayer MoS$_{2}$ by substitutional doping: A first-principles study},\
  \href {https://doi.org/https://doi.org/10.1016/j.physleta.2013.03.034}
  {\bibfield  {journal} {\bibinfo  {journal} {Phys. Lett. A}\ }\textbf
  {\bibinfo {volume} {377}},\ \bibinfo {pages} {1362} (\bibinfo {year}
  {2013})}\BibitemShut {NoStop}%
\bibitem [{\citenamefont {{Keong Koh}}\ \emph {et~al.}(2012)\citenamefont
  {{Keong Koh}}, \citenamefont {Chiu}, \citenamefont {Lim}, \citenamefont
  {Zhang},\ and\ \citenamefont {Pan}}]{DFT-2HMoS2-Hydrogenation-3}%
  \BibitemOpen
  \bibfield  {author} {\bibinfo {author} {\bibfnamefont {E.~W.}\ \bibnamefont
  {{Keong Koh}}}, \bibinfo {author} {\bibfnamefont {C.~H.}\ \bibnamefont
  {Chiu}}, \bibinfo {author} {\bibfnamefont {Y.~K.}\ \bibnamefont {Lim}},
  \bibinfo {author} {\bibfnamefont {Y.-W.}\ \bibnamefont {Zhang}},\ and\
  \bibinfo {author} {\bibfnamefont {H.}~\bibnamefont {Pan}},\ }\bibinfo {title}
  {Hydrogen adsorption on and diffusion through MoS$_{2}$ monolayer:
  First-principles study},\ \href
  {https://doi.org/https://doi.org/10.1016/j.ijhydene.2012.07.069} {\bibfield
  {journal} {\bibinfo  {journal} {Int. J. Hydrogen Energy}\ }\textbf {\bibinfo
  {volume} {37}},\ \bibinfo {pages} {14323} (\bibinfo {year}
  {2012})}\BibitemShut {NoStop}%
\bibitem [{\citenamefont {Huang}\ \emph {et~al.}(2013)\citenamefont {Huang},
  \citenamefont {Xiang},\ and\ \citenamefont
  {Wei}}]{DFT-Silicene-monolayer-single-surface-adsorp-F}%
  \BibitemOpen
  \bibfield  {author} {\bibinfo {author} {\bibfnamefont {B.}~\bibnamefont
  {Huang}}, \bibinfo {author} {\bibfnamefont {H.~J.}\ \bibnamefont {Xiang}},\
  and\ \bibinfo {author} {\bibfnamefont {S.-H.}\ \bibnamefont {Wei}},\
  }\bibinfo {title} {Chemical functionalization of silicene: spontaneous
  structural transition and exotic electronic properties},\ \href
  {https://doi.org/10.1103/PhysRevLett.111.145502} {\bibfield  {journal}
  {\bibinfo  {journal} {Phys. Rev. Lett.}\ }\textbf {\bibinfo {volume} {111}},\
  \bibinfo {pages} {145502} (\bibinfo {year} {2013})}\BibitemShut {NoStop}%
\bibitem [{\citenamefont {Xu}\ \emph {et~al.}(2013)\citenamefont {Xu},
  \citenamefont {Yan}, \citenamefont {Zhang}, \citenamefont {Wang},
  \citenamefont {Xu}, \citenamefont {Tang}, \citenamefont {Duan},\ and\
  \citenamefont {Zhang}}]{DFT-Silicene-monolayer-two-side-Functionalization}%
  \BibitemOpen
  \bibfield  {author} {\bibinfo {author} {\bibfnamefont {Y.}~\bibnamefont
  {Xu}}, \bibinfo {author} {\bibfnamefont {B.}~\bibnamefont {Yan}}, \bibinfo
  {author} {\bibfnamefont {H.-J.}\ \bibnamefont {Zhang}}, \bibinfo {author}
  {\bibfnamefont {J.}~\bibnamefont {Wang}}, \bibinfo {author} {\bibfnamefont
  {G.}~\bibnamefont {Xu}}, \bibinfo {author} {\bibfnamefont {P.}~\bibnamefont
  {Tang}}, \bibinfo {author} {\bibfnamefont {W.}~\bibnamefont {Duan}},\ and\
  \bibinfo {author} {\bibfnamefont {S.-C.}\ \bibnamefont {Zhang}},\ }\bibinfo
  {title} {Large-gap quantum spin Hall insulators in tin films},\ \href
  {https://doi.org/10.1103/PhysRevLett.111.136804} {\bibfield  {journal}
  {\bibinfo  {journal} {Phys. Rev. Lett.}\ }\textbf {\bibinfo {volume} {111}},\
  \bibinfo {pages} {136804} (\bibinfo {year} {2013})}\BibitemShut {NoStop}%
\bibitem [{\citenamefont {Houssa}\ \emph {et~al.}(2011)\citenamefont {Houssa},
  \citenamefont {Scalise}, \citenamefont {Sankaran}, \citenamefont {Pourtois},
  \citenamefont {Afanas'ev},\ and\ \citenamefont
  {Stesmans}}]{DFT-Hydrogenation-Si-Ge}%
  \BibitemOpen
  \bibfield  {author} {\bibinfo {author} {\bibfnamefont {M.}~\bibnamefont
  {Houssa}}, \bibinfo {author} {\bibfnamefont {E.}~\bibnamefont {Scalise}},
  \bibinfo {author} {\bibfnamefont {K.}~\bibnamefont {Sankaran}}, \bibinfo
  {author} {\bibfnamefont {G.}~\bibnamefont {Pourtois}}, \bibinfo {author}
  {\bibfnamefont {V.~V.}\ \bibnamefont {Afanas'ev}},\ and\ \bibinfo {author}
  {\bibfnamefont {A.}~\bibnamefont {Stesmans}},\ }\bibinfo {title} {Electronic
  properties of hydrogenated silicene and germanene},\ \href
  {https://doi.org/10.1063/1.3595682} {\bibfield  {journal} {\bibinfo
  {journal} {Appl. Phys. Lett.}\ }\textbf {\bibinfo {volume} {98}},\ \bibinfo
  {pages} {223107} (\bibinfo {year} {2011})}\BibitemShut {NoStop}%
\bibitem [{\citenamefont {Rassekh}\ \emph {et~al.}(2020)\citenamefont
  {Rassekh}, \citenamefont {He}, \citenamefont {{Farjami Shayesteh}},\ and\
  \citenamefont {Palacios}}]{DFT-CrI3-Tc-enhanced-Functionalization-O}%
  \BibitemOpen
  \bibfield  {author} {\bibinfo {author} {\bibfnamefont {M.}~\bibnamefont
  {Rassekh}}, \bibinfo {author} {\bibfnamefont {J.}~\bibnamefont {He}},
  \bibinfo {author} {\bibfnamefont {S.}~\bibnamefont {{Farjami Shayesteh}}},\
  and\ \bibinfo {author} {\bibfnamefont {J.~J.}\ \bibnamefont {Palacios}},\
  }\bibinfo {title} {Remarkably enhanced Curie temperature in monolayer
  CrI$_{3}$ by hydrogen and oxygen adsorption: A first-principles
  calculations},\ \href
  {https://doi.org/https://doi.org/10.1016/j.commatsci.2020.109820} {\bibfield
  {journal} {\bibinfo  {journal} {Comput. Mater. Sci.}\ }\textbf {\bibinfo
  {volume} {183}},\ \bibinfo {pages} {109820} (\bibinfo {year}
  {2020})}\BibitemShut {NoStop}%
\bibitem [{\citenamefont {He}\ \emph {et~al.}(2019)\citenamefont {He},
  \citenamefont {Ding}, \citenamefont {Zhong}, \citenamefont {Li},
  \citenamefont {Li},\ and\ \citenamefont
  {Zhang}}]{DFT-cgt-MolecularAdsorption-Tc-enhanced}%
  \BibitemOpen
  \bibfield  {author} {\bibinfo {author} {\bibfnamefont {J.}~\bibnamefont
  {He}}, \bibinfo {author} {\bibfnamefont {G.}~\bibnamefont {Ding}}, \bibinfo
  {author} {\bibfnamefont {C.}~\bibnamefont {Zhong}}, \bibinfo {author}
  {\bibfnamefont {S.}~\bibnamefont {Li}}, \bibinfo {author} {\bibfnamefont
  {D.}~\bibnamefont {Li}},\ and\ \bibinfo {author} {\bibfnamefont
  {G.}~\bibnamefont {Zhang}},\ }\bibinfo {title} {Remarkably enhanced
  ferromagnetism in a super-exchange governed
  Cr$_{\text{2}}$Ge$_{\text{2}}$Te$_{\text{6}}$ monolayer via molecular
  adsorption},\ \href {https://doi.org/10.1039/C8TC05530K} {\bibfield
  {journal} {\bibinfo  {journal} {J. Mater. Chem. C}\ }\textbf {\bibinfo
  {volume} {7}},\ \bibinfo {pages} {5084} (\bibinfo {year} {2019})}\BibitemShut
  {NoStop}%
\bibitem [{\citenamefont {Song}\ \emph {et~al.}(2019)\citenamefont {Song},
  \citenamefont {Xiao}, \citenamefont {Li}, \citenamefont {Lu}, \citenamefont
  {Jiang}, \citenamefont {Li}, \citenamefont {Chen},\ and\ \citenamefont
  {Zhong}}]{DFT-cgt-singlesite-adsorption-Tc-enhanced}%
  \BibitemOpen
  \bibfield  {author} {\bibinfo {author} {\bibfnamefont {C.}~\bibnamefont
  {Song}}, \bibinfo {author} {\bibfnamefont {W.}~\bibnamefont {Xiao}}, \bibinfo
  {author} {\bibfnamefont {L.}~\bibnamefont {Li}}, \bibinfo {author}
  {\bibfnamefont {Y.}~\bibnamefont {Lu}}, \bibinfo {author} {\bibfnamefont
  {P.}~\bibnamefont {Jiang}}, \bibinfo {author} {\bibfnamefont
  {C.}~\bibnamefont {Li}}, \bibinfo {author} {\bibfnamefont {A.}~\bibnamefont
  {Chen}},\ and\ \bibinfo {author} {\bibfnamefont {Z.}~\bibnamefont {Zhong}},\
  }\bibinfo {title} {Tunable band gap and enhanced ferromagnetism by surface
  adsorption in monolayer Cr$_{\text{2}}$Ge$_{\text{2}}$Te$_{\text{6}}$},\
  \href {https://doi.org/10.1103/PhysRevB.99.214435} {\bibfield  {journal}
  {\bibinfo  {journal} {Phys. Rev. B}\ }\textbf {\bibinfo {volume} {99}},\
  \bibinfo {pages} {214435} (\bibinfo {year} {2019})}\BibitemShut {NoStop}%
\bibitem [{\citenamefont {Zhou}\ \emph {et~al.}(2009)\citenamefont {Zhou},
  \citenamefont {Wang}, \citenamefont {Sun}, \citenamefont {Chen},
  \citenamefont {Kawazoe},\ and\ \citenamefont
  {Jena}}]{DFT-Graphene-Semihydrogenation-Ferromagnetism}%
  \BibitemOpen
  \bibfield  {author} {\bibinfo {author} {\bibfnamefont {J.}~\bibnamefont
  {Zhou}}, \bibinfo {author} {\bibfnamefont {Q.}~\bibnamefont {Wang}}, \bibinfo
  {author} {\bibfnamefont {Q.}~\bibnamefont {Sun}}, \bibinfo {author}
  {\bibfnamefont {X.~S.}\ \bibnamefont {Chen}}, \bibinfo {author}
  {\bibfnamefont {Y.}~\bibnamefont {Kawazoe}},\ and\ \bibinfo {author}
  {\bibfnamefont {P.}~\bibnamefont {Jena}},\ }\bibinfo {title} {Ferromagnetism
  in semihydrogenated graphene sheet},\ \href
  {https://doi.org/10.1021/nl9020733} {\bibfield  {journal} {\bibinfo
  {journal} {Nano Lett.}\ }\textbf {\bibinfo {volume} {9}},\ \bibinfo {pages}
  {3867} (\bibinfo {year} {2009})}\BibitemShut {NoStop}%
\bibitem [{\citenamefont {Zhou}\ \emph {et~al.}(2010)\citenamefont {Zhou},
  \citenamefont {Wang}, \citenamefont {Sun},\ and\ \citenamefont
  {Jena}}]{DFT-BN-Hydrogenation-monolayer}%
  \BibitemOpen
  \bibfield  {author} {\bibinfo {author} {\bibfnamefont {J.}~\bibnamefont
  {Zhou}}, \bibinfo {author} {\bibfnamefont {Q.}~\bibnamefont {Wang}}, \bibinfo
  {author} {\bibfnamefont {Q.}~\bibnamefont {Sun}},\ and\ \bibinfo {author}
  {\bibfnamefont {P.}~\bibnamefont {Jena}},\ }\bibinfo {title} {Electronic and
  magnetic properties of a BN sheet decorated with hydrogen and fluorine},\
  \href {https://doi.org/10.1103/PhysRevB.81.085442} {\bibfield  {journal}
  {\bibinfo  {journal} {Phys. Rev. B}\ }\textbf {\bibinfo {volume} {81}},\
  \bibinfo {pages} {085442} (\bibinfo {year} {2010})}\BibitemShut {NoStop}%
\bibitem [{\citenamefont {Ma}\ \emph {et~al.}(2015)\citenamefont {Ma},
  \citenamefont {Li}, \citenamefont {Kou}, \citenamefont {Yan}, \citenamefont
  {Niu}, \citenamefont {Dai},\ and\ \citenamefont
  {Heine}}]{DFT-IIIBi-monolayer-Hydrogenated-fluorinated-TI}%
  \BibitemOpen
  \bibfield  {author} {\bibinfo {author} {\bibfnamefont {Y.}~\bibnamefont
  {Ma}}, \bibinfo {author} {\bibfnamefont {X.}~\bibnamefont {Li}}, \bibinfo
  {author} {\bibfnamefont {L.}~\bibnamefont {Kou}}, \bibinfo {author}
  {\bibfnamefont {B.}~\bibnamefont {Yan}}, \bibinfo {author} {\bibfnamefont
  {C.}~\bibnamefont {Niu}}, \bibinfo {author} {\bibfnamefont {Y.}~\bibnamefont
  {Dai}},\ and\ \bibinfo {author} {\bibfnamefont {T.}~\bibnamefont {Heine}},\
  }\bibinfo {title} {Two-dimensional inversion-asymmetric topological
  insulators in functionalized III-Bi bilayers},\ \href
  {https://doi.org/10.1103/PhysRevB.91.235306} {\bibfield  {journal} {\bibinfo
  {journal} {Phys. Rev. B}\ }\textbf {\bibinfo {volume} {91}},\ \bibinfo
  {pages} {235306} (\bibinfo {year} {2015})}\BibitemShut {NoStop}%
\bibitem [{\citenamefont {Crisostomo}\ \emph {et~al.}(2015)\citenamefont
  {Crisostomo}, \citenamefont {Yao}, \citenamefont {Huang}, \citenamefont
  {Hsu}, \citenamefont {Chuang}, \citenamefont {Lin}, \citenamefont {Albao},\
  and\ \citenamefont {Bansil}}]{DFT-monolayer-Hydrogenated-TI}%
  \BibitemOpen
  \bibfield  {author} {\bibinfo {author} {\bibfnamefont {C.~P.}\ \bibnamefont
  {Crisostomo}}, \bibinfo {author} {\bibfnamefont {L.-Z.}\ \bibnamefont {Yao}},
  \bibinfo {author} {\bibfnamefont {Z.-Q.}\ \bibnamefont {Huang}}, \bibinfo
  {author} {\bibfnamefont {C.-H.}\ \bibnamefont {Hsu}}, \bibinfo {author}
  {\bibfnamefont {F.-C.}\ \bibnamefont {Chuang}}, \bibinfo {author}
  {\bibfnamefont {H.}~\bibnamefont {Lin}}, \bibinfo {author} {\bibfnamefont
  {M.~A.}\ \bibnamefont {Albao}},\ and\ \bibinfo {author} {\bibfnamefont
  {A.}~\bibnamefont {Bansil}},\ }\bibinfo {title} {Robust large gap
  two-dimensional topological insulators in hydrogenated III-V buckled
  honeycombs},\ \href {https://doi.org/10.1021/acs.nanolett.5b02293} {\bibfield
   {journal} {\bibinfo  {journal} {Nano Lett.}\ }\textbf {\bibinfo {volume}
  {15}},\ \bibinfo {pages} {6568} (\bibinfo {year} {2015})}\BibitemShut
  {NoStop}%
\bibitem [{\citenamefont {Li}\ \emph {et~al.}(2015)\citenamefont {Li},
  \citenamefont {Zhang}, \citenamefont {Chen},\ and\ \citenamefont
  {Zhao}}]{DFT-monolayer-GaBiCl2-TI-5strain}%
  \BibitemOpen
  \bibfield  {author} {\bibinfo {author} {\bibfnamefont {L.}~\bibnamefont
  {Li}}, \bibinfo {author} {\bibfnamefont {X.}~\bibnamefont {Zhang}}, \bibinfo
  {author} {\bibfnamefont {X.}~\bibnamefont {Chen}},\ and\ \bibinfo {author}
  {\bibfnamefont {M.}~\bibnamefont {Zhao}},\ }\bibinfo {title} {Giant
  topological nontrivial band gaps in chloridized gallium bismuthide},\ \href
  {https://doi.org/10.1021/nl504493d} {\bibfield  {journal} {\bibinfo
  {journal} {Nano Lett.}\ }\textbf {\bibinfo {volume} {15}},\ \bibinfo {pages}
  {1296} (\bibinfo {year} {2015})}\BibitemShut {NoStop}%
\bibitem [{\citenamefont {Liang}\ \emph {et~al.}(2017)\citenamefont {Liang},
  \citenamefont {Khazaei}, \citenamefont {Ranjbar}, \citenamefont {Arai},
  \citenamefont {Yunoki}, \citenamefont {Kawazoe}, \citenamefont {Weng},\ and\
  \citenamefont {Fang}}]{DFT-Ti3N2F2-MXene-TI-monolayer}%
  \BibitemOpen
  \bibfield  {author} {\bibinfo {author} {\bibfnamefont {Y.}~\bibnamefont
  {Liang}}, \bibinfo {author} {\bibfnamefont {M.}~\bibnamefont {Khazaei}},
  \bibinfo {author} {\bibfnamefont {A.}~\bibnamefont {Ranjbar}}, \bibinfo
  {author} {\bibfnamefont {M.}~\bibnamefont {Arai}}, \bibinfo {author}
  {\bibfnamefont {S.}~\bibnamefont {Yunoki}}, \bibinfo {author} {\bibfnamefont
  {Y.}~\bibnamefont {Kawazoe}}, \bibinfo {author} {\bibfnamefont
  {H.}~\bibnamefont {Weng}},\ and\ \bibinfo {author} {\bibfnamefont
  {Z.}~\bibnamefont {Fang}},\ }\bibinfo {title} {Theoretical prediction of
  two-dimensional functionalized MXene nitrides as topological insulators},\
  \href {https://doi.org/10.1103/PhysRevB.96.195414} {\bibfield  {journal}
  {\bibinfo  {journal} {Phys. Rev. B}\ }\textbf {\bibinfo {volume} {96}},\
  \bibinfo {pages} {195414} (\bibinfo {year} {2017})}\BibitemShut {NoStop}%
\bibitem [{\citenamefont {Weng}\ \emph {et~al.}(2015)\citenamefont {Weng},
  \citenamefont {Ranjbar}, \citenamefont {Liang}, \citenamefont {Song},
  \citenamefont {Khazaei}, \citenamefont {Yunoki}, \citenamefont {Arai},
  \citenamefont {Kawazoe}, \citenamefont {Fang},\ and\ \citenamefont
  {Dai}}]{DFT-M2CO2-MXene-TI-monolayer}%
  \BibitemOpen
  \bibfield  {author} {\bibinfo {author} {\bibfnamefont {H.}~\bibnamefont
  {Weng}}, \bibinfo {author} {\bibfnamefont {A.}~\bibnamefont {Ranjbar}},
  \bibinfo {author} {\bibfnamefont {Y.}~\bibnamefont {Liang}}, \bibinfo
  {author} {\bibfnamefont {Z.}~\bibnamefont {Song}}, \bibinfo {author}
  {\bibfnamefont {M.}~\bibnamefont {Khazaei}}, \bibinfo {author} {\bibfnamefont
  {S.}~\bibnamefont {Yunoki}}, \bibinfo {author} {\bibfnamefont
  {M.}~\bibnamefont {Arai}}, \bibinfo {author} {\bibfnamefont {Y.}~\bibnamefont
  {Kawazoe}}, \bibinfo {author} {\bibfnamefont {Z.}~\bibnamefont {Fang}},\ and\
  \bibinfo {author} {\bibfnamefont {X.}~\bibnamefont {Dai}},\ }\bibinfo {title}
  {Large-gap two-dimensional topological insulator in oxygen functionalized
  MXene},\ \href {https://doi.org/10.1103/PhysRevB.92.075436} {\bibfield
  {journal} {\bibinfo  {journal} {Phys. Rev. B}\ }\textbf {\bibinfo {volume}
  {92}},\ \bibinfo {pages} {075436} (\bibinfo {year} {2015})}\BibitemShut
  {NoStop}%
\bibitem [{\citenamefont {Wang}\ \emph
  {et~al.}(2016{\natexlab{a}})\citenamefont {Wang}, \citenamefont {Ji},
  \citenamefont {Zhang}, \citenamefont {Li}, \citenamefont {Li}, \citenamefont
  {Wang}, \citenamefont {Li},\ and\ \citenamefont
  {Yan}}]{DFT-TI-monolayer-dumbbellStanene-Functionalization}%
  \BibitemOpen
  \bibfield  {author} {\bibinfo {author} {\bibfnamefont {Y.-P.}\ \bibnamefont
  {Wang}}, \bibinfo {author} {\bibfnamefont {W.-X.}\ \bibnamefont {Ji}},
  \bibinfo {author} {\bibfnamefont {C.-W.}\ \bibnamefont {Zhang}}, \bibinfo
  {author} {\bibfnamefont {P.}~\bibnamefont {Li}}, \bibinfo {author}
  {\bibfnamefont {F.}~\bibnamefont {Li}}, \bibinfo {author} {\bibfnamefont
  {P.-J.}\ \bibnamefont {Wang}}, \bibinfo {author} {\bibfnamefont {S.-S.}\
  \bibnamefont {Li}},\ and\ \bibinfo {author} {\bibfnamefont {S.-S.}\
  \bibnamefont {Yan}},\ }\bibinfo {title} {Large-gap quantum spin Hall state in
  functionalized dumbbell stanene},\ \href {https://doi.org/10.1063/1.4942380}
  {\bibfield  {journal} {\bibinfo  {journal} {Appl. Phys. Lett.}\ }\textbf
  {\bibinfo {volume} {108}},\ \bibinfo {pages} {073104} (\bibinfo {year}
  {2016}{\natexlab{a}})}\BibitemShut {NoStop}%
\bibitem [{\citenamefont {Song}\ \emph {et~al.}(2014)\citenamefont {Song},
  \citenamefont {Liu}, \citenamefont {Yang}, \citenamefont {Han}, \citenamefont
  {Ye}, \citenamefont {Fu}, \citenamefont {Yang}, \citenamefont {Niu},
  \citenamefont {Lu},\ and\ \citenamefont
  {Yao}}]{DFT-TI-monolayer-BiX/SbX-HFClBr}%
  \BibitemOpen
  \bibfield  {author} {\bibinfo {author} {\bibfnamefont {Z.}~\bibnamefont
  {Song}}, \bibinfo {author} {\bibfnamefont {C.-C.}\ \bibnamefont {Liu}},
  \bibinfo {author} {\bibfnamefont {J.}~\bibnamefont {Yang}}, \bibinfo {author}
  {\bibfnamefont {J.}~\bibnamefont {Han}}, \bibinfo {author} {\bibfnamefont
  {M.}~\bibnamefont {Ye}}, \bibinfo {author} {\bibfnamefont {B.}~\bibnamefont
  {Fu}}, \bibinfo {author} {\bibfnamefont {Y.}~\bibnamefont {Yang}}, \bibinfo
  {author} {\bibfnamefont {Q.}~\bibnamefont {Niu}}, \bibinfo {author}
  {\bibfnamefont {J.}~\bibnamefont {Lu}},\ and\ \bibinfo {author}
  {\bibfnamefont {Y.}~\bibnamefont {Yao}},\ }\bibinfo {title} {Quantum spin
  Hall insulators and quantum valley Hall insulators of BiX/SbX (X=H, F, Cl and
  Br) monolayers with a record bulk band gap},\ \href
  {https://doi.org/10.1038/am.2014.113} {\bibfield  {journal} {\bibinfo
  {journal} {NPG Asia Mater.}\ }\textbf {\bibinfo {volume} {6}},\ \bibinfo
  {pages} {e147} (\bibinfo {year} {2014})}\BibitemShut {NoStop}%
\bibitem [{\citenamefont {Zhou}\ \emph
  {et~al.}(2018{\natexlab{a}})\citenamefont {Zhou}, \citenamefont {Liu},
  \citenamefont {Zhao},\ and\ \citenamefont
  {Yao}}]{DFT-III-monochalcogenides-Functionalization-O-TI}%
  \BibitemOpen
  \bibfield  {author} {\bibinfo {author} {\bibfnamefont {S.}~\bibnamefont
  {Zhou}}, \bibinfo {author} {\bibfnamefont {C.-C.}\ \bibnamefont {Liu}},
  \bibinfo {author} {\bibfnamefont {J.}~\bibnamefont {Zhao}},\ and\ \bibinfo
  {author} {\bibfnamefont {Y.}~\bibnamefont {Yao}},\ }\bibinfo {title}
  {Monolayer group-III monochalcogenides by oxygen functionalization: a
  promising class of two-dimensional topological insulators},\ \href
  {https://doi.org/10.1038/s41535-018-0089-0} {\bibfield  {journal} {\bibinfo
  {journal} {npj Quantum Mater.}\ }\textbf {\bibinfo {volume} {3}},\ \bibinfo
  {pages} {16} (\bibinfo {year} {2018}{\natexlab{a}})}\BibitemShut {NoStop}%
\bibitem [{\citenamefont {Chen}\ \emph {et~al.}(2010)\citenamefont {Chen},
  \citenamefont {Li}, \citenamefont {Yu}, \citenamefont {Li}, \citenamefont
  {Zhang}, \citenamefont {Zhou},\ and\ \citenamefont
  {Chen}}]{DFT-BN-Hydrogenation-monolayer-2}%
  \BibitemOpen
  \bibfield  {author} {\bibinfo {author} {\bibfnamefont {W.}~\bibnamefont
  {Chen}}, \bibinfo {author} {\bibfnamefont {Y.}~\bibnamefont {Li}}, \bibinfo
  {author} {\bibfnamefont {G.}~\bibnamefont {Yu}}, \bibinfo {author}
  {\bibfnamefont {C.-Z.}\ \bibnamefont {Li}}, \bibinfo {author} {\bibfnamefont
  {S.~B.}\ \bibnamefont {Zhang}}, \bibinfo {author} {\bibfnamefont
  {Z.}~\bibnamefont {Zhou}},\ and\ \bibinfo {author} {\bibfnamefont
  {Z.}~\bibnamefont {Chen}},\ }\bibinfo {title} {Hydrogenation: A simple
  approach to realize semiconductor-half-metal-metal transition in boron
  nitride nanoribbons},\ \href {https://doi.org/10.1021/ja908475v} {\bibfield
  {journal} {\bibinfo  {journal} {J. Am. Chem. Soc.}\ }\textbf {\bibinfo
  {volume} {132}},\ \bibinfo {pages} {1699} (\bibinfo {year}
  {2010})}\BibitemShut {NoStop}%
\bibitem [{\citenamefont {Zhou}\ \emph
  {et~al.}(2018{\natexlab{b}})\citenamefont {Zhou}, \citenamefont {Sun},
  \citenamefont {Zhang},\ and\ \citenamefont
  {Guo}}]{DFT-NbSeH2-monolayer-magnetism}%
  \BibitemOpen
  \bibfield  {author} {\bibinfo {author} {\bibfnamefont {X.}~\bibnamefont
  {Zhou}}, \bibinfo {author} {\bibfnamefont {X.}~\bibnamefont {Sun}}, \bibinfo
  {author} {\bibfnamefont {Z.}~\bibnamefont {Zhang}},\ and\ \bibinfo {author}
  {\bibfnamefont {W.}~\bibnamefont {Guo}},\ }\bibinfo {title} {Ferromagnetism
  in a semiconducting Janus NbSe hydride monolayer},\ \href
  {https://doi.org/10.1039/C8TC03016B} {\bibfield  {journal} {\bibinfo
  {journal} {J. Mater. Chem. C}\ }\textbf {\bibinfo {volume} {6}},\ \bibinfo
  {pages} {9675} (\bibinfo {year} {2018}{\natexlab{b}})}\BibitemShut {NoStop}%
\bibitem [{\citenamefont {Crisostomo}\ \emph {et~al.}(2017)\citenamefont
  {Crisostomo}, \citenamefont {Huang}, \citenamefont {Hsu}, \citenamefont
  {Chuang}, \citenamefont {Lin},\ and\ \citenamefont
  {Bansil}}]{DFT-IIIBi-monolayer-QAHE-Functionalization-magnetism}%
  \BibitemOpen
  \bibfield  {author} {\bibinfo {author} {\bibfnamefont {C.~P.}\ \bibnamefont
  {Crisostomo}}, \bibinfo {author} {\bibfnamefont {Z.-Q.}\ \bibnamefont
  {Huang}}, \bibinfo {author} {\bibfnamefont {C.-H.}\ \bibnamefont {Hsu}},
  \bibinfo {author} {\bibfnamefont {F.-C.}\ \bibnamefont {Chuang}}, \bibinfo
  {author} {\bibfnamefont {H.}~\bibnamefont {Lin}},\ and\ \bibinfo {author}
  {\bibfnamefont {A.}~\bibnamefont {Bansil}},\ }\bibinfo {title} {Chemically
  induced large-gap quantum anomalous Hall insulator states in III-Bi
  honeycombs},\ \href {https://doi.org/10.1038/s41524-017-0044-9} {\bibfield
  {journal} {\bibinfo  {journal} {npj Comput. Mater.}\ }\textbf {\bibinfo
  {volume} {3}},\ \bibinfo {pages} {39} (\bibinfo {year} {2017})}\BibitemShut
  {NoStop}%
\bibitem [{\citenamefont {Lu}\ \emph {et~al.}(2018)\citenamefont {Lu},
  \citenamefont {Wang}, \citenamefont {Chen},\ and\ \citenamefont
  {Liu}}]{DFT-IIIBi-monolayer-QSHE-Functionalization}%
  \BibitemOpen
  \bibfield  {author} {\bibinfo {author} {\bibfnamefont {Q.}~\bibnamefont
  {Lu}}, \bibinfo {author} {\bibfnamefont {B.}~\bibnamefont {Wang}}, \bibinfo
  {author} {\bibfnamefont {X.-R.}\ \bibnamefont {Chen}},\ and\ \bibinfo
  {author} {\bibfnamefont {W.-M.}\ \bibnamefont {Liu}},\ }\bibinfo {title}
  {Robust large-gap quantum spin Hall insulators in methyl-functionalized
  III-Bi buckled honeycombs},\ \href
  {https://doi.org/10.1103/PhysRevMaterials.2.014005} {\bibfield  {journal}
  {\bibinfo  {journal} {Phys. Rev. Mater.}\ }\textbf {\bibinfo {volume} {2}},\
  \bibinfo {pages} {014005} (\bibinfo {year} {2018})}\BibitemShut {NoStop}%
\bibitem [{\citenamefont {Jin}\ and\ \citenamefont
  {Jhi}(2015)}]{DFT-Bi-Sb-111-monolayer-QAHE-Functionalization-magnetism}%
  \BibitemOpen
  \bibfield  {author} {\bibinfo {author} {\bibfnamefont {K.-H.}\ \bibnamefont
  {Jin}}\ and\ \bibinfo {author} {\bibfnamefont {S.-H.}\ \bibnamefont {Jhi}},\
  }\bibinfo {title} {Quantum anomalous Hall and quantum spin-Hall phases in
  flattened Bi and Sb bilayers},\ \href {https://doi.org/10.1038/srep08426}
  {\bibfield  {journal} {\bibinfo  {journal} {Sci. Rep.}\ }\textbf {\bibinfo
  {volume} {5}},\ \bibinfo {pages} {8426} (\bibinfo {year} {2015})}\BibitemShut
  {NoStop}%
\bibitem [{\citenamefont {Wu}\ \emph {et~al.}(2014)\citenamefont {Wu},
  \citenamefont {Shan},\ and\ \citenamefont
  {Yan}}]{DFT-Sn-monolayer-halfadsorp-HighTemp-QAHE-Functionalization-magnetism}%
  \BibitemOpen
  \bibfield  {author} {\bibinfo {author} {\bibfnamefont {S.-C.}\ \bibnamefont
  {Wu}}, \bibinfo {author} {\bibfnamefont {G.}~\bibnamefont {Shan}},\ and\
  \bibinfo {author} {\bibfnamefont {B.}~\bibnamefont {Yan}},\ }\bibinfo {title}
  {Prediction of near-room-temperature quantum anomalous Hall effect on
  honeycomb materials},\ \href {https://doi.org/10.1103/PhysRevLett.113.256401}
  {\bibfield  {journal} {\bibinfo  {journal} {Phys. Rev. Lett.}\ }\textbf
  {\bibinfo {volume} {113}},\ \bibinfo {pages} {256401} (\bibinfo {year}
  {2014})}\BibitemShut {NoStop}%
\bibitem [{\citenamefont {Jin}\ \emph {et~al.}(2022)\citenamefont {Jin},
  \citenamefont {Xia}, \citenamefont {Shi}, \citenamefont {Hu}, \citenamefont
  {Claessen}, \citenamefont {Hanke}, \citenamefont {Thomale},\ and\
  \citenamefont {Li}}]{DFT-FunctionalizationBi-III-QAHE}%
  \BibitemOpen
  \bibfield  {author} {\bibinfo {author} {\bibfnamefont {S.}~\bibnamefont
  {Jin}}, \bibinfo {author} {\bibfnamefont {Y.}~\bibnamefont {Xia}}, \bibinfo
  {author} {\bibfnamefont {W.}~\bibnamefont {Shi}}, \bibinfo {author}
  {\bibfnamefont {J.}~\bibnamefont {Hu}}, \bibinfo {author} {\bibfnamefont
  {R.}~\bibnamefont {Claessen}}, \bibinfo {author} {\bibfnamefont
  {W.}~\bibnamefont {Hanke}}, \bibinfo {author} {\bibfnamefont
  {R.}~\bibnamefont {Thomale}},\ and\ \bibinfo {author} {\bibfnamefont
  {G.}~\bibnamefont {Li}},\ }\bibinfo {title} {Large-gap quantum anomalous Hall
  states induced by functionalizing buckled Bi-III
  monolayer/Al$_{\text{2}}$O$_{\text{3}}$},\ \href
  {https://doi.org/10.1103/PhysRevB.106.125151} {\bibfield  {journal} {\bibinfo
   {journal} {Phys. Rev. B}\ }\textbf {\bibinfo {volume} {106}},\ \bibinfo
  {pages} {125151} (\bibinfo {year} {2022})}\BibitemShut {NoStop}%
\bibitem [{\citenamefont {Makaremi}\ \emph {et~al.}(2017)\citenamefont
  {Makaremi}, \citenamefont {Mortazavi},\ and\ \citenamefont
  {Singh}}]{DFT-C3N-monolayer-magnetism}%
  \BibitemOpen
  \bibfield  {author} {\bibinfo {author} {\bibfnamefont {M.}~\bibnamefont
  {Makaremi}}, \bibinfo {author} {\bibfnamefont {B.}~\bibnamefont
  {Mortazavi}},\ and\ \bibinfo {author} {\bibfnamefont {C.~V.}\ \bibnamefont
  {Singh}},\ }\bibinfo {title} {Adsorption of metallic, metalloidic, and
  nonmetallic adatoms on two-dimensional C$_{\text{3}}$N},\ \href
  {https://doi.org/10.1021/acs.jpcc.7b04511} {\bibfield  {journal} {\bibinfo
  {journal} {J. Phys. Chem. C.}\ }\textbf {\bibinfo {volume} {121}},\ \bibinfo
  {pages} {18575} (\bibinfo {year} {2017})}\BibitemShut {NoStop}%
\bibitem [{\citenamefont {Wu}\ \emph {et~al.}(2024)\citenamefont {Wu},
  \citenamefont {Guo}, \citenamefont {Wu}, \citenamefont {Li},\ and\
  \citenamefont {Wu}}]{DFT-MoSe2F2-LargeStrain}%
  \BibitemOpen
  \bibfield  {author} {\bibinfo {author} {\bibfnamefont {J.}~\bibnamefont
  {Wu}}, \bibinfo {author} {\bibfnamefont {R.}~\bibnamefont {Guo}}, \bibinfo
  {author} {\bibfnamefont {D.}~\bibnamefont {Wu}}, \bibinfo {author}
  {\bibfnamefont {X.}~\bibnamefont {Li}},\ and\ \bibinfo {author}
  {\bibfnamefont {X.}~\bibnamefont {Wu}},\ }\bibinfo {title} {Turning
  nonmagnetic two-dimensional molybdenum disulfides into room-temperature
  ferromagnets by the synergistic effect of lattice stretching and charge
  injection},\ \href {https://doi.org/10.1021/acs.jpclett.3c03478} {\bibfield
  {journal} {\bibinfo  {journal} {J. Phys. Chem. Lett.}\ }\textbf {\bibinfo
  {volume} {15}},\ \bibinfo {pages} {2293} (\bibinfo {year}
  {2024})}\BibitemShut {NoStop}%
\bibitem [{\citenamefont {Zhang}\ \emph {et~al.}(2024)\citenamefont {Zhang},
  \citenamefont {Zhou}, \citenamefont {Wang}, \citenamefont {Wang},
  \citenamefont {Li},\ and\ \citenamefont {Li}}]{DFT-MoSe2F2-QAHE}%
  \BibitemOpen
  \bibfield  {author} {\bibinfo {author} {\bibfnamefont {Z.}~\bibnamefont
  {Zhang}}, \bibinfo {author} {\bibfnamefont {Z.}~\bibnamefont {Zhou}},
  \bibinfo {author} {\bibfnamefont {X.}~\bibnamefont {Wang}}, \bibinfo {author}
  {\bibfnamefont {H.}~\bibnamefont {Wang}}, \bibinfo {author} {\bibfnamefont
  {X.}~\bibnamefont {Li}},\ and\ \bibinfo {author} {\bibfnamefont
  {X.}~\bibnamefont {Li}},\ }\bibinfo {title} {Quantum anomalous Hall state in
  a fluorinated 1T-MoSe$_{\text{2}}$ monolayer},\ \href
  {https://doi.org/10.1103/PhysRevB.109.L081406} {\bibfield  {journal}
  {\bibinfo  {journal} {Phys. Rev. B}\ }\textbf {\bibinfo {volume} {109}},\
  \bibinfo {pages} {L081406} (\bibinfo {year} {2024})}\BibitemShut {NoStop}%
\bibitem [{\citenamefont {Zhang}\ \emph {et~al.}(2016)\citenamefont {Zhang},
  \citenamefont {Zhou}, \citenamefont {Zhang}, \citenamefont {Zhao},
  \citenamefont {Yao},\ and\ \citenamefont {Yang}}]{DFT-Sn-monolayer-H-QAHE}%
  \BibitemOpen
  \bibfield  {author} {\bibinfo {author} {\bibfnamefont {H.}~\bibnamefont
  {Zhang}}, \bibinfo {author} {\bibfnamefont {T.}~\bibnamefont {Zhou}},
  \bibinfo {author} {\bibfnamefont {J.}~\bibnamefont {Zhang}}, \bibinfo
  {author} {\bibfnamefont {B.}~\bibnamefont {Zhao}}, \bibinfo {author}
  {\bibfnamefont {Y.}~\bibnamefont {Yao}},\ and\ \bibinfo {author}
  {\bibfnamefont {Z.}~\bibnamefont {Yang}},\ }\bibinfo {title} {Quantum
  anomalous Hall effect in stanene on a nonmagnetic substrate},\ \href
  {https://doi.org/10.1103/PhysRevB.94.235409} {\bibfield  {journal} {\bibinfo
  {journal} {Phys. Rev. B}\ }\textbf {\bibinfo {volume} {94}},\ \bibinfo
  {pages} {235409} (\bibinfo {year} {2016})}\BibitemShut {NoStop}%
\bibitem [{\citenamefont {Chen}\ \emph {et~al.}(2025)\citenamefont {Chen},
  \citenamefont {Chen}, \citenamefont {Zhao}, \citenamefont {Hu}, \citenamefont
  {Yuan},\ and\ \citenamefont {Ren}}]{DFT-MoTe2F2-QAHE-QuadraticBdCrossing}%
  \BibitemOpen
  \bibfield  {author} {\bibinfo {author} {\bibfnamefont {F.}~\bibnamefont
  {Chen}}, \bibinfo {author} {\bibfnamefont {H.}~\bibnamefont {Chen}}, \bibinfo
  {author} {\bibfnamefont {X.}~\bibnamefont {Zhao}}, \bibinfo {author}
  {\bibfnamefont {G.}~\bibnamefont {Hu}}, \bibinfo {author} {\bibfnamefont
  {X.}~\bibnamefont {Yuan}},\ and\ \bibinfo {author} {\bibfnamefont
  {J.}~\bibnamefont {Ren}},\ }\bibinfo {title} {Quadratic band crossing induced
  quantum anomalous Hall effect in monolayer MoTe$_{\text{2}}$F$_{\text{2}}$},\
  \href {https://doi.org/10.1103/PhysRevB.111.075133} {\bibfield  {journal}
  {\bibinfo  {journal} {Phys. Rev. B}\ }\textbf {\bibinfo {volume} {111}},\
  \bibinfo {pages} {075133} (\bibinfo {year} {2025})}\BibitemShut {NoStop}%
\bibitem [{\citenamefont {Boukhvalov}\ \emph {et~al.}(2008)\citenamefont
  {Boukhvalov}, \citenamefont {Katsnelson},\ and\ \citenamefont
  {Lichtenstein}}]{DFT-Graphene-Hydrogenation-magnetism-distortion}%
  \BibitemOpen
  \bibfield  {author} {\bibinfo {author} {\bibfnamefont {D.~W.}\ \bibnamefont
  {Boukhvalov}}, \bibinfo {author} {\bibfnamefont {M.~I.}\ \bibnamefont
  {Katsnelson}},\ and\ \bibinfo {author} {\bibfnamefont {A.~I.}\ \bibnamefont
  {Lichtenstein}},\ }\bibinfo {title} {Hydrogen on graphene: Electronic
  structure, total energy, structural distortions and magnetism from
  first-principles calculations},\ \href
  {https://doi.org/10.1103/PhysRevB.77.035427} {\bibfield  {journal} {\bibinfo
  {journal} {Phys. Rev. B}\ }\textbf {\bibinfo {volume} {77}},\ \bibinfo
  {pages} {035427} (\bibinfo {year} {2008})}\BibitemShut {NoStop}%
\bibitem [{\citenamefont {Sofo}\ \emph {et~al.}(2007)\citenamefont {Sofo},
  \citenamefont {Chaudhari},\ and\ \citenamefont
  {Barber}}]{DFT-Hydrogenation-graphene-gapopen}%
  \BibitemOpen
  \bibfield  {author} {\bibinfo {author} {\bibfnamefont {J.~O.}\ \bibnamefont
  {Sofo}}, \bibinfo {author} {\bibfnamefont {A.~S.}\ \bibnamefont
  {Chaudhari}},\ and\ \bibinfo {author} {\bibfnamefont {G.~D.}\ \bibnamefont
  {Barber}},\ }\bibinfo {title} {Graphane: A two-dimensional hydrocarbon},\
  \href {https://doi.org/10.1103/PhysRevB.75.153401} {\bibfield  {journal}
  {\bibinfo  {journal} {Phys. Rev. B}\ }\textbf {\bibinfo {volume} {75}},\
  \bibinfo {pages} {153401} (\bibinfo {year} {2007})}\BibitemShut {NoStop}%
\bibitem [{\citenamefont {Elias}\ \emph {et~al.}(2009)\citenamefont {Elias},
  \citenamefont {Nair}, \citenamefont {Mohiuddin}, \citenamefont {Morozov},
  \citenamefont {Blake}, \citenamefont {Halsall}, \citenamefont {Ferrari},
  \citenamefont {Boukhvalov}, \citenamefont {Katsnelson}, \citenamefont
  {Geim},\ and\ \citenamefont {Novoselov}}]{EXP-Hydrogenation-graphene}%
  \BibitemOpen
  \bibfield  {author} {\bibinfo {author} {\bibfnamefont {D.~C.}\ \bibnamefont
  {Elias}}, \bibinfo {author} {\bibfnamefont {R.~R.}\ \bibnamefont {Nair}},
  \bibinfo {author} {\bibfnamefont {T.~M.~G.}\ \bibnamefont {Mohiuddin}},
  \bibinfo {author} {\bibfnamefont {S.~V.}\ \bibnamefont {Morozov}}, \bibinfo
  {author} {\bibfnamefont {P.}~\bibnamefont {Blake}}, \bibinfo {author}
  {\bibfnamefont {M.~P.}\ \bibnamefont {Halsall}}, \bibinfo {author}
  {\bibfnamefont {A.~C.}\ \bibnamefont {Ferrari}}, \bibinfo {author}
  {\bibfnamefont {D.~W.}\ \bibnamefont {Boukhvalov}}, \bibinfo {author}
  {\bibfnamefont {M.~I.}\ \bibnamefont {Katsnelson}}, \bibinfo {author}
  {\bibfnamefont {A.~K.}\ \bibnamefont {Geim}},\ and\ \bibinfo {author}
  {\bibfnamefont {K.~S.}\ \bibnamefont {Novoselov}},\ }\bibinfo {title}
  {Control of graphene's properties by reversible hydrogenation: Evidence for
  graphane},\ \href {https://doi.org/10.1126/science.1167130} {\bibfield
  {journal} {\bibinfo  {journal} {Science}\ }\textbf {\bibinfo {volume}
  {323}},\ \bibinfo {pages} {610} (\bibinfo {year} {2009})}\BibitemShut
  {NoStop}%
\bibitem [{\citenamefont {Balog}\ \emph {et~al.}(2010)\citenamefont {Balog},
  \citenamefont {Jørgensen}, \citenamefont {Nilsson}, \citenamefont
  {Andersen}, \citenamefont {Rienks}, \citenamefont {Bianchi}, \citenamefont
  {Fanetti}, \citenamefont {Lægsgaard}, \citenamefont {Baraldi}, \citenamefont
  {Lizzit}, \citenamefont {Sljivancanin}, \citenamefont {Besenbacher},
  \citenamefont {Hammer}, \citenamefont {Pedersen}, \citenamefont {Hofmann},\
  and\ \citenamefont {Hornekær}}]{EXP-Hydrogenation-graphene-gapopen}%
  \BibitemOpen
  \bibfield  {author} {\bibinfo {author} {\bibfnamefont {R.}~\bibnamefont
  {Balog}}, \bibinfo {author} {\bibfnamefont {B.}~\bibnamefont {Jørgensen}},
  \bibinfo {author} {\bibfnamefont {L.}~\bibnamefont {Nilsson}}, \bibinfo
  {author} {\bibfnamefont {M.}~\bibnamefont {Andersen}}, \bibinfo {author}
  {\bibfnamefont {E.}~\bibnamefont {Rienks}}, \bibinfo {author} {\bibfnamefont
  {M.}~\bibnamefont {Bianchi}}, \bibinfo {author} {\bibfnamefont
  {M.}~\bibnamefont {Fanetti}}, \bibinfo {author} {\bibfnamefont
  {E.}~\bibnamefont {Lægsgaard}}, \bibinfo {author} {\bibfnamefont
  {A.}~\bibnamefont {Baraldi}}, \bibinfo {author} {\bibfnamefont
  {S.}~\bibnamefont {Lizzit}}, \bibinfo {author} {\bibfnamefont
  {Z.}~\bibnamefont {Sljivancanin}}, \bibinfo {author} {\bibfnamefont
  {F.}~\bibnamefont {Besenbacher}}, \bibinfo {author} {\bibfnamefont
  {B.}~\bibnamefont {Hammer}}, \bibinfo {author} {\bibfnamefont {T.~G.}\
  \bibnamefont {Pedersen}}, \bibinfo {author} {\bibfnamefont {P.}~\bibnamefont
  {Hofmann}},\ and\ \bibinfo {author} {\bibfnamefont {L.}~\bibnamefont
  {Hornekær}},\ }\bibinfo {title} {Bandgap opening in graphene induced by
  patterned hydrogen adsorption},\ \href {https://doi.org/10.1038/nmat2710}
  {\bibfield  {journal} {\bibinfo  {journal} {Nat. Mater.}\ }\textbf {\bibinfo
  {volume} {9}},\ \bibinfo {pages} {315} (\bibinfo {year} {2010})}\BibitemShut
  {NoStop}%
\bibitem [{\citenamefont {Balog}\ \emph {et~al.}(2009)\citenamefont {Balog},
  \citenamefont {Jørgensen}, \citenamefont {Wells}, \citenamefont
  {Lægsgaard}, \citenamefont {Hofmann}, \citenamefont {Besenbacher},\ and\
  \citenamefont {Hornekær}}]{EXP-Graphene-Adsorption-SingleSide}%
  \BibitemOpen
  \bibfield  {author} {\bibinfo {author} {\bibfnamefont {R.}~\bibnamefont
  {Balog}}, \bibinfo {author} {\bibfnamefont {B.}~\bibnamefont {Jørgensen}},
  \bibinfo {author} {\bibfnamefont {J.}~\bibnamefont {Wells}}, \bibinfo
  {author} {\bibfnamefont {E.}~\bibnamefont {Lægsgaard}}, \bibinfo {author}
  {\bibfnamefont {P.}~\bibnamefont {Hofmann}}, \bibinfo {author} {\bibfnamefont
  {F.}~\bibnamefont {Besenbacher}},\ and\ \bibinfo {author} {\bibfnamefont
  {L.}~\bibnamefont {Hornekær}},\ }\bibinfo {title} {Atomic hydrogen adsorbate
  structures on graphene},\ \href {https://doi.org/10.1021/ja902714h}
  {\bibfield  {journal} {\bibinfo  {journal} {J. Am. Chem. Soc.}\ }\textbf
  {\bibinfo {volume} {131}},\ \bibinfo {pages} {8744} (\bibinfo {year}
  {2009})}\BibitemShut {NoStop}%
\bibitem [{\citenamefont {Balakrishnan}\ \emph {et~al.}(2013)\citenamefont
  {Balakrishnan}, \citenamefont {Kok Wai~Koon}, \citenamefont {Jaiswal},
  \citenamefont {Castro~Neto},\ and\ \citenamefont
  {{\"O}zyilmaz}}]{EXP-Graphene-Hydrogenation-EnhacnedSOC}%
  \BibitemOpen
  \bibfield  {author} {\bibinfo {author} {\bibfnamefont {J.}~\bibnamefont
  {Balakrishnan}}, \bibinfo {author} {\bibfnamefont {G.}~\bibnamefont {Kok
  Wai~Koon}}, \bibinfo {author} {\bibfnamefont {M.}~\bibnamefont {Jaiswal}},
  \bibinfo {author} {\bibfnamefont {A.~H.}\ \bibnamefont {Castro~Neto}},\ and\
  \bibinfo {author} {\bibfnamefont {B.}~\bibnamefont {{\"O}zyilmaz}},\
  }\bibinfo {title} {Colossal enhancement of spin--orbit coupling in weakly
  hydrogenated graphene},\ \href {https://doi.org/10.1038/nphys2576} {\bibfield
   {journal} {\bibinfo  {journal} {Nat. Phys.}\ }\textbf {\bibinfo {volume}
  {9}},\ \bibinfo {pages} {284} (\bibinfo {year} {2013})}\BibitemShut {NoStop}%
\bibitem [{\citenamefont {Sessi}\ \emph {et~al.}(2009)\citenamefont {Sessi},
  \citenamefont {Guest}, \citenamefont {Bode},\ and\ \citenamefont
  {Guisinger}}]{EXP-Graphene-Hydrogenation-Passivation-RoomTemp}%
  \BibitemOpen
  \bibfield  {author} {\bibinfo {author} {\bibfnamefont {P.}~\bibnamefont
  {Sessi}}, \bibinfo {author} {\bibfnamefont {J.~R.}\ \bibnamefont {Guest}},
  \bibinfo {author} {\bibfnamefont {M.}~\bibnamefont {Bode}},\ and\ \bibinfo
  {author} {\bibfnamefont {N.~P.}\ \bibnamefont {Guisinger}},\ }\bibinfo
  {title} {Patterning graphene at the nanometer scale via hydrogen
  desorption},\ \href {https://doi.org/10.1021/nl902605t} {\bibfield  {journal}
  {\bibinfo  {journal} {Nano Lett.}\ }\textbf {\bibinfo {volume} {9}},\
  \bibinfo {pages} {4343} (\bibinfo {year} {2009})}\BibitemShut {NoStop}%
\bibitem [{\citenamefont {González-Herrero}\ \emph {et~al.}(2016)\citenamefont
  {González-Herrero}, \citenamefont {Gómez-Rodríguez}, \citenamefont
  {Mallet}, \citenamefont {Moaied}, \citenamefont {Palacios}, \citenamefont
  {Salgado}, \citenamefont {Ugeda}, \citenamefont {Veuillen}, \citenamefont
  {Yndurain},\ and\ \citenamefont
  {Brihuega}}]{EXP-Graphene-Hydrogenation-magnetism}%
  \BibitemOpen
  \bibfield  {author} {\bibinfo {author} {\bibfnamefont {H.}~\bibnamefont
  {González-Herrero}}, \bibinfo {author} {\bibfnamefont {J.~M.}\ \bibnamefont
  {Gómez-Rodríguez}}, \bibinfo {author} {\bibfnamefont {P.}~\bibnamefont
  {Mallet}}, \bibinfo {author} {\bibfnamefont {M.}~\bibnamefont {Moaied}},
  \bibinfo {author} {\bibfnamefont {J.~J.}\ \bibnamefont {Palacios}}, \bibinfo
  {author} {\bibfnamefont {C.}~\bibnamefont {Salgado}}, \bibinfo {author}
  {\bibfnamefont {M.~M.}\ \bibnamefont {Ugeda}}, \bibinfo {author}
  {\bibfnamefont {J.-Y.}\ \bibnamefont {Veuillen}}, \bibinfo {author}
  {\bibfnamefont {F.}~\bibnamefont {Yndurain}},\ and\ \bibinfo {author}
  {\bibfnamefont {I.}~\bibnamefont {Brihuega}},\ }\bibinfo {title}
  {Atomic-scale control of graphene magnetism by using hydrogen atoms},\ \href
  {https://doi.org/10.1126/science.aad8038} {\bibfield  {journal} {\bibinfo
  {journal} {Science}\ }\textbf {\bibinfo {volume} {352}},\ \bibinfo {pages}
  {437} (\bibinfo {year} {2016})}\BibitemShut {NoStop}%
\bibitem [{\citenamefont {Zhang}\ and\ \citenamefont
  {Feng}(2012)}]{EXP-BN-Hydrogenation-Oneside}%
  \BibitemOpen
  \bibfield  {author} {\bibinfo {author} {\bibfnamefont {H.~X.}\ \bibnamefont
  {Zhang}}\ and\ \bibinfo {author} {\bibfnamefont {P.~X.}\ \bibnamefont
  {Feng}},\ }\bibinfo {title} {Controlling bandgap of rippled hexagonal boron
  nitride membranes via plasma treatment},\ \href
  {https://doi.org/10.1021/am201435z} {\bibfield  {journal} {\bibinfo
  {journal} {ACS Appl. Mater. Interfaces}\ }\textbf {\bibinfo {volume} {4}},\
  \bibinfo {pages} {30} (\bibinfo {year} {2012})}\BibitemShut {NoStop}%
\bibitem [{\citenamefont {Koswattage}\ \emph {et~al.}(2011)\citenamefont
  {Koswattage}, \citenamefont {Shimoyama}, \citenamefont {Baba}, \citenamefont
  {Sekiguchi},\ and\ \citenamefont
  {Nakagawa}}]{EXP-BN-Hydrogenation-OneSide-ClusterDiscussion}%
  \BibitemOpen
  \bibfield  {author} {\bibinfo {author} {\bibfnamefont {K.~R.}\ \bibnamefont
  {Koswattage}}, \bibinfo {author} {\bibfnamefont {I.}~\bibnamefont
  {Shimoyama}}, \bibinfo {author} {\bibfnamefont {Y.}~\bibnamefont {Baba}},
  \bibinfo {author} {\bibfnamefont {T.}~\bibnamefont {Sekiguchi}},\ and\
  \bibinfo {author} {\bibfnamefont {K.}~\bibnamefont {Nakagawa}},\ }\bibinfo
  {title} {Selective adsorption of atomic hydrogen on a h-BN thin film},\ \href
  {https://doi.org/10.1063/1.3605497} {\bibfield  {journal} {\bibinfo
  {journal} {J. Chem. Phys.}\ }\textbf {\bibinfo {volume} {135}},\ \bibinfo
  {pages} {014706} (\bibinfo {year} {2011})}\BibitemShut {NoStop}%
\bibitem [{\citenamefont {Cuenca}\ \emph {et~al.}(2020)\citenamefont {Cuenca},
  \citenamefont {Mandal}, \citenamefont {Morgan}, \citenamefont {Snowball},
  \citenamefont {Porch},\ and\ \citenamefont
  {Williams}}]{EXP-BN-Hydrogenation-OneSide-2020}%
  \BibitemOpen
  \bibfield  {author} {\bibinfo {author} {\bibfnamefont {J.~A.}\ \bibnamefont
  {Cuenca}}, \bibinfo {author} {\bibfnamefont {S.}~\bibnamefont {Mandal}},
  \bibinfo {author} {\bibfnamefont {D.~J.}\ \bibnamefont {Morgan}}, \bibinfo
  {author} {\bibfnamefont {M.}~\bibnamefont {Snowball}}, \bibinfo {author}
  {\bibfnamefont {A.}~\bibnamefont {Porch}},\ and\ \bibinfo {author}
  {\bibfnamefont {O.~A.}\ \bibnamefont {Williams}},\ }\bibinfo {title}
  {Dielectric spectroscopy of hydrogen-treated hexagonal boron nitride
  ceramics},\ \href {https://doi.org/10.1021/acsaelm.9b00767} {\bibfield
  {journal} {\bibinfo  {journal} {ACS Appl. Electron. Mater.}\ }\textbf
  {\bibinfo {volume} {2}},\ \bibinfo {pages} {1193} (\bibinfo {year}
  {2020})}\BibitemShut {NoStop}%
\bibitem [{\citenamefont {Ohtomo}\ \emph {et~al.}(2017)\citenamefont {Ohtomo},
  \citenamefont {Yamauchi}, \citenamefont {Sun}, \citenamefont {Kuzubov},
  \citenamefont {Mikhaleva}, \citenamefont {Avramov}, \citenamefont {Entani},
  \citenamefont {Matsumoto}, \citenamefont {Naramoto},\ and\ \citenamefont
  {Sakai}}]{EXP-BN-Hydrogenation-OneSide-OnNi111}%
  \BibitemOpen
  \bibfield  {author} {\bibinfo {author} {\bibfnamefont {M.}~\bibnamefont
  {Ohtomo}}, \bibinfo {author} {\bibfnamefont {Y.}~\bibnamefont {Yamauchi}},
  \bibinfo {author} {\bibfnamefont {X.}~\bibnamefont {Sun}}, \bibinfo {author}
  {\bibfnamefont {A.~A.}\ \bibnamefont {Kuzubov}}, \bibinfo {author}
  {\bibfnamefont {N.~S.}\ \bibnamefont {Mikhaleva}}, \bibinfo {author}
  {\bibfnamefont {P.~V.}\ \bibnamefont {Avramov}}, \bibinfo {author}
  {\bibfnamefont {S.}~\bibnamefont {Entani}}, \bibinfo {author} {\bibfnamefont
  {Y.}~\bibnamefont {Matsumoto}}, \bibinfo {author} {\bibfnamefont
  {H.}~\bibnamefont {Naramoto}},\ and\ \bibinfo {author} {\bibfnamefont
  {S.}~\bibnamefont {Sakai}},\ }\bibinfo {title} {Direct observation of
  site-selective hydrogenation and spin-polarization in hydrogenated hexagonal
  boron nitride on Ni(111)},\ \href {https://doi.org/10.1039/C6NR06308J}
  {\bibfield  {journal} {\bibinfo  {journal} {Nanoscale}\ }\textbf {\bibinfo
  {volume} {9}},\ \bibinfo {pages} {2369} (\bibinfo {year} {2017})}\BibitemShut
  {NoStop}%
\bibitem [{\citenamefont {Vogg}\ \emph {et~al.}(2000)\citenamefont {Vogg},
  \citenamefont {Brandt},\ and\ \citenamefont
  {Stutzmann}}]{EXP-Germanene-Hydrogenation-Bulk-2}%
  \BibitemOpen
  \bibfield  {author} {\bibinfo {author} {\bibfnamefont {G.}~\bibnamefont
  {Vogg}}, \bibinfo {author} {\bibfnamefont {M.~S.}\ \bibnamefont {Brandt}},\
  and\ \bibinfo {author} {\bibfnamefont {M.}~\bibnamefont {Stutzmann}},\
  }\bibinfo {title} {Polygermyne—A prototype system for layered germanium
  polymers},\ \href
  {https://doi.org/https://doi.org/10.1002/1521-4095(200009)12:17<1278::AID-ADMA1278>3.0.CO;2-Y}
  {\bibfield  {journal} {\bibinfo  {journal} {Adv. Mater.}\ }\textbf {\bibinfo
  {volume} {12}},\ \bibinfo {pages} {1278} (\bibinfo {year}
  {2000})}\BibitemShut {NoStop}%
\bibitem [{\citenamefont {Cultrara}\ \emph {et~al.}(2018)\citenamefont
  {Cultrara}, \citenamefont {Wang}, \citenamefont {Arguilla}, \citenamefont
  {Scudder}, \citenamefont {Jiang}, \citenamefont {Windl}, \citenamefont
  {Bobev},\ and\ \citenamefont
  {Goldberger}}]{EXP-Germanene-Hydrogenation-Bulk-3}%
  \BibitemOpen
  \bibfield  {author} {\bibinfo {author} {\bibfnamefont {N.~D.}\ \bibnamefont
  {Cultrara}}, \bibinfo {author} {\bibfnamefont {Y.}~\bibnamefont {Wang}},
  \bibinfo {author} {\bibfnamefont {M.~Q.}\ \bibnamefont {Arguilla}}, \bibinfo
  {author} {\bibfnamefont {M.~R.}\ \bibnamefont {Scudder}}, \bibinfo {author}
  {\bibfnamefont {S.}~\bibnamefont {Jiang}}, \bibinfo {author} {\bibfnamefont
  {W.}~\bibnamefont {Windl}}, \bibinfo {author} {\bibfnamefont
  {S.}~\bibnamefont {Bobev}},\ and\ \bibinfo {author} {\bibfnamefont {J.~E.}\
  \bibnamefont {Goldberger}},\ }\bibinfo {title} {Synthesis of 1T, 2H, and 6R
  germanane polytypes},\ \href {https://doi.org/10.1021/acs.chemmater.7b04990}
  {\bibfield  {journal} {\bibinfo  {journal} {Chem. Mater.}\ }\textbf {\bibinfo
  {volume} {30}},\ \bibinfo {pages} {1335} (\bibinfo {year}
  {2018})}\BibitemShut {NoStop}%
\bibitem [{\citenamefont {Nakamura}\ and\ \citenamefont
  {Nakano}(2018)}]{EXP-Germanene-Hydrogenation-FewLayers}%
  \BibitemOpen
  \bibfield  {author} {\bibinfo {author} {\bibfnamefont {D.}~\bibnamefont
  {Nakamura}}\ and\ \bibinfo {author} {\bibfnamefont {H.}~\bibnamefont
  {Nakano}},\ }\bibinfo {title} {Liquid-phase exfoliation of germanane based on
  Hansen solubility parameters},\ \href
  {https://doi.org/10.1021/acs.chemmater.8b02153} {\bibfield  {journal}
  {\bibinfo  {journal} {Chem. Mater.}\ }\textbf {\bibinfo {volume} {30}},\
  \bibinfo {pages} {5333} (\bibinfo {year} {2018})}\BibitemShut {NoStop}%
\bibitem [{\citenamefont {Liu}\ \emph {et~al.}(2019{\natexlab{a}})\citenamefont
  {Liu}, \citenamefont {Wang}, \citenamefont {Sun}, \citenamefont {Dai},\ and\
  \citenamefont {Huang}}]{EXP-Germanene-Hydrogenation-Bulk-4}%
  \BibitemOpen
  \bibfield  {author} {\bibinfo {author} {\bibfnamefont {Z.}~\bibnamefont
  {Liu}}, \bibinfo {author} {\bibfnamefont {Z.}~\bibnamefont {Wang}}, \bibinfo
  {author} {\bibfnamefont {Q.}~\bibnamefont {Sun}}, \bibinfo {author}
  {\bibfnamefont {Y.}~\bibnamefont {Dai}},\ and\ \bibinfo {author}
  {\bibfnamefont {B.}~\bibnamefont {Huang}},\ }\bibinfo {title}
  {Methyl-terminated germanane GeCH$_{\text{3}}$ synthesized by solvothermal
  method with improved photocatalytic properties},\ \href
  {https://doi.org/https://doi.org/10.1016/j.apsusc.2018.10.228} {\bibfield
  {journal} {\bibinfo  {journal} {Appl. Surf. Sci.}\ }\textbf {\bibinfo
  {volume} {467-468}},\ \bibinfo {pages} {881} (\bibinfo {year}
  {2019}{\natexlab{a}})}\BibitemShut {NoStop}%
\bibitem [{\citenamefont {Liu}\ \emph {et~al.}(2019{\natexlab{b}})\citenamefont
  {Liu}, \citenamefont {Dai}, \citenamefont {Zheng},\ and\ \citenamefont
  {Huang}}]{EXP-Germanene-Hydrogenation-Bulk-5}%
  \BibitemOpen
  \bibfield  {author} {\bibinfo {author} {\bibfnamefont {Z.}~\bibnamefont
  {Liu}}, \bibinfo {author} {\bibfnamefont {Y.}~\bibnamefont {Dai}}, \bibinfo
  {author} {\bibfnamefont {Z.}~\bibnamefont {Zheng}},\ and\ \bibinfo {author}
  {\bibfnamefont {B.}~\bibnamefont {Huang}},\ }\bibinfo {title}
  {Covalently-terminated germanane GeH and GeCH$_{\text{3}}$ for hydrogen
  generation from catalytic hydrolysis of ammonia borane under visible light
  irradiation},\ \href
  {https://doi.org/https://doi.org/10.1016/j.catcom.2018.09.016} {\bibfield
  {journal} {\bibinfo  {journal} {Catal. Commun.}\ }\textbf {\bibinfo {volume}
  {118}},\ \bibinfo {pages} {46} (\bibinfo {year}
  {2019}{\natexlab{b}})}\BibitemShut {NoStop}%
\bibitem [{\citenamefont {Jiang}\ \emph {et~al.}(2014)\citenamefont {Jiang},
  \citenamefont {Butler}, \citenamefont {Bianco}, \citenamefont {Restrepo},
  \citenamefont {Windl},\ and\ \citenamefont
  {Goldberger}}]{EXP-Germanene-Hydrogenation-Bulk-6}%
  \BibitemOpen
  \bibfield  {author} {\bibinfo {author} {\bibfnamefont {S.}~\bibnamefont
  {Jiang}}, \bibinfo {author} {\bibfnamefont {S.}~\bibnamefont {Butler}},
  \bibinfo {author} {\bibfnamefont {E.}~\bibnamefont {Bianco}}, \bibinfo
  {author} {\bibfnamefont {O.~D.}\ \bibnamefont {Restrepo}}, \bibinfo {author}
  {\bibfnamefont {W.}~\bibnamefont {Windl}},\ and\ \bibinfo {author}
  {\bibfnamefont {J.~E.}\ \bibnamefont {Goldberger}},\ }\bibinfo {title}
  {Improving the stability and optical properties of germanane via one-step
  covalent methyl-termination},\ \href {https://doi.org/10.1038/ncomms4389}
  {\bibfield  {journal} {\bibinfo  {journal} {Nat. Commun.}\ }\textbf {\bibinfo
  {volume} {5}},\ \bibinfo {pages} {3389} (\bibinfo {year} {2014})}\BibitemShut
  {NoStop}%
\bibitem [{\citenamefont {Bianco}\ \emph {et~al.}(2013)\citenamefont {Bianco},
  \citenamefont {Butler}, \citenamefont {Jiang}, \citenamefont {Restrepo},
  \citenamefont {Windl},\ and\ \citenamefont
  {Goldberger}}]{EXP-Ge-Hydrogenation-lattice-change}%
  \BibitemOpen
  \bibfield  {author} {\bibinfo {author} {\bibfnamefont {E.}~\bibnamefont
  {Bianco}}, \bibinfo {author} {\bibfnamefont {S.}~\bibnamefont {Butler}},
  \bibinfo {author} {\bibfnamefont {S.}~\bibnamefont {Jiang}}, \bibinfo
  {author} {\bibfnamefont {O.~D.}\ \bibnamefont {Restrepo}}, \bibinfo {author}
  {\bibfnamefont {W.}~\bibnamefont {Windl}},\ and\ \bibinfo {author}
  {\bibfnamefont {J.~E.}\ \bibnamefont {Goldberger}},\ }\bibinfo {title}
  {Stability and exfoliation of germanane: A germanium graphane analogue},\
  \href {https://doi.org/10.1021/nn4009406} {\bibfield  {journal} {\bibinfo
  {journal} {ACS Nano}\ }\textbf {\bibinfo {volume} {7}},\ \bibinfo {pages}
  {4414} (\bibinfo {year} {2013})}\BibitemShut {NoStop}%
\bibitem [{\citenamefont {Qiu}\ \emph {et~al.}(2015{\natexlab{a}})\citenamefont
  {Qiu}, \citenamefont {Fu}, \citenamefont {Xu}, \citenamefont {Oreshkin},
  \citenamefont {Shao}, \citenamefont {Li}, \citenamefont {Meng}, \citenamefont
  {Chen},\ and\ \citenamefont {Wu}}]{EXP-Silicene-Hydrogenation-oneside-STM}%
  \BibitemOpen
  \bibfield  {author} {\bibinfo {author} {\bibfnamefont {J.}~\bibnamefont
  {Qiu}}, \bibinfo {author} {\bibfnamefont {H.}~\bibnamefont {Fu}}, \bibinfo
  {author} {\bibfnamefont {Y.}~\bibnamefont {Xu}}, \bibinfo {author}
  {\bibfnamefont {A.~I.}\ \bibnamefont {Oreshkin}}, \bibinfo {author}
  {\bibfnamefont {T.}~\bibnamefont {Shao}}, \bibinfo {author} {\bibfnamefont
  {H.}~\bibnamefont {Li}}, \bibinfo {author} {\bibfnamefont {S.}~\bibnamefont
  {Meng}}, \bibinfo {author} {\bibfnamefont {L.}~\bibnamefont {Chen}},\ and\
  \bibinfo {author} {\bibfnamefont {K.}~\bibnamefont {Wu}},\ }\bibinfo {title}
  {Ordered and reversible hydrogenation of silicene},\ \href
  {https://doi.org/10.1103/PhysRevLett.114.126101} {\bibfield  {journal}
  {\bibinfo  {journal} {Phys. Rev. Lett.}\ }\textbf {\bibinfo {volume} {114}},\
  \bibinfo {pages} {126101} (\bibinfo {year} {2015}{\natexlab{a}})}\BibitemShut
  {NoStop}%
\bibitem [{\citenamefont {Qiu}\ \emph {et~al.}(2015{\natexlab{b}})\citenamefont
  {Qiu}, \citenamefont {Fu}, \citenamefont {Xu}, \citenamefont {Zhou},
  \citenamefont {Meng}, \citenamefont {Li}, \citenamefont {Chen},\ and\
  \citenamefont {Wu}}]{EXP-Silicene-Hydrogenation-oneside-2}%
  \BibitemOpen
  \bibfield  {author} {\bibinfo {author} {\bibfnamefont {J.}~\bibnamefont
  {Qiu}}, \bibinfo {author} {\bibfnamefont {H.}~\bibnamefont {Fu}}, \bibinfo
  {author} {\bibfnamefont {Y.}~\bibnamefont {Xu}}, \bibinfo {author}
  {\bibfnamefont {Q.}~\bibnamefont {Zhou}}, \bibinfo {author} {\bibfnamefont
  {S.}~\bibnamefont {Meng}}, \bibinfo {author} {\bibfnamefont {H.}~\bibnamefont
  {Li}}, \bibinfo {author} {\bibfnamefont {L.}~\bibnamefont {Chen}},\ and\
  \bibinfo {author} {\bibfnamefont {K.}~\bibnamefont {Wu}},\ }\bibinfo {title}
  {From silicene to half-silicane by hydrogenation},\ \href
  {https://doi.org/10.1021/acsnano.5b04722} {\bibfield  {journal} {\bibinfo
  {journal} {ACS Nano}\ }\textbf {\bibinfo {volume} {9}},\ \bibinfo {pages}
  {11192} (\bibinfo {year} {2015}{\natexlab{b}})}\BibitemShut {NoStop}%
\bibitem [{\citenamefont {Wang}\ \emph
  {et~al.}(2016{\natexlab{b}})\citenamefont {Wang}, \citenamefont {Olovsson},\
  and\ \citenamefont
  {Uhrberg}}]{EXP-Silicene-Hydrogenation-Oneside-2016-APRES-HighCoverage50}%
  \BibitemOpen
  \bibfield  {author} {\bibinfo {author} {\bibfnamefont {W.}~\bibnamefont
  {Wang}}, \bibinfo {author} {\bibfnamefont {W.}~\bibnamefont {Olovsson}},\
  and\ \bibinfo {author} {\bibfnamefont {R.~I.~G.}\ \bibnamefont {Uhrberg}},\
  }\bibinfo {title} {Band structure of hydrogenated silicene on Ag(111):
  Evidence for half-silicane},\ \href
  {https://doi.org/10.1103/PhysRevB.93.081406} {\bibfield  {journal} {\bibinfo
  {journal} {Phys. Rev. B}\ }\textbf {\bibinfo {volume} {93}},\ \bibinfo
  {pages} {081406} (\bibinfo {year} {2016}{\natexlab{b}})}\BibitemShut
  {NoStop}%
\bibitem [{\citenamefont {Dahn}\ \emph {et~al.}(1993)\citenamefont {Dahn},
  \citenamefont {Way}, \citenamefont {Fuller},\ and\ \citenamefont
  {Tse}}]{EXP-Silicene-Hydrogenation-Si6H6-twoside}%
  \BibitemOpen
  \bibfield  {author} {\bibinfo {author} {\bibfnamefont {J.~R.}\ \bibnamefont
  {Dahn}}, \bibinfo {author} {\bibfnamefont {B.~M.}\ \bibnamefont {Way}},
  \bibinfo {author} {\bibfnamefont {E.}~\bibnamefont {Fuller}},\ and\ \bibinfo
  {author} {\bibfnamefont {J.~S.}\ \bibnamefont {Tse}},\ }\bibinfo {title}
  {Structure of siloxene and layered polysilane
  (Si$_{\text{6}}$H$_{\text{6}}$)},\ \href
  {https://doi.org/10.1103/PhysRevB.48.17872} {\bibfield  {journal} {\bibinfo
  {journal} {Phys. Rev. B}\ }\textbf {\bibinfo {volume} {48}},\ \bibinfo
  {pages} {17872} (\bibinfo {year} {1993})}\BibitemShut {NoStop}%
\bibitem [{\citenamefont {Yamanaka}\ \emph {et~al.}(1996)\citenamefont
  {Yamanaka}, \citenamefont {Matsu-ura},\ and\ \citenamefont
  {Ishikawa}}]{EXP-Silicene-Hydrogenation-Twoside-1996}%
  \BibitemOpen
  \bibfield  {author} {\bibinfo {author} {\bibfnamefont {S.}~\bibnamefont
  {Yamanaka}}, \bibinfo {author} {\bibfnamefont {H.}~\bibnamefont
  {Matsu-ura}},\ and\ \bibinfo {author} {\bibfnamefont {M.}~\bibnamefont
  {Ishikawa}},\ }\bibinfo {title} {New deintercalation reaction of calcium from
  calcium disilicide. Synthesis of layered polysilane},\ \href
  {https://doi.org/https://doi.org/10.1016/0025-5408(95)00195-6} {\bibfield
  {journal} {\bibinfo  {journal} {Mater. Res. Bull.}\ }\textbf {\bibinfo
  {volume} {31}},\ \bibinfo {pages} {307} (\bibinfo {year} {1996})}\BibitemShut
  {NoStop}%
\bibitem [{\citenamefont {Dettlaff-Weglikowska}\ \emph
  {et~al.}(1997)\citenamefont {Dettlaff-Weglikowska}, \citenamefont {H\"onle},
  \citenamefont {Molassioti-Dohms}, \citenamefont {Finkbeiner},\ and\
  \citenamefont {Weber}}]{EXP-Silicene-Hydrogenation-Twoside-1997}%
  \BibitemOpen
  \bibfield  {author} {\bibinfo {author} {\bibfnamefont {U.}~\bibnamefont
  {Dettlaff-Weglikowska}}, \bibinfo {author} {\bibfnamefont {W.}~\bibnamefont
  {H\"onle}}, \bibinfo {author} {\bibfnamefont {A.}~\bibnamefont
  {Molassioti-Dohms}}, \bibinfo {author} {\bibfnamefont {S.}~\bibnamefont
  {Finkbeiner}},\ and\ \bibinfo {author} {\bibfnamefont {J.}~\bibnamefont
  {Weber}},\ }\bibinfo {title} {Structure and optical properties of the planar
  silicon compounds polysilane and W\"ohler siloxene},\ \href
  {https://doi.org/10.1103/PhysRevB.56.13132} {\bibfield  {journal} {\bibinfo
  {journal} {Phys. Rev. B}\ }\textbf {\bibinfo {volume} {56}},\ \bibinfo
  {pages} {13132} (\bibinfo {year} {1997})}\BibitemShut {NoStop}%
\bibitem [{\citenamefont {Nakano}\ \emph {et~al.}(2012)\citenamefont {Nakano},
  \citenamefont {Nakano}, \citenamefont {Nakanishi}, \citenamefont {Tanaka},
  \citenamefont {Sugiyama}, \citenamefont {Ikuno}, \citenamefont {Okamoto},\
  and\ \citenamefont {Ohta}}]{EXP-Silicene-Hydrogenation-Bulk}%
  \BibitemOpen
  \bibfield  {author} {\bibinfo {author} {\bibfnamefont {H.}~\bibnamefont
  {Nakano}}, \bibinfo {author} {\bibfnamefont {M.}~\bibnamefont {Nakano}},
  \bibinfo {author} {\bibfnamefont {K.}~\bibnamefont {Nakanishi}}, \bibinfo
  {author} {\bibfnamefont {D.}~\bibnamefont {Tanaka}}, \bibinfo {author}
  {\bibfnamefont {Y.}~\bibnamefont {Sugiyama}}, \bibinfo {author}
  {\bibfnamefont {T.}~\bibnamefont {Ikuno}}, \bibinfo {author} {\bibfnamefont
  {H.}~\bibnamefont {Okamoto}},\ and\ \bibinfo {author} {\bibfnamefont
  {T.}~\bibnamefont {Ohta}},\ }\bibinfo {title} {Preparation of alkyl-modified
  silicon nanosheets by hydrosilylation of layered polysilane
  (Si$_{\text{6}}$H$_{\text{6}}$)},\ \href {https://doi.org/10.1021/ja212086n}
  {\bibfield  {journal} {\bibinfo  {journal} {J. Am. Chem. Soc.}\ }\textbf
  {\bibinfo {volume} {134}},\ \bibinfo {pages} {5452} (\bibinfo {year}
  {2012})}\BibitemShut {NoStop}%
\bibitem [{\citenamefont {Qiu}\ \emph {et~al.}(2022)\citenamefont {Qiu},
  \citenamefont {Wang}, \citenamefont {Wang}, \citenamefont {Yao},
  \citenamefont {Meng},\ and\ \citenamefont
  {Liu}}]{EXP-Silicene-Hydrogenation-2022}%
  \BibitemOpen
  \bibfield  {author} {\bibinfo {author} {\bibfnamefont {J.}~\bibnamefont
  {Qiu}}, \bibinfo {author} {\bibfnamefont {H.}~\bibnamefont {Wang}}, \bibinfo
  {author} {\bibfnamefont {J.}~\bibnamefont {Wang}}, \bibinfo {author}
  {\bibfnamefont {X.}~\bibnamefont {Yao}}, \bibinfo {author} {\bibfnamefont
  {S.}~\bibnamefont {Meng}},\ and\ \bibinfo {author} {\bibfnamefont
  {Y.}~\bibnamefont {Liu}},\ }\bibinfo {title} {Revealing the hydrogenated
  structure of
  silicene-$(\ensuremath{\surd}13\ifmmode\times\else\texttimes\fi{}\ensuremath{\surd}13)R13.{9}^{\ensuremath{\circ}}$
  by tip-induced dehydrogenation},\ \href
  {https://doi.org/10.1103/PhysRevB.106.184102} {\bibfield  {journal} {\bibinfo
   {journal} {Phys. Rev. B}\ }\textbf {\bibinfo {volume} {106}},\ \bibinfo
  {pages} {184102} (\bibinfo {year} {2022})}\BibitemShut {NoStop}%
\bibitem [{\citenamefont {Deng}\ \emph {et~al.}(2022)\citenamefont {Deng},
  \citenamefont {Zhao}, \citenamefont {Park}, \citenamefont {Yan},
  \citenamefont {Sobczak}, \citenamefont {Lakra}, \citenamefont {Buzi},\ and\
  \citenamefont
  {Krusin-Elbaum}}]{EXP-Hydrogenation-Bi2Te3-revisible-preventroxidation}%
  \BibitemOpen
  \bibfield  {author} {\bibinfo {author} {\bibfnamefont {H.}~\bibnamefont
  {Deng}}, \bibinfo {author} {\bibfnamefont {L.}~\bibnamefont {Zhao}}, \bibinfo
  {author} {\bibfnamefont {K.}~\bibnamefont {Park}}, \bibinfo {author}
  {\bibfnamefont {J.}~\bibnamefont {Yan}}, \bibinfo {author} {\bibfnamefont
  {K.}~\bibnamefont {Sobczak}}, \bibinfo {author} {\bibfnamefont
  {A.}~\bibnamefont {Lakra}}, \bibinfo {author} {\bibfnamefont
  {E.}~\bibnamefont {Buzi}},\ and\ \bibinfo {author} {\bibfnamefont
  {L.}~\bibnamefont {Krusin-Elbaum}},\ }\bibinfo {title} {Topological surface
  currents accessed through reversible hydrogenation of the three-dimensional
  bulk},\ \href {https://doi.org/10.1038/s41467-022-29957-3} {\bibfield
  {journal} {\bibinfo  {journal} {Nat. Commun.}\ }\textbf {\bibinfo {volume}
  {13}},\ \bibinfo {pages} {2308} (\bibinfo {year} {2022})}\BibitemShut
  {NoStop}%
\bibitem [{\citenamefont {Wang}\ \emph {et~al.}(2020)\citenamefont {Wang},
  \citenamefont {Xia}, \citenamefont {Gou}, \citenamefont {Cheng},
  \citenamefont {Xu}, \citenamefont {Chen},\ and\ \citenamefont
  {Wu}}]{EXP-Sn2Bi-Hydrogenation-semiconductor}%
  \BibitemOpen
  \bibfield  {author} {\bibinfo {author} {\bibfnamefont {X.}~\bibnamefont
  {Wang}}, \bibinfo {author} {\bibfnamefont {B.}~\bibnamefont {Xia}}, \bibinfo
  {author} {\bibfnamefont {J.}~\bibnamefont {Gou}}, \bibinfo {author}
  {\bibfnamefont {P.}~\bibnamefont {Cheng}}, \bibinfo {author} {\bibfnamefont
  {Y.}~\bibnamefont {Xu}}, \bibinfo {author} {\bibfnamefont {L.}~\bibnamefont
  {Chen}},\ and\ \bibinfo {author} {\bibfnamefont {K.}~\bibnamefont {Wu}},\
  }\bibinfo {title} {Symmetry breaking and reversible hydrogenation of
  two-dimensional semiconductor Sn$_{\text{2}}$Bi},\ \href
  {https://doi.org/10.1088/0256-307X/37/6/066802} {\bibfield  {journal}
  {\bibinfo  {journal} {Chin. Phys. Lett.}\ }\textbf {\bibinfo {volume} {37}},\
  \bibinfo {pages} {066802} (\bibinfo {year} {2020})}\BibitemShut {NoStop}%
\bibitem [{\citenamefont {Han}\ \emph {et~al.}(2013)\citenamefont {Han},
  \citenamefont {Hwang}, \citenamefont {Kim}, \citenamefont {Yun},
  \citenamefont {Lee}, \citenamefont {Park}, \citenamefont {Ryu}, \citenamefont
  {Park}, \citenamefont {Yoo}, \citenamefont {Yoon}, \citenamefont {Hong},
  \citenamefont {Kim},\ and\ \citenamefont
  {Park}}]{EXP-2H-MoS2-Hydrogenation-Ferromagnetism}%
  \BibitemOpen
  \bibfield  {author} {\bibinfo {author} {\bibfnamefont {S.~W.}\ \bibnamefont
  {Han}}, \bibinfo {author} {\bibfnamefont {Y.~H.}\ \bibnamefont {Hwang}},
  \bibinfo {author} {\bibfnamefont {S.-H.}\ \bibnamefont {Kim}}, \bibinfo
  {author} {\bibfnamefont {W.~S.}\ \bibnamefont {Yun}}, \bibinfo {author}
  {\bibfnamefont {J.~D.}\ \bibnamefont {Lee}}, \bibinfo {author} {\bibfnamefont
  {M.~G.}\ \bibnamefont {Park}}, \bibinfo {author} {\bibfnamefont
  {S.}~\bibnamefont {Ryu}}, \bibinfo {author} {\bibfnamefont {J.~S.}\
  \bibnamefont {Park}}, \bibinfo {author} {\bibfnamefont {D.-H.}\ \bibnamefont
  {Yoo}}, \bibinfo {author} {\bibfnamefont {S.-P.}\ \bibnamefont {Yoon}},
  \bibinfo {author} {\bibfnamefont {S.~C.}\ \bibnamefont {Hong}}, \bibinfo
  {author} {\bibfnamefont {K.~S.}\ \bibnamefont {Kim}},\ and\ \bibinfo {author}
  {\bibfnamefont {Y.~S.}\ \bibnamefont {Park}},\ }\bibinfo {title} {Controlling
  ferromagnetic easy axis in a layered MoS$_{\text{2}}$ single crystal},\ \href
  {https://doi.org/10.1103/PhysRevLett.110.247201} {\bibfield  {journal}
  {\bibinfo  {journal} {Phys. Rev. Lett.}\ }\textbf {\bibinfo {volume} {110}},\
  \bibinfo {pages} {247201} (\bibinfo {year} {2013})}\BibitemShut {NoStop}%
\bibitem [{\citenamefont {Han}\ \emph {et~al.}(2015)\citenamefont {Han},
  \citenamefont {Yun}, \citenamefont {Lee}, \citenamefont {Hwang},
  \citenamefont {Baik}, \citenamefont {Shin}, \citenamefont {Lee},
  \citenamefont {Park},\ and\ \citenamefont
  {Kim}}]{EXP-2H-MoS2-Hydrogenation-PhaseTransition-Twoside}%
  \BibitemOpen
  \bibfield  {author} {\bibinfo {author} {\bibfnamefont {S.~W.}\ \bibnamefont
  {Han}}, \bibinfo {author} {\bibfnamefont {W.~S.}\ \bibnamefont {Yun}},
  \bibinfo {author} {\bibfnamefont {J.~D.}\ \bibnamefont {Lee}}, \bibinfo
  {author} {\bibfnamefont {Y.~H.}\ \bibnamefont {Hwang}}, \bibinfo {author}
  {\bibfnamefont {J.}~\bibnamefont {Baik}}, \bibinfo {author} {\bibfnamefont
  {H.~J.}\ \bibnamefont {Shin}}, \bibinfo {author} {\bibfnamefont {W.~G.}\
  \bibnamefont {Lee}}, \bibinfo {author} {\bibfnamefont {Y.~S.}\ \bibnamefont
  {Park}},\ and\ \bibinfo {author} {\bibfnamefont {K.~S.}\ \bibnamefont
  {Kim}},\ }\bibinfo {title} {Hydrogenation-induced atomic stripes on the
  2H-MoS$_{\text{2}}$ surface},\ \href
  {https://doi.org/10.1103/PhysRevB.92.241303} {\bibfield  {journal} {\bibinfo
  {journal} {Phys. Rev. B}\ }\textbf {\bibinfo {volume} {92}},\ \bibinfo
  {pages} {241303} (\bibinfo {year} {2015})}\BibitemShut {NoStop}%
\bibitem [{\citenamefont {Sundberg}\ \emph {et~al.}(1991)\citenamefont
  {Sundberg}, \citenamefont {Moyes},\ and\ \citenamefont
  {Tomkinson}}]{EXP-1991-MoS2-Hydrogenation-uptake0.23}%
  \BibitemOpen
  \bibfield  {author} {\bibinfo {author} {\bibfnamefont {P.}~\bibnamefont
  {Sundberg}}, \bibinfo {author} {\bibfnamefont {R.~B.}\ \bibnamefont
  {Moyes}},\ and\ \bibinfo {author} {\bibfnamefont {J.}~\bibnamefont
  {Tomkinson}},\ }\bibinfo {title} {Inelastic neutron scattering spectroscopy
  of hydrogen adsorbed on powdered-MoS$_{\text{2}}$, MoS$_{\text{2}}$-alumina
  and nickel-promoted MOS$_{\text{2}}$},\ \href
  {https://doi.org/https://doi.org/10.1002/bscb.19911001123} {\bibfield
  {journal} {\bibinfo  {journal} {Bull. Soc. Chim. Belg.}\ }\textbf {\bibinfo
  {volume} {100}},\ \bibinfo {pages} {967} (\bibinfo {year}
  {1991})}\BibitemShut {NoStop}%
\bibitem [{\citenamefont {Wright}\ \emph {et~al.}(1980)\citenamefont {Wright},
  \citenamefont {Sampson}, \citenamefont {Fraser}, \citenamefont {Moyes},
  \citenamefont {Wells},\ and\ \citenamefont
  {Riekel}}]{EXP-1980-MoS2-Hydrogenation-uptake0.37}%
  \BibitemOpen
  \bibfield  {author} {\bibinfo {author} {\bibfnamefont {C.~J.}\ \bibnamefont
  {Wright}}, \bibinfo {author} {\bibfnamefont {C.}~\bibnamefont {Sampson}},
  \bibinfo {author} {\bibfnamefont {D.}~\bibnamefont {Fraser}}, \bibinfo
  {author} {\bibfnamefont {R.~B.}\ \bibnamefont {Moyes}}, \bibinfo {author}
  {\bibfnamefont {P.~B.}\ \bibnamefont {Wells}},\ and\ \bibinfo {author}
  {\bibfnamefont {C.}~\bibnamefont {Riekel}},\ }\bibinfo {title} {Hydrogen
  sorption by molybdenum sulphide catalysts},\ \href
  {https://doi.org/10.1039/F19807601585} {\bibfield  {journal} {\bibinfo
  {journal} {J.C.S. Faraday I}\ }\textbf {\bibinfo {volume} {76}},\ \bibinfo
  {pages} {1585} (\bibinfo {year} {1980})}\BibitemShut {NoStop}%
\bibitem [{\citenamefont {Pierucci}\ \emph {et~al.}(2017)\citenamefont
  {Pierucci}, \citenamefont {Henck}, \citenamefont {Ben~Aziza}, \citenamefont
  {Naylor}, \citenamefont {Balan}, \citenamefont {Rault}, \citenamefont
  {Silly}, \citenamefont {Dappe}, \citenamefont {Bertran}, \citenamefont
  {Le~Fèvre}, \citenamefont {Sirotti}, \citenamefont {Johnson},\ and\
  \citenamefont {Ouerghi}}]{EXP-MoS2-hydrogenation-np-type-Vacancy}%
  \BibitemOpen
  \bibfield  {author} {\bibinfo {author} {\bibfnamefont {D.}~\bibnamefont
  {Pierucci}}, \bibinfo {author} {\bibfnamefont {H.}~\bibnamefont {Henck}},
  \bibinfo {author} {\bibfnamefont {Z.}~\bibnamefont {Ben~Aziza}}, \bibinfo
  {author} {\bibfnamefont {C.~H.}\ \bibnamefont {Naylor}}, \bibinfo {author}
  {\bibfnamefont {A.}~\bibnamefont {Balan}}, \bibinfo {author} {\bibfnamefont
  {J.~E.}\ \bibnamefont {Rault}}, \bibinfo {author} {\bibfnamefont {M.~G.}\
  \bibnamefont {Silly}}, \bibinfo {author} {\bibfnamefont {Y.~J.}\ \bibnamefont
  {Dappe}}, \bibinfo {author} {\bibfnamefont {F.}~\bibnamefont {Bertran}},
  \bibinfo {author} {\bibfnamefont {P.}~\bibnamefont {Le~Fèvre}}, \bibinfo
  {author} {\bibfnamefont {F.}~\bibnamefont {Sirotti}}, \bibinfo {author}
  {\bibfnamefont {A.~T.~C.}\ \bibnamefont {Johnson}},\ and\ \bibinfo {author}
  {\bibfnamefont {A.}~\bibnamefont {Ouerghi}},\ }\bibinfo {title} {Tunable
  doping in hydrogenated single layered molybdenum disulfide},\ \href
  {https://doi.org/10.1021/acsnano.6b07661} {\bibfield  {journal} {\bibinfo
  {journal} {ACS Nano}\ }\textbf {\bibinfo {volume} {11}},\ \bibinfo {pages}
  {1755} (\bibinfo {year} {2017})}\BibitemShut {NoStop}%
\bibitem [{\citenamefont {Ma}\ \emph {et~al.}(2017)\citenamefont {Ma},
  \citenamefont {Yoon}, \citenamefont {Jang}, \citenamefont {Jeong},
  \citenamefont {Kim}, \citenamefont {Nayak},\ and\ \citenamefont
  {Shin}}]{EXP-MoSe2-Hydrogenation}%
  \BibitemOpen
  \bibfield  {author} {\bibinfo {author} {\bibfnamefont {K.~Y.}\ \bibnamefont
  {Ma}}, \bibinfo {author} {\bibfnamefont {S.~I.}\ \bibnamefont {Yoon}},
  \bibinfo {author} {\bibfnamefont {A.-R.}\ \bibnamefont {Jang}}, \bibinfo
  {author} {\bibfnamefont {H.~Y.}\ \bibnamefont {Jeong}}, \bibinfo {author}
  {\bibfnamefont {Y.-J.}\ \bibnamefont {Kim}}, \bibinfo {author} {\bibfnamefont
  {P.~K.}\ \bibnamefont {Nayak}},\ and\ \bibinfo {author} {\bibfnamefont
  {H.~S.}\ \bibnamefont {Shin}},\ }\bibinfo {title} {Hydrogenation of monolayer
  molybdenum diselenide via hydrogen plasma treatment},\ \href
  {https://doi.org/10.1039/C7TC02592K} {\bibfield  {journal} {\bibinfo
  {journal} {J. Mater. Chem. C}\ }\textbf {\bibinfo {volume} {5}},\ \bibinfo
  {pages} {11294} (\bibinfo {year} {2017})}\BibitemShut {NoStop}%
\bibitem [{\citenamefont {Blöchl}(1994)}]{PAWmethod}%
  \BibitemOpen
  \bibfield  {author} {\bibinfo {author} {\bibfnamefont {P.~E.}\ \bibnamefont
  {Blöchl}},\ }\bibinfo {title} {Projector augmented-wave method},\ \href
  {https://doi.org/10.1103/PhysRevB.50.17953} {\bibfield  {journal} {\bibinfo
  {journal} {Phys. Rev. B}\ }\textbf {\bibinfo {volume} {50}},\ \bibinfo
  {pages} {17953} (\bibinfo {year} {1994})}\BibitemShut {NoStop}%
\bibitem [{\citenamefont {Kresse}\ and\ \citenamefont
  {Furthmüller}(1996)}]{VASP}%
  \BibitemOpen
  \bibfield  {author} {\bibinfo {author} {\bibfnamefont {G.}~\bibnamefont
  {Kresse}}\ and\ \bibinfo {author} {\bibfnamefont {J.}~\bibnamefont
  {Furthmüller}},\ }\bibinfo {title} {Efficient iterative schemes for ab
  initio total-energy calculations using a plane-wave basis set},\ \href
  {https://doi.org/10.1103/PhysRevB.54.11169} {\bibfield  {journal} {\bibinfo
  {journal} {Phys. Rev. B}\ }\textbf {\bibinfo {volume} {54}},\ \bibinfo
  {pages} {11169} (\bibinfo {year} {1996})}\BibitemShut {NoStop}%
\bibitem [{\citenamefont {Perdew}\ \emph {et~al.}(1996)\citenamefont {Perdew},
  \citenamefont {Burke},\ and\ \citenamefont {Ernzerhof}}]{PBE}%
  \BibitemOpen
  \bibfield  {author} {\bibinfo {author} {\bibfnamefont {J.~P.}\ \bibnamefont
  {Perdew}}, \bibinfo {author} {\bibfnamefont {K.}~\bibnamefont {Burke}},\ and\
  \bibinfo {author} {\bibfnamefont {M.}~\bibnamefont {Ernzerhof}},\ }\bibinfo
  {title} {Generalized gradient approximation made simple},\ \href
  {https://doi.org/10.1103/PhysRevLett.77.3865} {\bibfield  {journal} {\bibinfo
   {journal} {Phys. Rev. Lett.}\ }\textbf {\bibinfo {volume} {77}},\ \bibinfo
  {pages} {3865} (\bibinfo {year} {1996})}\BibitemShut {NoStop}%
\bibitem [{\citenamefont {Monkhorst}\ and\ \citenamefont
  {Pack}(1976)}]{M-PKpoints}%
  \BibitemOpen
  \bibfield  {author} {\bibinfo {author} {\bibfnamefont {H.~J.}\ \bibnamefont
  {Monkhorst}}\ and\ \bibinfo {author} {\bibfnamefont {J.~D.}\ \bibnamefont
  {Pack}},\ }\bibinfo {title} {Special points for {Brillouin-zone}
  integrations},\ \href {https://doi.org/10.1103/PhysRevB.13.5188} {\bibfield
  {journal} {\bibinfo  {journal} {Phys. Rev. B}\ }\textbf {\bibinfo {volume}
  {13}},\ \bibinfo {pages} {5188} (\bibinfo {year} {1976})}\BibitemShut
  {NoStop}%
\bibitem [{\citenamefont {Dudarev}\ \emph {et~al.}(1998)\citenamefont
  {Dudarev}, \citenamefont {Botton}, \citenamefont {Savrasov}, \citenamefont
  {Humphreys},\ and\ \citenamefont {Sutton}}]{LDAUTYPE=2}%
  \BibitemOpen
  \bibfield  {author} {\bibinfo {author} {\bibfnamefont {S.~L.}\ \bibnamefont
  {Dudarev}}, \bibinfo {author} {\bibfnamefont {G.~A.}\ \bibnamefont {Botton}},
  \bibinfo {author} {\bibfnamefont {S.~Y.}\ \bibnamefont {Savrasov}}, \bibinfo
  {author} {\bibfnamefont {C.~J.}\ \bibnamefont {Humphreys}},\ and\ \bibinfo
  {author} {\bibfnamefont {A.~P.}\ \bibnamefont {Sutton}},\ }\bibinfo {title}
  {Electron-energy-loss spectra and the structural stability of nickel oxide:
  An {LSDA+U} study},\ \href {https://doi.org/10.1103/PhysRevB.57.1505}
  {\bibfield  {journal} {\bibinfo  {journal} {Phys. Rev. B}\ }\textbf {\bibinfo
  {volume} {57}},\ \bibinfo {pages} {1505} (\bibinfo {year}
  {1998})}\BibitemShut {NoStop}%
\bibitem [{\citenamefont {Wolff}(1989)}]{MonteCarloMethod1}%
  \BibitemOpen
  \bibfield  {author} {\bibinfo {author} {\bibfnamefont {U.}~\bibnamefont
  {Wolff}},\ }\bibinfo {title} {Collective Monte Carlo updating for spin
  systems},\ \href {https://doi.org/10.1103/PhysRevLett.62.361} {\bibfield
  {journal} {\bibinfo  {journal} {Phys. Rev. Lett.}\ }\textbf {\bibinfo
  {volume} {62}},\ \bibinfo {pages} {361} (\bibinfo {year} {1989})}\BibitemShut
  {NoStop}%
\bibitem [{\citenamefont {Alzate-Cardona}\ \emph {et~al.}(2019)\citenamefont
  {Alzate-Cardona}, \citenamefont {Sabogal-Suárez}, \citenamefont {Evans},\
  and\ \citenamefont {Restrepo-Parra}}]{MonteCarloMethod2}%
  \BibitemOpen
  \bibfield  {author} {\bibinfo {author} {\bibfnamefont {J.~D.}\ \bibnamefont
  {Alzate-Cardona}}, \bibinfo {author} {\bibfnamefont {D.}~\bibnamefont
  {Sabogal-Suárez}}, \bibinfo {author} {\bibfnamefont {R.~F.~L.}\ \bibnamefont
  {Evans}},\ and\ \bibinfo {author} {\bibfnamefont {E.}~\bibnamefont
  {Restrepo-Parra}},\ }\bibinfo {title} {Optimal phase space sampling for Monte
  Carlo simulations of Heisenberg spin systems},\ \href
  {https://doi.org/10.1088/1361-648X/aaf852} {\bibfield  {journal} {\bibinfo
  {journal} {J. Phys. Condens. Matter}\ }\textbf {\bibinfo {volume} {31}},\
  \bibinfo {pages} {095802} (\bibinfo {year} {2019})}\BibitemShut {NoStop}%
\bibitem [{\citenamefont {Pizzi}\ \emph {et~al.}(2020)\citenamefont {Pizzi},
  \citenamefont {Vitale}, \citenamefont {Arita}, \citenamefont {Blügel},
  \citenamefont {Freimuth}, \citenamefont {Géranton}, \citenamefont
  {Gibertini}, \citenamefont {Gresch}, \citenamefont {Johnson}, \citenamefont
  {Koretsune}, \citenamefont {Ibañez-Azpiroz}, \citenamefont {Lee},
  \citenamefont {Lihm}, \citenamefont {Marchand}, \citenamefont {Marrazzo},
  \citenamefont {Mokrousov}, \citenamefont {Mustafa}, \citenamefont {Nohara},
  \citenamefont {Nomura}, \citenamefont {Paulatto}, \citenamefont {Poncé},
  \citenamefont {Ponweiser}, \citenamefont {Qiao}, \citenamefont {Thöle},
  \citenamefont {Tsirkin}, \citenamefont {Wierzbowska}, \citenamefont
  {Marzari}, \citenamefont {Vanderbilt}, \citenamefont {Souza}, \citenamefont
  {Mostofi},\ and\ \citenamefont {Yates}}]{Wannier90}%
  \BibitemOpen
  \bibfield  {author} {\bibinfo {author} {\bibfnamefont {G.}~\bibnamefont
  {Pizzi}}, \bibinfo {author} {\bibfnamefont {V.}~\bibnamefont {Vitale}},
  \bibinfo {author} {\bibfnamefont {R.}~\bibnamefont {Arita}}, \bibinfo
  {author} {\bibfnamefont {S.}~\bibnamefont {Blügel}}, \bibinfo {author}
  {\bibfnamefont {F.}~\bibnamefont {Freimuth}}, \bibinfo {author}
  {\bibfnamefont {G.}~\bibnamefont {Géranton}}, \bibinfo {author}
  {\bibfnamefont {M.}~\bibnamefont {Gibertini}}, \bibinfo {author}
  {\bibfnamefont {D.}~\bibnamefont {Gresch}}, \bibinfo {author} {\bibfnamefont
  {C.}~\bibnamefont {Johnson}}, \bibinfo {author} {\bibfnamefont
  {T.}~\bibnamefont {Koretsune}}, \bibinfo {author} {\bibfnamefont
  {J.}~\bibnamefont {Ibañez-Azpiroz}}, \bibinfo {author} {\bibfnamefont
  {H.}~\bibnamefont {Lee}}, \bibinfo {author} {\bibfnamefont {J.-M.}\
  \bibnamefont {Lihm}}, \bibinfo {author} {\bibfnamefont {D.}~\bibnamefont
  {Marchand}}, \bibinfo {author} {\bibfnamefont {A.}~\bibnamefont {Marrazzo}},
  \bibinfo {author} {\bibfnamefont {Y.}~\bibnamefont {Mokrousov}}, \bibinfo
  {author} {\bibfnamefont {J.~I.}\ \bibnamefont {Mustafa}}, \bibinfo {author}
  {\bibfnamefont {Y.}~\bibnamefont {Nohara}}, \bibinfo {author} {\bibfnamefont
  {Y.}~\bibnamefont {Nomura}}, \bibinfo {author} {\bibfnamefont
  {L.}~\bibnamefont {Paulatto}}, \bibinfo {author} {\bibfnamefont
  {S.}~\bibnamefont {Poncé}}, \bibinfo {author} {\bibfnamefont
  {T.}~\bibnamefont {Ponweiser}}, \bibinfo {author} {\bibfnamefont
  {J.}~\bibnamefont {Qiao}}, \bibinfo {author} {\bibfnamefont {F.}~\bibnamefont
  {Thöle}}, \bibinfo {author} {\bibfnamefont {S.~S.}\ \bibnamefont {Tsirkin}},
  \bibinfo {author} {\bibfnamefont {M.}~\bibnamefont {Wierzbowska}}, \bibinfo
  {author} {\bibfnamefont {N.}~\bibnamefont {Marzari}}, \bibinfo {author}
  {\bibfnamefont {D.}~\bibnamefont {Vanderbilt}}, \bibinfo {author}
  {\bibfnamefont {I.}~\bibnamefont {Souza}}, \bibinfo {author} {\bibfnamefont
  {A.~A.}\ \bibnamefont {Mostofi}},\ and\ \bibinfo {author} {\bibfnamefont
  {J.~R.}\ \bibnamefont {Yates}},\ }\bibinfo {title} {Wannier90 as a community
  code: new features and applications},\ \href
  {https://doi.org/10.1088/1361-648X/ab51ff} {\bibfield  {journal} {\bibinfo
  {journal} {J. Phys.: Condens. Matter.}\ }\textbf {\bibinfo {volume} {32}},\
  \bibinfo {pages} {165902} (\bibinfo {year} {2020})}\BibitemShut {NoStop}%
\bibitem [{\citenamefont {Marzari}\ \emph {et~al.}(2012)\citenamefont
  {Marzari}, \citenamefont {Mostofi}, \citenamefont {Yates}, \citenamefont
  {Souza},\ and\ \citenamefont
  {Vanderbilt}}]{Review-Maximally-Localized-Wannier90}%
  \BibitemOpen
  \bibfield  {author} {\bibinfo {author} {\bibfnamefont {N.}~\bibnamefont
  {Marzari}}, \bibinfo {author} {\bibfnamefont {A.~A.}\ \bibnamefont
  {Mostofi}}, \bibinfo {author} {\bibfnamefont {J.~R.}\ \bibnamefont {Yates}},
  \bibinfo {author} {\bibfnamefont {I.}~\bibnamefont {Souza}},\ and\ \bibinfo
  {author} {\bibfnamefont {D.}~\bibnamefont {Vanderbilt}},\ }\bibinfo {title}
  {Maximally localized Wannier functions: Theory and applications},\ \href
  {https://doi.org/10.1103/RevModPhys.84.1419} {\bibfield  {journal} {\bibinfo
  {journal} {Rev. Mod. Phys.}\ }\textbf {\bibinfo {volume} {84}},\ \bibinfo
  {pages} {1419} (\bibinfo {year} {2012})}\BibitemShut {NoStop}%
\bibitem [{\citenamefont {Wu}\ \emph {et~al.}(2018)\citenamefont {Wu},
  \citenamefont {Zhang}, \citenamefont {Song}, \citenamefont {Troyer},\ and\
  \citenamefont {Soluyanov}}]{WannierTools}%
  \BibitemOpen
  \bibfield  {author} {\bibinfo {author} {\bibfnamefont {Q.}~\bibnamefont
  {Wu}}, \bibinfo {author} {\bibfnamefont {S.}~\bibnamefont {Zhang}}, \bibinfo
  {author} {\bibfnamefont {H.-F.}\ \bibnamefont {Song}}, \bibinfo {author}
  {\bibfnamefont {M.}~\bibnamefont {Troyer}},\ and\ \bibinfo {author}
  {\bibfnamefont {A.~A.}\ \bibnamefont {Soluyanov}},\ }\bibinfo {title}
  {WannierTools: An open-source software package for novel topological
  materials},\ \href
  {https://doi.org/https://doi.org/10.1016/j.cpc.2017.09.033} {\bibfield
  {journal} {\bibinfo  {journal} {Comput. Phys. Commun.}\ }\textbf {\bibinfo
  {volume} {224}},\ \bibinfo {pages} {405} (\bibinfo {year}
  {2018})}\BibitemShut {NoStop}%
\bibitem [{\citenamefont {Ye}\ \emph {et~al.}(2014)\citenamefont {Ye},
  \citenamefont {Quhe}, \citenamefont {Zheng}, \citenamefont {Ni},
  \citenamefont {Wang}, \citenamefont {Yuan}, \citenamefont {Tse},
  \citenamefont {Shi}, \citenamefont {Gao},\ and\ \citenamefont
  {Lu}}]{DFT-Ge-monolayer-adsorption-energy}%
  \BibitemOpen
  \bibfield  {author} {\bibinfo {author} {\bibfnamefont {M.}~\bibnamefont
  {Ye}}, \bibinfo {author} {\bibfnamefont {R.}~\bibnamefont {Quhe}}, \bibinfo
  {author} {\bibfnamefont {J.}~\bibnamefont {Zheng}}, \bibinfo {author}
  {\bibfnamefont {Z.}~\bibnamefont {Ni}}, \bibinfo {author} {\bibfnamefont
  {Y.}~\bibnamefont {Wang}}, \bibinfo {author} {\bibfnamefont {Y.}~\bibnamefont
  {Yuan}}, \bibinfo {author} {\bibfnamefont {G.}~\bibnamefont {Tse}}, \bibinfo
  {author} {\bibfnamefont {J.}~\bibnamefont {Shi}}, \bibinfo {author}
  {\bibfnamefont {Z.}~\bibnamefont {Gao}},\ and\ \bibinfo {author}
  {\bibfnamefont {J.}~\bibnamefont {Lu}},\ }\bibinfo {title} {Tunable band gap
  in germanene by surface adsorption},\ \href
  {https://doi.org/https://doi.org/10.1016/j.physe.2013.12.016} {\bibfield
  {journal} {\bibinfo  {journal} {Physica E.}\ }\textbf {\bibinfo {volume}
  {59}},\ \bibinfo {pages} {60} (\bibinfo {year} {2014})}\BibitemShut {NoStop}%
\bibitem [{\citenamefont {Shi}\ \emph {et~al.}(2013)\citenamefont {Shi},
  \citenamefont {Pan}, \citenamefont {Zhang},\ and\ \citenamefont
  {Yakobson}}]{DFT-MoS2-adsorption-energy}%
  \BibitemOpen
  \bibfield  {author} {\bibinfo {author} {\bibfnamefont {H.}~\bibnamefont
  {Shi}}, \bibinfo {author} {\bibfnamefont {H.}~\bibnamefont {Pan}}, \bibinfo
  {author} {\bibfnamefont {Y.-W.}\ \bibnamefont {Zhang}},\ and\ \bibinfo
  {author} {\bibfnamefont {B.~I.}\ \bibnamefont {Yakobson}},\ }\bibinfo {title}
  {Strong ferromagnetism in hydrogenated monolayer MoS$_{\text{2}}$ tuned by
  strain},\ \href {https://doi.org/10.1103/PhysRevB.88.205305} {\bibfield
  {journal} {\bibinfo  {journal} {Phys. Rev. B}\ }\textbf {\bibinfo {volume}
  {88}},\ \bibinfo {pages} {205305} (\bibinfo {year} {2013})}\BibitemShut
  {NoStop}%
\bibitem [{\citenamefont {Kresse}\ and\ \citenamefont
  {Hafner}(2000)}]{DFT-H-H-Dissociation-energy-4.48eV}%
  \BibitemOpen
  \bibfield  {author} {\bibinfo {author} {\bibfnamefont {G.}~\bibnamefont
  {Kresse}}\ and\ \bibinfo {author} {\bibfnamefont {J.}~\bibnamefont
  {Hafner}},\ }\bibinfo {title} {First-principles study of the adsorption of
  atomic H on Ni (111), (100) and (110)},\ \href
  {https://doi.org/https://doi.org/10.1016/S0039-6028(00)00457-X} {\bibfield
  {journal} {\bibinfo  {journal} {Surf. Sci.}\ }\textbf {\bibinfo {volume}
  {459}},\ \bibinfo {pages} {287} (\bibinfo {year} {2000})}\BibitemShut
  {NoStop}%
\bibitem [{\citenamefont {{See Supplemental Material at [URL will be inserted
  by publisher] for additional calculations results.}}()}]{Supplement}%
  \BibitemOpen
  \bibfield  {author} {\bibinfo {author} {\bibnamefont {{See Supplemental
  Material at [URL will be inserted by publisher] for additional calculations
  results.}}},\ }\href@noop {} {\ }\BibitemShut {NoStop}%
\bibitem [{\citenamefont {Xiang}\ \emph {et~al.}(2013)\citenamefont {Xiang},
  \citenamefont {Lee}, \citenamefont {Koo}, \citenamefont {Gong},\ and\
  \citenamefont {Whangbo}}]{Method-Energy-Mapping}%
  \BibitemOpen
  \bibfield  {author} {\bibinfo {author} {\bibfnamefont {H.}~\bibnamefont
  {Xiang}}, \bibinfo {author} {\bibfnamefont {C.}~\bibnamefont {Lee}}, \bibinfo
  {author} {\bibfnamefont {H.-J.}\ \bibnamefont {Koo}}, \bibinfo {author}
  {\bibfnamefont {X.}~\bibnamefont {Gong}},\ and\ \bibinfo {author}
  {\bibfnamefont {M.-H.}\ \bibnamefont {Whangbo}},\ }\bibinfo {title} {Magnetic
  properties and energy-mapping analysis},\ \href
  {https://doi.org/10.1039/C2DT31662E} {\bibfield  {journal} {\bibinfo
  {journal} {Dalton Trans.}\ }\textbf {\bibinfo {volume} {42}},\ \bibinfo
  {pages} {823} (\bibinfo {year} {2013})}\BibitemShut {NoStop}%
\bibitem [{\citenamefont {Anderson}(1950)}]{Model-Anderson-Superexchange-0}%
  \BibitemOpen
  \bibfield  {author} {\bibinfo {author} {\bibfnamefont {P.~W.}\ \bibnamefont
  {Anderson}},\ }\bibinfo {title} {Antiferromagnetism. Theory of superexchange
  interaction},\ \href {https://doi.org/10.1103/PhysRev.79.350} {\bibfield
  {journal} {\bibinfo  {journal} {Phys. Rev.}\ }\textbf {\bibinfo {volume}
  {79}},\ \bibinfo {pages} {350} (\bibinfo {year} {1950})}\BibitemShut
  {NoStop}%
\bibitem [{\citenamefont {Anderson}(1959)}]{Model-Anderson-Superexchange}%
  \BibitemOpen
  \bibfield  {author} {\bibinfo {author} {\bibfnamefont {P.~W.}\ \bibnamefont
  {Anderson}},\ }\bibinfo {title} {New approach to the theory of superexchange
  interactions},\ \href {https://doi.org/10.1103/PhysRev.115.2} {\bibfield
  {journal} {\bibinfo  {journal} {Phys. Rev.}\ }\textbf {\bibinfo {volume}
  {115}},\ \bibinfo {pages} {2} (\bibinfo {year} {1959})}\BibitemShut {NoStop}%
\bibitem [{\citenamefont {Goodenough}(1955)}]{Model-Anderson-Superexchange-2}%
  \BibitemOpen
  \bibfield  {author} {\bibinfo {author} {\bibfnamefont {J.~B.}\ \bibnamefont
  {Goodenough}},\ }\bibinfo {title} {Theory of the role of covalence in the
  perovskite-type manganites [La, M(II)]MnO$_{\text{3}}$},\ \href
  {https://doi.org/10.1103/PhysRev.100.564} {\bibfield  {journal} {\bibinfo
  {journal} {Phys. Rev.}\ }\textbf {\bibinfo {volume} {100}},\ \bibinfo {pages}
  {564} (\bibinfo {year} {1955})}\BibitemShut {NoStop}%
\bibitem [{\citenamefont {Kanamori}(1960)}]{Model-Anderson-Superexchange-3}%
  \BibitemOpen
  \bibfield  {author} {\bibinfo {author} {\bibfnamefont {J.}~\bibnamefont
  {Kanamori}},\ }\bibinfo {title} {Crystal distortion in magnetic compounds},\
  \href {https://doi.org/10.1063/1.1984590} {\bibfield  {journal} {\bibinfo
  {journal} {J. Appl. Phys.}\ }\textbf {\bibinfo {volume} {31}},\ \bibinfo
  {pages} {S14} (\bibinfo {year} {1960})}\BibitemShut {NoStop}%
\bibitem [{\citenamefont {Qian}\ and\ \citenamefont
  {Dai}(2017)}]{Qian2017Ferromagnetism}%
  \BibitemOpen
  \bibfield  {author} {\bibinfo {author} {\bibfnamefont {K.-M.}\ \bibnamefont
  {Qian}}\ and\ \bibinfo {author} {\bibfnamefont {D.-S.}\ \bibnamefont {Dai}},\
  }\href@noop {} {\bibinfo {title} {Ferromagnetism: Volume I}}\ (\bibinfo
  {publisher} {Science Press},\ \bibinfo {address} {Beijing},\ \bibinfo {year}
  {2017})\ pp.\ \bibinfo {pages} {193--198}\BibitemShut {NoStop}%
\bibitem [{\citenamefont {Varshalovich}\ \emph {et~al.}(1988)\citenamefont
  {Varshalovich}, \citenamefont {Moskalev},\ and\ \citenamefont
  {Khersonskii}}]{Wigner9j}%
  \BibitemOpen
  \bibfield  {author} {\bibinfo {author} {\bibfnamefont {D.~A.}\ \bibnamefont
  {Varshalovich}}, \bibinfo {author} {\bibfnamefont {A.~N.}\ \bibnamefont
  {Moskalev}},\ and\ \bibinfo {author} {\bibfnamefont {V.~K.}\ \bibnamefont
  {Khersonskii}},\ }\href {https://doi.org/10.1142/0270} {\bibinfo {title}
  {Quantum theory of angular momentum}}\ (\bibinfo  {publisher} {World
  Scientific},\ \bibinfo {year} {1988})\BibitemShut {NoStop}%
\bibitem [{\citenamefont {Schrieffer}\ and\ \citenamefont
  {Wolff}(1966)}]{Model-Anderson-SW}%
  \BibitemOpen
  \bibfield  {author} {\bibinfo {author} {\bibfnamefont {J.~R.}\ \bibnamefont
  {Schrieffer}}\ and\ \bibinfo {author} {\bibfnamefont {P.~A.}\ \bibnamefont
  {Wolff}},\ }\bibinfo {title} {Relation between the Anderson and Kondo
  Hamiltonians},\ \href {https://doi.org/10.1103/PhysRev.149.491} {\bibfield
  {journal} {\bibinfo  {journal} {Phys. Rev.}\ }\textbf {\bibinfo {volume}
  {149}},\ \bibinfo {pages} {491} (\bibinfo {year} {1966})}\BibitemShut
  {NoStop}%
\end{thebibliography}

\end{document}